\documentclass[3p,times]{elsarticle}

\usepackage{amssymb}                       
\usepackage{amsmath}                       
\usepackage{amsthm}                        
\usepackage{listings}                      
\usepackage[cal=boondox]{mathalfa} 
\usepackage{lineno,hyperref}
\modulolinenumbers[5]

\usepackage[table]{xcolor}
\usepackage{algorithm}  
\usepackage{algpseudocode} 
\usepackage{mathrsfs}
\usepackage{textcomp}
\usepackage{gensymb}
\usepackage{float}
\usepackage{graphicx}
\usepackage{subcaption}
\usepackage{physics} 

\journal{Computer Physics Communications}

\newcommand{\be}{\begin{equation}}
\newcommand{\ee}{\end{equation}}
\newcommand{\bi}{\begin{itemize}}
\newcommand{\ei}{\end{itemize}}

\newcommand{\beq}{\begin{equation}}
\newcommand{\eeq}{\end{equation}}

\begin{document}

\begin{frontmatter}

\title{Implicit High-Order Flux Reconstruction Solver for High-Speed Compressible Flows}

\author[vkiaddress]{Ray Vandenhoeck\corref{mycorrespondingauthor}}
\cortext[mycorrespondingauthor]{Corresponding author}
\ead{ray.vandenhoeck@kuleuven.be}

\author[kuladdress]{Andrea Lani}
\ead{andrea.lani@kuleuven.be}

\address[vkiaddress]{Department of Mechanical Engineering, KU Leuven, Celestijnenlaan 300, 3001, Leuven, Belgium;\\ Aeronautics \& Aerospace Department, Von Karman Institute for Fluid Dynamics, Waterloosesteenweg 72, 1640 Sint-Genesius-Rode, Belgium}
\address[kuladdress]{Centre for Mathematical Plasma Astrophysics, KU Leuven, Celestijnenlaan 200B, 3001, Leuven, Belgium;\\ Aeronautics \& Aerospace Department, Von Karman Institute for Fluid Dynamics, Waterloosesteenweg 72, 1640 Sint-Genesius-Rode, Belgium} 

\begin{abstract}
The present paper addresses the development and implementation of the first high-order Flux Reconstruction (FR) solver for high-speed flows within the open-source COOLFluiD (Computational Object-Oriented Libraries for Fluid Dynamics) platform. The resulting solver is fully implicit and able to simulate compressible flow problems governed by either the Euler or the Navier-Stokes equations in two and three dimensions. Furthermore, it can run in parallel on multiple CPU-cores and is designed to handle unstructured grids consisting of both straight and curved edged quadrilateral or hexahedral elements. While most of the implementation relies on state-of-the-art FR algorithms, an improved and more case-independent shock capturing scheme has been developed in order to tackle the first viscous hypersonic simulations using the FR method. Extensive verification of the FR solver has been performed through the use of reproducible benchmark test cases with flow speeds ranging from subsonic to hypersonic, up to Mach 17.6. The obtained results have been favorably compared to those available in literature. Furthermore, so-called ``super-accuracy'' is retrieved for certain cases when solving the Euler equations. The strengths of the FR solver in terms of computational accuracy per degree of freedom are also illustrated. Finally, the influence of the characterizing parameters of the FR method as well as the the influence of the novel shock capturing scheme on the accuracy of the developed solver is discussed. 
\end{abstract}


\end{frontmatter}


\section*{Introduction}
The field of Computational Fluid Dynamics (CFD) has been progressing significantly in recent years as a result of the need for increasingly complex simulations in a multitude of fields such as aerospace, automotive, wind engineering, atmospheric flows, etc. for a broad range of industrial and scientific applications. Depending on the spatial discretization of the Partial Differential Equations (PDEs) describing flow dynamics, the numerical methods developed in this field can be categorized into Finite Volume (FV), Finite Difference (FD), and Finite Element (FE) schemes. For each family of schemes, commercial codes used in the industry have largely been based on low-order methods which provide a second-order accuracy in space \cite{sheshadri2016analysis}. Despite being generally robust, intuitive, and reliable, these codes have proven to be insufficiently accurate when applied to complex problems as encountered in the domains of turbulence, aero-acoustics, plasma flows, etc. As a result of their high numerical dissipation, low-order methods are not suitable for problems dealing with the propagation of waves and vortex-dominated flows, as encountered for example around helicopters.

High-order methods can lead to considerably more accurate solutions for a comparable computational cost, i.e. for approximately the same number of degrees of freedom. Besides being computationally more efficient in terms of order of accuracy per degree of freedom, they also provide a lower numerical dissipation \cite{sheshadri2016analysis}. However, high-order methods are more complicated to implement as compared to low-order methods and are often less robust. The fact that robust high-order mesh generators are not readily available has also limited their use for industrial applications \cite{vincent2011facilitating}. Nonetheless, high-order methods have drawn considerable attention among researchers over the past couple of years due to their higher accuracy and their capability to deal with complex geometries \cite{huynh2014high}. Furthermore, the generation of high-order curved meshes has become increasingly supported by mesh generators such as the open-source Gmsh \cite{geuzaine2009gmsh}. This has resulted in several successful attempts to develop high-order extensions of the existing FV, FD and FE schemes.

FD methods have been extended to higher orders in a straightforward manner by widening the stencil. In this regard, Compact Finite Difference \cite{lele1992compact} schemes have proven to provide higher orders of accuracy. Nonetheless, all FD methods exhibit the same weakness: they are incapable to readily deal with complex geometries, unless utilized within an overset framework which complicates mesh generation \cite{pulliam2011high,henshaw2006moving,lani2013variable}. FV methods, on the other hand, do not have this shortcoming. As a consequence, several high-order extensions have been developed, such as the $k$-Exact method \cite{barth1990higher,barth2013high,delanaye1999quadratic}, the Essential Non-Oscillatory (ENO) \cite{harten1987uniformly,abgrall1994essentially,ollivier1997quasi}, and Weighted Essential Non-Oscillatory (WENO) \cite{liu1994weighted,hu1999weighted,friedrich1998weighted} schemes. These methods do not generally use a compact stencil and their utility for capturing complex fluid flow phenomena remains limited as the computational complexity rapidly increases at higher orders.

Due to the drawbacks of the high-order FV and FD extensions, several FE type methods that can attain arbitrarily high orders of accuracy on unstructured grids have been developed. In this respect, the Discontinuous Galerkin (DG) method was first proposed in 1973 by Reed and Hill \cite{reed1973triangularmesh}. Various extensions to the DG method have been proposed, such as Local Discontinuous Galerkin (LDG) \cite{cockburn1998local}, Compact Discontinuous Galerkin (CDG) \cite{peraire2008compact} and Hybrid Discontinuous Galerkin (HDG) \cite{cockburn2009unified}.

Another new high-order FE method is the so-called Spectral Difference (SD) method which was originally introduced by Kopriva and Kolias \cite{kopriva1996conservative} and has been expanded upon in several works such as \cite{liu2006spectral,liang2009spectral,liang2009large}. The main difference with the DG method is that the SD method solves the strong form of the PDE, whereas the DG method solves the weak or variational form.

Among the various developed methods, the Flux Reconstruction (FR) or Correction Procedure via Reconstruction (CPR) formulation \cite{huynh2007flux,huynh2009reconstruction} is one of the most recent and promising family of schemes. The FR method was first proposed by Huynh in 2007 \cite{huynh2007flux}. This approach provides a unifying framework for several existing high-order schemes, such as SD and DG, while being simpler and more computationally efficient than the original versions \cite{huynh2013flux}. The relation between FR and DG has been studied thoroughly by Allaneau and Jameson \cite{allaneau2011connections}, De Grazia et al. \cite{de2014connections}, and Zwanenburg et al. \cite{zwanenburg2016equivalence}. Furthermore, the FR approach is naturally adaptable to HPC architectures such as Graphical Processing Units (GPUs). Examples of massively parallel GPU implementations of the FR approach are given by Manuel et al. \cite{lopez2014verification}, Vincent et al. \cite{vincent2015pyfr}, and Witherden et al. \cite{witherden2015heterogeneous}. The FR methods have been extended to work on any element types, including simplex element types in 2D and 3D \cite{huynh2011high,castonguay2012new}, and are thus suitable to handle arbitrarily complex geometries.

The FR formulation proposed by Huynh is applicable to 1D advection problems and can be extended to quadrilateral elements in 2D through tensor products of the one-dimensional base and correction functions \cite{huynh2007flux}. In 2009, Huynh extended the FR method to diffusion problems. In this case, the flux also depends on the gradient of the solution, and the reconstruction procedure is applied to both the solution and the flux \cite{huynh2009reconstruction}.

Gao and Wang proposed the Lifting Collocation Penalty (LCP) method for advection-diffusion problems in 2009 \cite{gao2009high,wang2009unifying}. This class of schemes is closely related to FR and the two methods are proven to be identical for certain cases in 1D \cite{yu2013connection}. The term Correction Procedure via Reconstruction (CPR) is used to refer to all LCP and FR schemes \cite{yu2013connection,haga2011high}. Furthermore, Haga, Gao and Wang have successfully applied the LCP schemes to non-linear advection-diffusion problems on meshes of quadrilateral and triangular elements in 2D \cite{gao2009high}, and prisms and tetrahedra in 3D \cite{haga2010high,haga2011high}. However, it remains difficult to identify stable LCP schemes, while this is more trivial for the FR approach \cite{castonguay2012new}. In 2011, Huynh proposed a method to extend FR to deal with advection-diffusion problems on triangles \cite{huynh2011high}.

In 2011, Vincent, Castonguay, and Jameson proposed a class of FR schemes for linear advection problems in 1D, called Vincent-Castonguay-Jameson-Huynh (VCJH) schemes \cite{vincent2011new}. This method is characterized by a single parameter that determines the analytical form of the correction functions. Vincent, Castonguay, and Jameson proved the stability of the 1D VCJH schemes for linear advection using a similar energy method to the one Jameson used in \cite{jameson2010proof} to prove the stability of an SD scheme for linear advection. As such, VCJH schemes are also referred to as Energy Stable Flux Reconstruction schemes (ESFR). Castonguay, Williams,  Vincent, Lopez, and Jameson developed a multi-GPU enabled FR solver using the VCJH scheme, which is presented in \cite{castonguay2011development}.

Vincent, Castonguay, and Jameson extended the FR method and VCJH schemes to triangular elements in \cite{castonguay2012new} and proved the energy stability of these schemes. In \cite{castonguay2011application}, VCJH schemes are extended to non-linear advection and used to solve the 2D Euler equations for meshes of simplex elements in a number of test cases. ESFR schemes are applied to the diffusion equations in \cite{ou2011high}. In \cite{williams2011extension}, the VCJH schemes are extended to deal with non-linear advection-diffusion problems for 2D mixed grids.

The present paper treats a novel FR solver that is fully implicit and able to handle hypersonic viscous test cases. To the authors' knowledge, this is the first FR solver for hypersonic compressible flows. A novel shock capturing scheme for FR is introduced. The solver was implemented within COOLFluiD\footnote{https://github.com/andrealani/COOLFluiD/wiki}, a world-class open-source framework for multi-physics modeling and simulations where different numerical techniques, physical models, post-processing algorithms can coexist and work together \cite{lani13,lani2006reusable,lani1}.

Most shock capturing schemes for high-order FE-type methods that are currently being studied, are either adapted to explicit time stepping methods or unable to handle hypersonic viscous flows. This imposes a severe limit to the Courant-Friedrichs-Lewy (CFL) number and gives rise to a need for a large amount of iterations and computational time in order to reach convergence. In order to alleviate this issue, the present shock capturing scheme was successfully implemented for both explicit and implicit time stepping methods and verified in hypersonic test cases. 

After giving an overview of the governing equations and the FR formulation in respectively section \ref{GoverningEqs} and \ref{sec:FR}, the proposed shock capturing scheme is presented in section \ref{sec:SC}. Subsequently the implementation of the solver is discussed in section \ref{sec:impl}. Finally, section \ref{verification} presents subsonic and hypersonic verification test cases both in inviscid and viscous regimes.

\section{Governing Equations}\label{GoverningEqs}
Compressible flows are described by the Navier-Stokes equations. This set of equations is derived from the basic conservation laws: conservation of mass, momentum and energy where the fluid is assumed to be a continuum. This results in the following system of governing equations for three-dimensional flows:
\begin{subequations}
\begin{align}
    \text{continuity:} & \qquad \frac{\partial \rho}{\partial t} + \div (\rho \mathbf{v}) = 0,
    \label{eq:Continuity}
    \\
    \text{x-momentum:} & \qquad \frac{\partial (\rho v_x)}{\partial t} + \div (\rho v_x \mathbf{v}) = \frac{\partial (\sigma_{xx} - p)}{\partial x} + \frac{\partial \tau_{yx}}{\partial y} + \frac{\partial \tau_{zx}}{\partial z},
    \label{eq:x-momentum}
    \\
    \text{y-momentum:} & \qquad \frac{\partial (\rho v_y)}{\partial t} + \div (\rho v_y \mathbf{v}) = \frac{\partial \tau_{xy}}{\partial x} + \frac{\partial (\sigma_{yy}-p)}{\partial y} + \frac{\partial \tau_{zy}}{\partial z},
    \label{eq:y-momentum}
    \\
    \text{z-momentum:} & \qquad \frac{\partial (\rho v_z)}{\partial t} + \div (\rho v_z \mathbf{v}) = \frac{\partial \tau_{xz}}{\partial x} + \frac{\partial \tau_{yz}}{\partial y} + \frac{\partial (\sigma_{zz}-p)}{\partial z},
    \label{eq:z-momentum}
    \\
    \begin{split}
    \text{energy:} & \qquad \frac{\partial (\rho e_t)}{\partial t} + \div (\rho e_t \mathbf{v}) = -\div (p \mathbf{v}) + \Bigg( \frac{\partial (v_x \sigma_{xx})}{\partial x} + \frac{\partial (v_x \tau_{yx})}{\partial y} \\
    & \qquad + \frac{\partial (v_x \tau_{zx})}{\partial z} + \frac{\partial (v_y \tau_{xy})}{\partial x} + \frac{\partial (v_y \sigma_{yy})}{\partial y} + \frac{\partial (v_y \tau_{zy})}{\partial z} + \frac{\partial (v_z \tau_{xz})}{\partial x} \\
    & \qquad + \frac{\partial (v_z \tau_{yz})}{\partial y} + \frac{\partial (v_z \sigma_{zz})}{\partial z} \Bigg) - \div \mathbcal{q},
    \label{eq:energy}
    \end{split}
\end{align}
\label{eq:NSEquations}
\end{subequations}
where $\sigma_{ii}$ denotes the normal viscous stress in the $i$-direction, and $\tau_{ij}$ denotes the shear viscous stress in the $j$-direction on a plane normal to the $i$-direction. $\rho$ denotes the density, $p$ the pressure and $\mathbf{v}$ the velocity vector. The heat flux vector $\mathbcal{q}$ is linearly proportional to the local temperature gradient as imposed by Fourier's law of heat conduction:
\begin{equation}
    \mathbcal{q} = -\kappa \grad T,
\end{equation}
where $\kappa$ represents the coefficient of thermal conductivity. Equations \ref{eq:Continuity} through \ref{eq:energy} are commonly known as the Navier-Stokes equations. When there is no viscosity and heat conduction present in the fluid, its flow is governed by the Euler equations. This implies that all viscous stresses $\sigma_{ii}$ and $\tau_{ij}$, as well as the thermal conductivity $\kappa$, are zero. 

Assuming a Newtonian fluid, the viscous stresses are expressed as a function of the rates of deformation:
\begin{equation}
    \sigma_{xx}=2\mu \frac{\partial v_x}{\partial x} + \lambda \div \mathbf{v}, \qquad \sigma_{yy}=2\mu \frac{\partial v_y}{\partial y} + \lambda \div \mathbf{v}, \qquad \sigma_{zz}=2\mu \frac{\partial v_z}{\partial z} + \lambda \div \mathbf{v},
\end{equation}
\begin{equation}
    \tau_{xy}=\tau_{yx}= \mu \left( \frac{\partial v_x}{\partial y} + \frac{\partial v_y}{\partial x} \right), \quad \tau_{xz}=\tau_{zx}= \mu \left( \frac{\partial v_x}{\partial z} + \frac{\partial v_z}{\partial x} \right),\quad \tau_{yz}=\tau_{zy}= \mu \left( \frac{\partial v_y}{\partial z} + \frac{\partial v_z}{\partial y} \right). 
\end{equation}
For gases, it is common practice to adopt Stokes' hypothesis, which states that $\lambda = -\frac{2}{3}\mu$.

\subsection{Equations of State}
The dynamics of three-dimensional flow are described by a system (\ref{eq:NSEquations}) of five partial differential equations with five unknowns, which can be expressed for instance as the conservative variables $(\rho,\rho v_x, \rho v_y, \rho v_z, \rho e_t)$. In order to evaluate the pressure $p$ and temperature $T$ based on these variables, two equations of state are considered. These relations can be obtained by assuming that the fluid is in thermodynamic equilibrium. By adopting the ideal gas model, which is valid for air over a wide range of thermodynamic conditions, the temperature and pressure are expressed as a function of the other unknown variables:
\begin{align}
    p &= (\gamma - 1) \rho \left(e_t-\frac{1}{2}(v_x^2+v_y^2+v_z^2)\right), \label{pressure}\\
    T &= \frac{p}{R \rho} = \frac{(\gamma - 1)}{R}\left(e_t-\frac{1}{2}(v_x^2+v_y^2+v_z^2)\right),
\end{align}
where $R$ is the specific ideal gas constant.

\subsection{Formulation in Terms of Convective and Diffusive Fluxes}

The system of equations \ref{eq:NSEquations} can be rewritten in the compact form, called the conservative form:
\begin{equation}\label{cons law}
    \frac{\partial \mathbf{u}}{\partial t} + \div \mathbf{f}(\mathbf{u},\grad\mathbf{u}) = \frac{\partial \mathbf{u}}{\partial t} + \div \mathbf{f_C}(\mathbf{u}) - \div \mathbf{f_D}(\mathbf{u}, \grad\mathbf{u})=0,
\end{equation}
where $\mathbf{u} = (\rho,\ \rho v_x,\ \rho v_y,\ \rho v_z,\ \rho e_t)^T$ denotes the set of conserved variables and $\mathbf{f}(\mathbf{u})$ the set of total flux vectors. Consequently $\div \mathbf{f}(\mathbf{u})$ denotes the set of the divergences of the flux vectors. The flux vectors can be split up into a convective flux $\mathbf{f}_{C}(\mathbf{u})$, and a diffusive flux $\mathbf{f}_{D}(\mathbf{u},\grad \mathbf{u})$. The convective and the diffusive vectors are defined as $\mathbf{f}_{C}(\mathbf{u}) = (\mathbf{f}_{x,C};\ \mathbf{f}_{y,C};\ \mathbf{f}_{z,C})^T$ and $\mathbf{f}_{D}(\mathbf{u}, \grad \mathbf{u}) = (\mathbf{f}_{x,D};\ \mathbf{f}_{y,D};\ \mathbf{f}_{z,D})^T$:
\begin{equation}
    \mathbf{f}_{x,C} = 
    \begin{pmatrix}
    \rho v_x \\ \rho v_x^2 + p \\ \rho v_x v_y \\ \rho v_x v_z \\ v_x(\rho e_t + p)
    \end{pmatrix},
    \qquad
    \mathbf{f}_{y,C} = 
    \begin{pmatrix}
    \rho v_y \\ \rho v_x v_y \\ \rho v_y^2 + p \\ \rho v_y v_z \\ v_y (\rho e_t + p)
    \end{pmatrix},
    \qquad
    \mathbf{f}_{z,C} = 
    \begin{pmatrix}
    \rho v_z \\ \rho v_x v_z \\ \rho v_y v_z \\ \rho v_z^2 + p\\ v_z (\rho e_t + p)
    \end{pmatrix},
\end{equation}

\begin{equation}
    \mathbf{f}_{x,D} = 
    \begin{pmatrix}
    0 \\ \sigma_{xx} \\ \tau_{yx} \\ \tau_{zx} \\ v_x \sigma_{xx} + v_y \tau_{yx} + v_z \tau_{zx} - \mathcal{q}_x
    \end{pmatrix},
    \quad
    \mathbf{f}_{y,D} = 
    \begin{pmatrix}
    0 \\ \tau_{xy} \\ \sigma_{yy} \\ \tau_{zy} \\ v_x \tau_{xy} + v_y \sigma_{yy} + v_z \tau_{zy} - \mathcal{q}_y
    \end{pmatrix},
    \quad
    \mathbf{f}_{z,D} = 
    \begin{pmatrix}
    0 \\ \tau_{xz} \\ \tau_{yz} \\ \sigma_{zz} \\ v_x \tau_{xz} + v_y \tau_{yz} + v_z \sigma_{zz} - \mathcal{q}_z
    \end{pmatrix}.
\end{equation}

\section{Flux Reconstruction Formulation}\label{sec:FR}

This section presents a concise review of the FR method in one dimension equivalent to the method presented in \cite{vincent2011new,castonguay2011development}. Consider solving the one-dimensional advection-diffusion problem on an arbitrary domain $\Omega$, given by:
\begin{equation}
    \frac{\partial u}{\partial t} + \frac{\partial f}{\partial x} = 0 \quad \text{with} \quad x \in \Omega = [x_L,x_R] \quad \text{and} \quad f = f\left(u,\frac{\partial u}{\partial x}\right).
    \label{eq:ConservationLaw}
\end{equation}
In equation \ref{eq:ConservationLaw}, $x$ represents the one-dimensional coordinate, $t$ is the time variable, $u(x,t)$ is a conservative variable, and $f$ is the flux which is a scalar for the 1D case. The Euler and Navier-Stokes equations can be written as a system of conservation equations, equivalent to expression \ref{eq:ConservationLaw}. The FR method is classified as a FE-type method. Hence, the first step of the procedure involves partitioning the spatial domain $\Omega$ into a finite number $N$ of non-overlapping, non-empty, open sub-domains $\Omega_n'$ such that:
\begin{equation}
     \Omega_n' = \{x\ |\ x_n < x < x_{n+1}\} \quad \text{with} \quad x_1 = x_L \quad \text{and} \quad x_{N+1} = x_R.
\end{equation}
The boundary of each sub-domain consists of the two end-points and is denoted by $\Gamma_n = \{x_n, x_{n+1}\}$. The closed domain corresponding to the union of the open sub-domain $\Omega_n'$ and its boundary $\Gamma_n$ is called an element and is denoted by  $\Omega_n = \Omega_n' \cup \Gamma_n $ such that:
\begin{equation}
    \bigcup\limits_{n=1}^{N} \Omega_n = \Omega
    \quad \text{and} \quad
    \bigcap\limits_{n=1}^{N} \Omega_n = \emptyset.
    \label{partition}
\end{equation}
The exact solution $u$ of the conservation law \ref{eq:ConservationLaw} is approximated by a function $u_n^\delta$ within each element $\Omega_n$. This function $u_n^\delta$ corresponds to a polynomial of degree $P$ within $\Omega_n$ and is identically zero on $\Omega \setminus \Omega_n$. Furthermore, it is important to note that these polynomials are generally discontinuous across elements. The approximate solution polynomial $u^\delta$ within the entire domain $\Omega$ is constructed through summation of the elemental solutions $u_n^\delta$. In the same manner, the exact flux $f$ is approximated by a polynomial of degree $P+1$ within each sub-domain $\Omega_n$. This elemental approximate flux is denoted by $f_n^\delta$ and is identically zero outside of $\Omega_n$. The  overall approximate flux $f^\delta$ in the entire domain $\Omega$ is given by the sum of the elemental approximate fluxes such that:
\begin{equation}
    u^\delta(x,t) = \sum_{n=1}^{N} u_n^\delta(x,t) \approx u(x,t) \quad \text{ and } \quad f^\delta(x,t) = \sum_{n=1}^{N} f_n^\delta(x,t) \approx f(x,t).
\end{equation}
Each element $\Omega_n$ together with the approximate solution $u_n^\delta$ and flux $f_n^\delta$ inside it, is transformed to a standard reference element $\Omega_S = \{\xi\ |\ -1 \leq \xi \leq 1 \}$. In this manner, the computations for all elements are done in the same reference domain $\Omega_S$. The transformation between $\Omega_n$ and $\Omega_S$ is carried out by means of a mapping function $\Theta_n(\xi)$. The solution $u_n^\delta$ within each sub-domain $\Omega_n$ can then be obtained by solving the transformed conservation equation \ref{eq:ConservationLaw} within the reference element $\Omega_S$:
\begin{equation}
    \frac{\partial \hat{u}^\delta}{\partial t} + \frac{\partial \hat{f}^\delta}{\partial \xi} = 0\quad \text{with} \quad \xi \in \Omega_S,
    \label{eq:ConservationLawTransform}
\end{equation}
with the transformed physical quantities given by:
\begin{equation}
    \hat{u}^\delta = \hat{u}^\delta(\xi,t) = J_n u_n^\delta(\Theta_n(\xi),t) 
    \quad \text{and} \quad
    \hat{f}^\delta = \hat{f}^\delta(\xi,t) = f_n^\delta(\Theta_n(\xi),t).
    \label{eq:VariablesTransform}
\end{equation}
In equation \ref{eq:VariablesTransform}, $J_n$ represents the Jacobian of the mapping $\Theta_n(\xi)$.

The FR framework for solving equation \ref{eq:ConservationLawTransform} within the reference element $\Omega_S$ consists of seven subsequent steps. In the first step, a specific form for the approximate solution $\hat{u}^{\delta}$ within $\Omega_S$ is defined. To this end, a set of $P + 1$ distinct solution points $\xi_i$ (with $i$ from 1 to $P+1$) are chosen within $\Omega_S$ at which the values of $\hat{u}^{\delta}$ are assumed to be known. The approximate solution $\hat{u}^{\delta}$ is then defined as a degree $P$ polynomial of the following form:
\begin{equation}
    \hat{u}^{\delta} = \sum_{i=1}^{P+1} \hat{u}_{i}^{\delta} \mathcal{l}_{i}(\xi),
    \label{ApproxSolution}
\end{equation}
where $\hat{u}_{i}^{\delta}$ is the value of $\hat{u}^{\delta}$ at the $i$-th solution point, and $\mathcal{l}_{i}(\xi)$ is the 1D Lagrange polynomial associated with the $i$-th solution point.

\begin{figure}[H]
  \centering
  \includegraphics[width=0.6\textwidth]{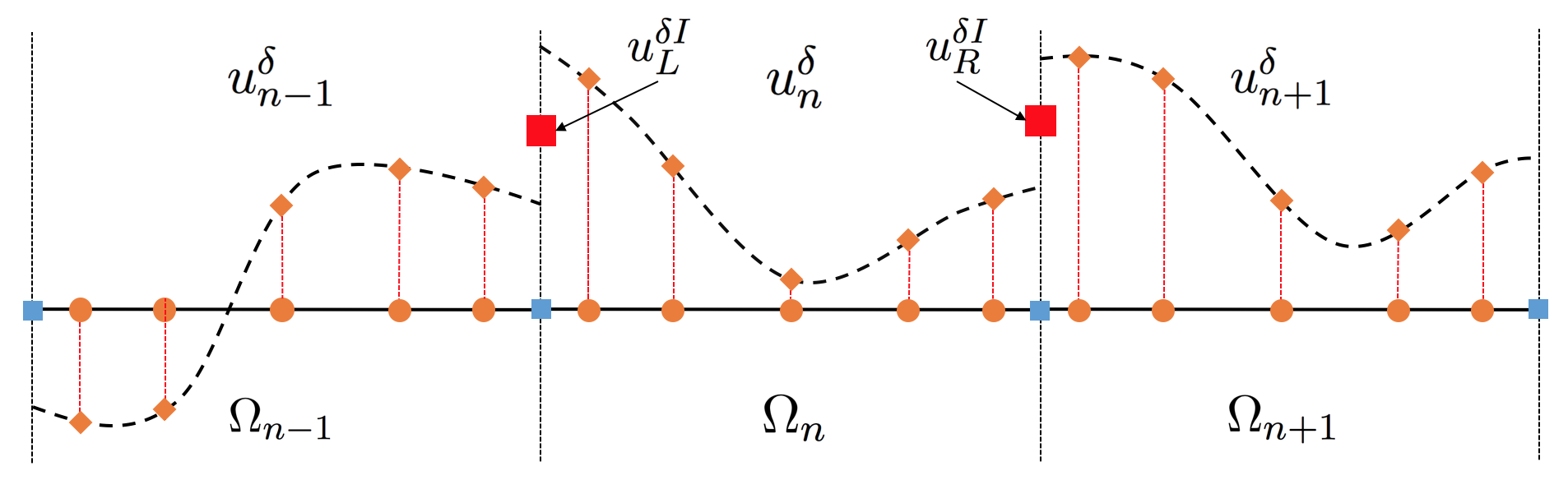}
  \caption{Computation of the common interface solutions $u_R^{\delta I}$ and $u_L^{\delta I}$ at the left and right boundary of $\Omega_n$}
  \label{fig:CommonInterfaceSolution}
\end{figure}

The second step consists of determining a common interface solution $\hat{u}^{\delta I}$ at the boundaries of the standard element $\Omega_S$, i.e. $\xi = \pm 1$. These two end-points of the standard element $\Omega_S$ are referred to as the flux points. As such, at the boundaries of the the standard element the flux points of two neighboring elements coincide. In order to compute the interface value $\hat{u}^{\delta I}$, the approximate solution $\hat{u}^{\delta}$ must be first evaluated in the flux points by means of equation \ref{ApproxSolution}. The approximate solutions evaluated at the left and right boundaries are denoted as $\hat{u}^{\delta}_L = \hat{u}^{\delta}(-1)$ and $\hat{u}^{\delta}_R = \hat{u}^{\delta}(1)$. Once this procedure has been performed for all the elements, the common interface solution $\hat{u}^{\delta I}$ can be computed at all flux points. Several schemes can be used to compute the interface solution, such as Bassi-Rebay 2 (BR2) \cite{bassi1997high} or simply an average of the left and right discontinuous solutions. The common interface solutions corresponding to the left and right ends of $\Omega_n$ are denoted by $u_{L}^{\delta I}$ and $u_{R}^{\delta I}$. Figure \ref{fig:CommonInterfaceSolution} illustrates the process of computing the common interface values.
The third step of the FR method consists of calculating a corrected solution gradient $\hat{q}^{\delta}$. The corrected solution gradient $\hat{q}^{\delta}$ is defined in terms of two correction functions $g_L(\xi)$ and $g_R(\xi)$ of degree $P+1$, which approximate the zero function within the reference element $\Omega_S$ and satisfy the following boundary conditions:
\begin{equation}\label{eq:h1}
    g_L(-1) = 1, \qquad g_R(-1) = 0, \qquad g_L(1) = 0, \qquad g_R(1) = 1.
\end{equation}
Furthermore, the correction function for the right boundary $g_R(\xi)$ is defined as the reflection of the correction function $g_L(\xi)$ with respect to the vertical axis in order to ensure symmetry of the correction process: $g_L(\xi) = g_R(-\xi)$.

\begin{figure}[H]
  \centering
  \includegraphics[width=0.3\textwidth]{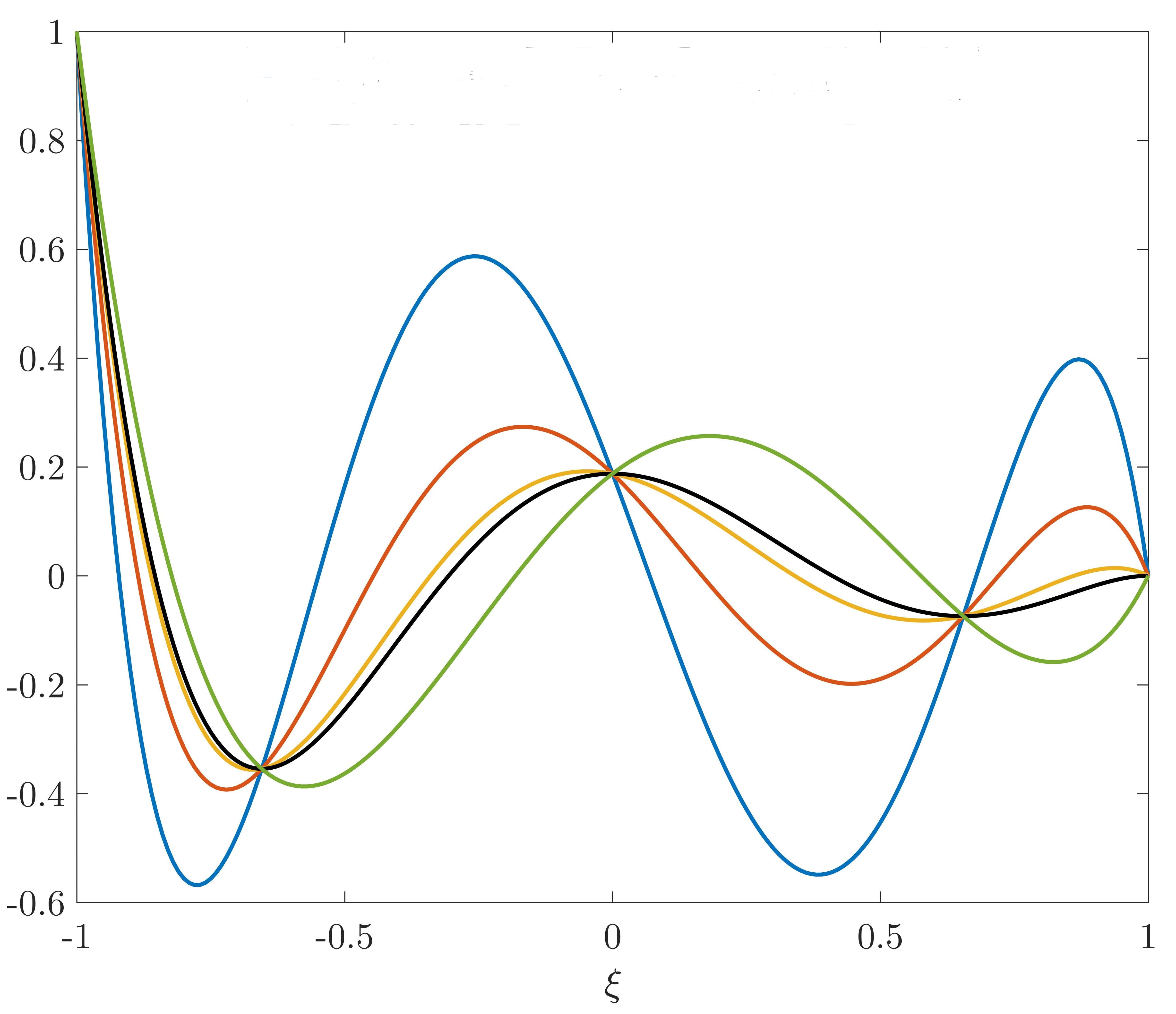}
  \caption{Left correction functions for different values of $c$ for $P=4$}
  \label{fig:corrFcts}
\end{figure}

The corrected solution gradient $\hat{q}^{\delta}$ is then computed via the following expression:
\begin{equation}
    \hat{q}^{\delta} = \frac{\partial \hat{u}^{\delta}}{\partial \xi} + (\hat{u}_{L}^{\delta I} - \hat{u}_{L}^{\delta})\frac{dg_L}{d\xi} + (\hat{u}_{R}^{\delta I} - \hat{u}_{R}^{\delta})\frac{dg_R}{d\xi},
    \label{CorrectedGrad}
\end{equation}
The VCJH scheme is used to define the exact polynomial expression of $g_L$ and $g_R$, as found in \cite{castonguay2011application}. Figure \ref{fig:corrFcts} presents the family of VCJH correction functions for $P=4$.
The fourth step consists of calculating the approximate discontinuous flux $\hat{f}^{\delta D}(\xi)$ on the reference element $\Omega_S$. To this end, the flux is evaluated at each solution point $\xi_i$ using the approximate solution $\hat{u}^\delta$, and the corrected solution gradient $\hat{q}^\delta$. The approximate discontinuous flux $\hat{f}^{\delta D}$ is constructed as a polynomial of degree $P$ in the same manner as the approximate solution $\hat{u}^{\delta}$ in equation \ref{ApproxSolution}.

The fifth step of the FR framework involves computing the transformed common interface fluxes $\hat{f}^{\delta I}$ at the flux points of the standard element $\Omega_S$, i.e. $\xi = \pm 1$. In order to compute these interface fluxes, it is necessary to first obtain the values of the approximate solution $\hat{u}^{\delta}$ and the corrected gradient $\hat{q}^{\delta}$ at the boundaries of the standard element by using equations \ref{ApproxSolution} and \ref{CorrectedGrad}, respectively. Once this procedure has been performed for the adjacent elements, the common interface flux can be computed. The common interface flux at the left and right end of the standard element $\Omega_S$ is denoted as $\hat{f}_{L}^{\delta I}$ and $\hat{f}_{R}^{\delta I}$, respectively. In the case of the Navier-Stokes equations, it is computed as the sum of a convective and a diffusive term. For the convective interface flux, an upwind biased approximate Riemann solver, like the Roe scheme \cite{roe1981approximate}, can be used. The diffusive interface flux is often calculated following the BR2 scheme \cite{bassi1997high}.

The flux must be continuous across element boundaries in order to obtain a conservative scheme. Therefore, the sixth stage of the FR framework consists of adding a correction flux function $\hat{f}^{\delta C}$ to the approximate discontinuous flux $\hat{f}^{\delta D}$ resulting in the continuous flux $\hat{f}^{\delta}$. The correction flux $\hat{f}^{\delta C}$ is a polynomial of degree $P+1$ defined in such a manner that the continuous flux $\hat{f}^{\delta}$ equals the common interface flux in the flux points, while approximating the discontinuous flux $\hat{f}^{\delta D}$ within the standard element $\Omega_S$. In order to satisfy the aforementioned requirements, $\hat{f}^{\delta C}$ is constructed by means of the degree $P+1$ correction functions $h_L(\xi)$ and $h_R(\xi)$. This correction procedure is performed within the standard element $\Omega_S$ as the correction functions are defined within this domain. The correction functions $h_L$ and $h_R$ are equivalent to the correction functions $g_L$ and $g_R$ used to define the corrected gradient $\hat{q}^{\delta}$ and have the same properties as expressed in \ref{eq:h1} .

\begin{figure}[H]
  \centering
  \includegraphics[width=0.3\textwidth]{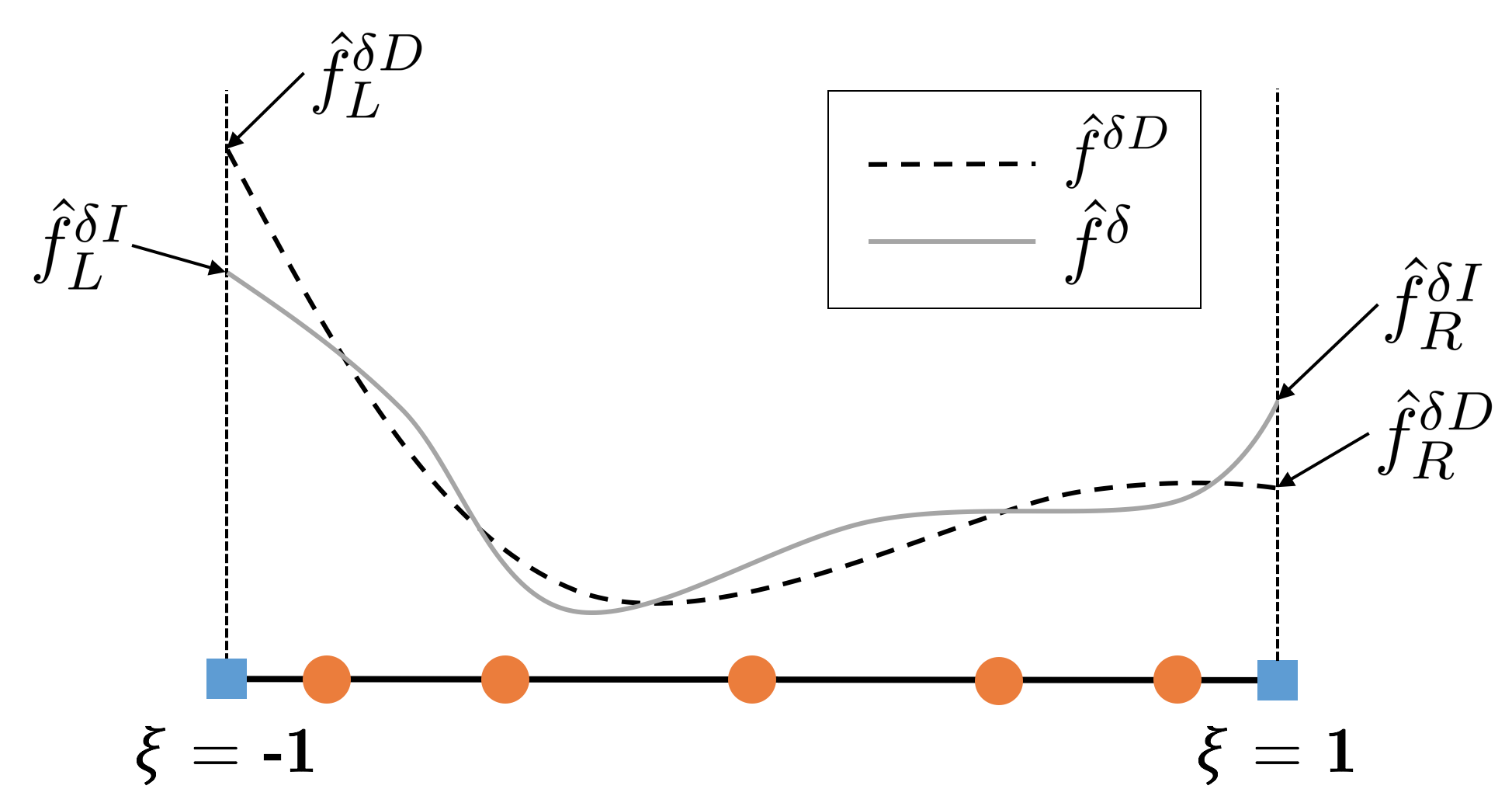}
  \caption{Correction procedure $\hat{f^{\delta D}}$ such that the values of the total flux $\hat{f}^{\delta}$ at the boundaries are equal to the common interface fluxes $\hat{f}^{\delta I}_L$ and $\hat{f}^{\delta I}_R$}.
  \label{fig:CorrectionDiscontFlux}
\end{figure}

The correction flux function $\hat{f}^{\delta C}$ takes the form:
\begin{equation}
    \hat{f}^{\delta C} = ( \hat{f}^{\delta I}_{L} - \hat{f}^{\delta D}_{L}) h_L(\xi) + ( \hat{f}^{\delta I}_{R} - \hat{f}^{\delta D}_{R}) h_R(\xi),
\end{equation}
where $\hat{f}^{\delta D}_{L} = \hat{f}^{\delta D}(-1)$ and $\hat{f}^{\delta D}_{R} = \hat{f}^{\delta D}(1)$ are the values of the approximate discontinuous flux at the left and right boundaries, evaluated from the Lagrange interpolation. The continuous flux $\hat{f}^{\delta}$ within the standard element $\Omega_S$ is then written as the sum of the approximated discontinuous flux and the correction flux: 
\begin{equation}
    \hat{f}^{\delta} = \hat{f}^{\delta D} + \hat{f}^{\delta C} = \hat{f}^{\delta D} + (\hat{f}^{\delta I}_{L} - \hat{f}^{\delta D}_{L}) h_L + ( \hat{f}^{\delta I}_{R} - \hat{f}^{\delta D}_{R}) h_R.
\end{equation}
A schematic of the flux correction procedure within the reference element $\Omega_S$ is given in figure \ref{fig:CorrectionDiscontFlux}.
The final step consists of computing the divergence of the total flux $\hat{f}^{\delta}$ at each solution point $\xi_i$, which is trivial as the derivatives of both the Lagrange polynomials and correction functions are exactly known. Finally, a time marching strategy such as a backward Euler method is then applied to advance the approximate solution $\hat{u}^{\delta}$ in time discretizing the following expression:
\begin{equation}
    \frac{d \hat{u}^{\delta}_i}{dt} = - \frac{\partial \hat{f}^{\delta}}{\partial \xi}(\xi_i).
\end{equation}

\subsection{FR in 2D and 3D}\label{sec:FR2D}

\begin{figure}[ht]
  \centering
  \includegraphics[width=0.4\textwidth]{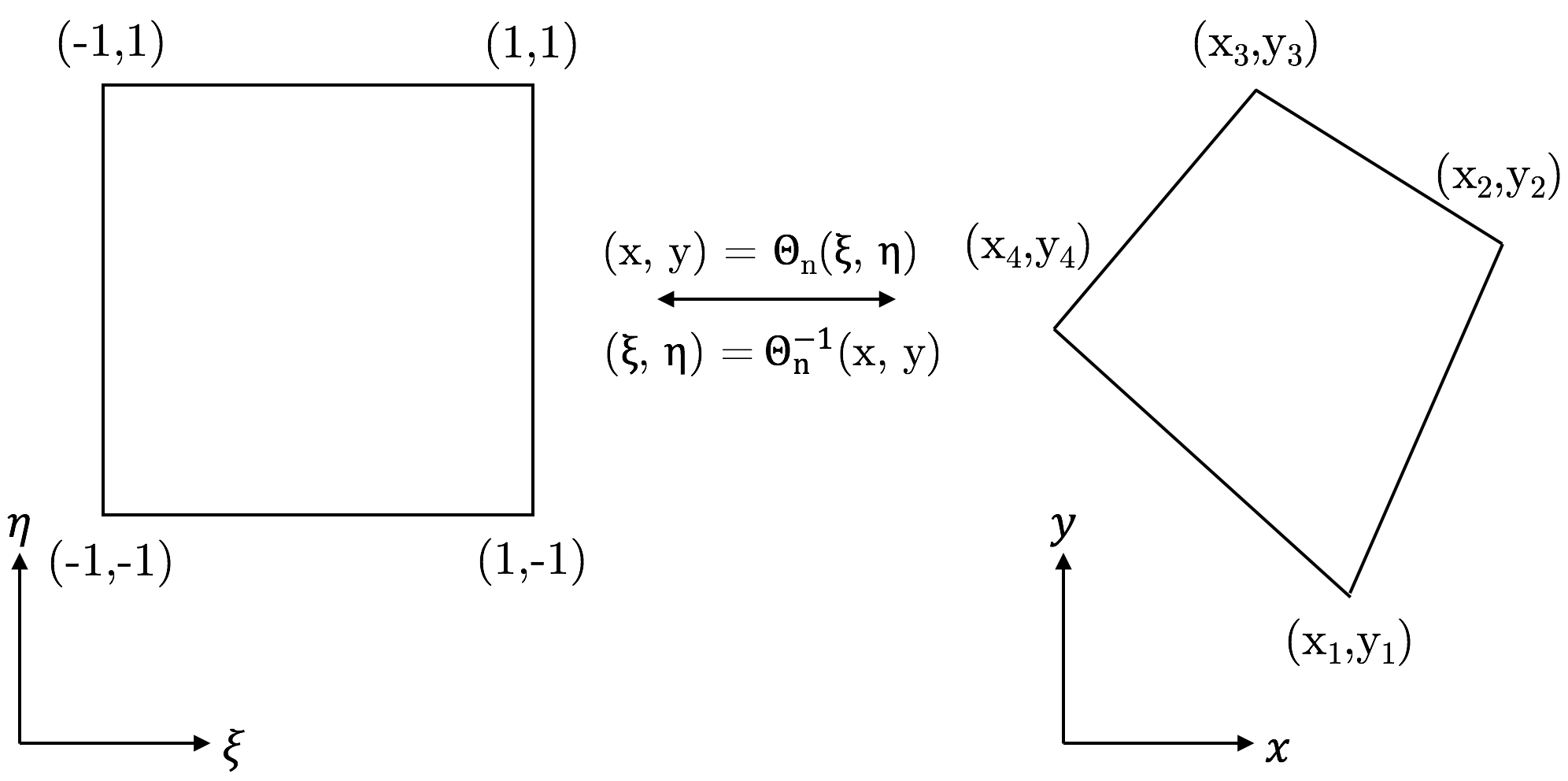}
  \caption{Mapping between the physical domain $(x,y)$ on the right and the reference element $(\xi,\eta)$ on the left}
  \label{fig:QuadrilateralMapping}
\end{figure}

As proposed by Huynh \cite{huynh2007flux}, the FR framework can easily be adapted to work on quadrilateral and hexahedral elements through the construction of a tensor product basis. A brief overview of the procedure for quadrilaterals is given. The method for hexahedra is analogous. Consider solving the 2D scalar conservation law within an arbitrary domain $\Omega$ given by:
\begin{equation}
    \frac{\partial u}{\partial t} + \boldsymbol{\nabla_{x}} \cdot \mathbf{f} = 0.
    \label{ConservationLaw2D}
\end{equation}
In equation \ref{ConservationLaw2D}, $\mathbf{f} = (f,g)$ represents the flux vector with $f(u,\boldsymbol{\nabla} u)$ and $g(u,\boldsymbol{\nabla} u)$ the fluxes in the $x$- and $y$-directions, respectively. Once again, the spatial domain $\Omega$ is partitioned into $N$ non-empty, non-overlapping, conforming quadrilateral elements $\Omega_n = \{x,y\ | \ x_n < x < x_{n+1} \ , \ y_n < y < y_{n+1}\}$ such that equation \ref{partition} is satisfied. Similar to the one-dimensional approach, each quadrilateral element in the physical domain $\Omega_n$ is mapped onto a standard element $\Omega_S = \{\xi,\eta \ | \ -1 \leq \xi \leq 1 \ , \ -1 \leq \eta \leq 1 \}$ in order to simplify the implementation. The mapping between the physical domain $\mathbf{x} = (x,y)$ and the computational domain $\boldsymbol{\xi} = (\xi,\eta)$ can be written in the form:
\begin{equation}
    \begin{pmatrix}
    x \\ y
    \end{pmatrix}
    = \mathbf{\Theta_n} (\xi,\eta) = \sum_{i=1}^{K}\mathbf{M}_i(\xi,\eta)
    \begin{pmatrix}
    x_i \\ y_i
    \end{pmatrix},
    \label{TransformationQuad}
\end{equation}
where $K$ is the number of points used to define the physical element, $(x_i,y_i)$ are the physical coordinates of these points, and $\mathbf{M}_i(\xi,\eta)$ are the shape functions defined on $\Omega_S$. A schematic representation of the mapping between the physical domain $\mathbf{x} = (x,y)$ and the reference domain $\boldsymbol{\xi} = (\xi,\eta)$ is given in figure \ref{fig:QuadrilateralMapping}. In this example, the four vertices $(x_i,y_i)$ are used to define the shape of the physical element, hence $K=4$.

\begin{figure}[ht]
  \centering
  \includegraphics[width=0.25\textwidth]{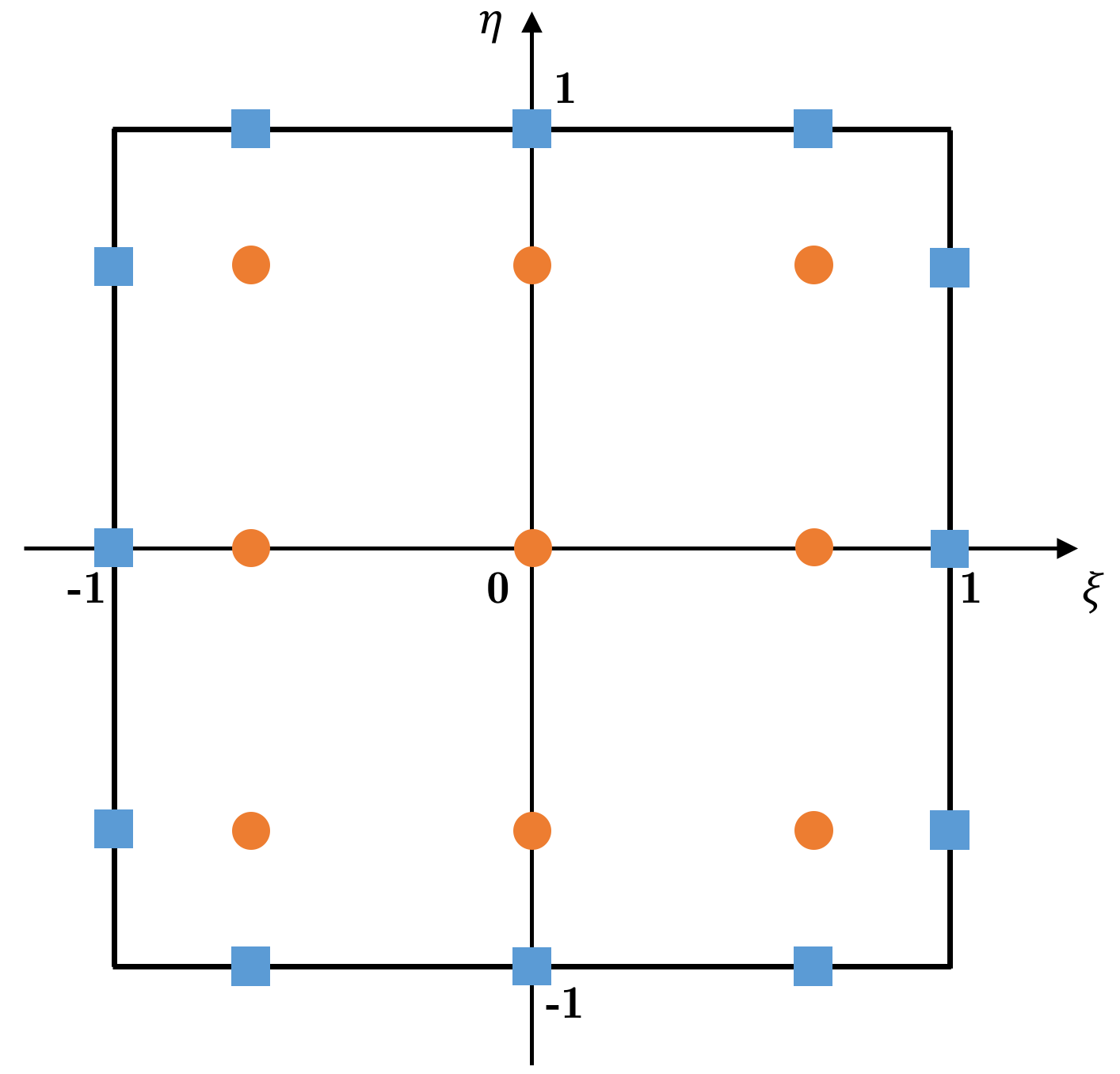}
  \caption{2D standard element $\Omega_S$ for $p=2$, solution points $\xi_{i}$ are represented by orange circles and flux points points by blue squares}
  \label{fig:StdElement2D}
\end{figure}

The Jacobian matrix of the mapping $\mathbf{\Theta_n}$ is denoted by $\mathbf{J_n}$ and the corresponding determinant by $J_n$. The Jacobian matrix $\mathbf{J}$ can be determined as follows:
\begin{equation}\label{jacobMapping}
    \mathbf{J} = \frac{\partial \boldsymbol{\Theta_n}}{\partial \boldsymbol{\xi}}.
\end{equation}
The solution $u^{\delta}_{n}$ within every sub-domain $\Omega_n$ is then found by solving the transformed version of the conservation equation \ref{ConservationLaw2D} within the standard element $\Omega_S$:
\begin{equation}
    \frac{\partial \hat{u}^{\delta}}{\partial t} + \mathbf{\nabla_{\boldsymbol{\xi}}} \cdot \mathbf{\hat{f}^{\delta}} = 0,\text{ where } \quad
    \begin{matrix}
    \hat{u}^{\delta}(\xi,\eta,t) &= J_n u^{\delta}_n (\Theta_n(\xi,\eta),t), \\
    \hat{\mathbf{f}}^{\delta}(\xi,\eta,t) &= (\hat{f}^{\delta}, \hat{g}^{\delta}) = J_n \mathbf{J}_n^{-1} \mathbf{f}^\delta.
    \end{matrix}
    \label{ConservationLaw2D_Transformed}
\end{equation}
$J_n$ and $\mathbf{J}_n^{-1}$ are determined from equation \ref{TransformationQuad} and \ref{jacobMapping}. An example of a quadrilateral standard element $\Omega_S$ for $P=2$ is given in figure \ref{fig:StdElement2D}. For quadrilateral elements, the number of solution points $N_s^{quad}$, represented by orange circles, is equal to $(P+1)^2$. A common choice for the solution points is the tensor product of the 1D Gauss-Legendre points. In the case of hexahedral elements, the number of solution points is $N_s^{hexa} = (P+1)^3$. Furthermore, a set of $P+1$ flux points, represented by blue squares, is distributed on each edge of the quadrilateral element. For hexahedral elements, $(P+1)^2$ flux points are distributed on each face. This results in a total number of flux points $N_f^{quad}=4(P+1)$ for quadrilateral elements, and $N_f^{hexa}=6(P+1)^2$ for hexahedral elements.

Similar to the one-dimensional approach, the approximate solution $\hat{u}^{\delta}$ within the reference element $\Omega_n$ is constructed as:
 \begin{equation}
    \hat{u}^{\delta} = \sum_{i=1}^{(P+1)} \sum_{j=1}^{(P+1)} \hat{u}_{i,j}^{\delta} \mathcal{l}_{i}(\xi) \mathcal{l}_{j}(\eta),
    \label{ApproxSolution_Quad}
\end{equation}
where $\hat{u}_{i,j}^{\delta}$ is the value of the approximate transformed solution $\hat{u}^{\delta}$ at the corresponding solution point $(\xi_{i},\eta_{j})$, and $\mathcal{l}_{i}(\xi)$ and $\mathcal{l}_{j}(\eta)$ are the 1D Lagrange polynomials associated with the solution points $\xi_{i}$ and $\eta_{j}$, respectively. By evaluating equation \ref{ApproxSolution_Quad} at the boundaries of $\Omega_n$, one can calculate the common interface solutions at the flux points in a similar way as was done for the one-dimensional approach. Once these values have been determined, the corrected gradient $\hat{\mathbf{q}}^{\delta} = (\hat{q}_{\xi},\hat{q}_{\eta})$ is constructed by using the one-dimensional correction functions $g_L$ and $g_R$. The $\xi$- and the $\eta$-components of the corrected gradient at each solution point are calculated independently as follows:
\begin{align}
    \hat{q}^{\delta}_{\xi}(\xi_i, \eta_j) &= \frac{\partial \hat{u}^{\delta}}{\partial \xi}(\xi_i,\eta_j) + (\hat{u}_{L}^{\delta I} - \hat{u}_{L}^{\delta})\frac{\partial g_L}{\partial \xi}(\xi_i) + (\hat{u}_{R}^{\delta I} - \hat{u}_{R}^{\delta})\frac{\partial g_R}{\partial \xi}(\xi_i), \\
    \hat{q}^{\delta}_{\eta}(\xi_i, \eta_j) &= \frac{\partial \hat{u}^{\delta}}{\partial \eta}(\xi_i,\eta_j) + (\hat{u}_{B}^{\delta I} - \hat{u}_{B}^{\delta})\frac{\partial g_B}{\partial \eta}(\eta_j) + (\hat{u}_{T}^{\delta I} - \hat{u}_{T}^{\delta})\frac{\partial g_T}{\partial \eta}(\eta_j),
    \label{CorrectedGradQuad}
\end{align}
where $\hat{u}^{\delta I}_{L}$, $\hat{u}^{\delta I}_{R}$, $\hat{u}^{\delta I}_{B}$, and $\hat{u}^{\delta I}_{T}$ are the transformed common interface solutions at the flux points located on the left ($\xi = -1$), right ($\xi = 1$), bottom ($\eta = -1$), and top ($\eta = 1$) edges respectively. The corrected gradient at each solution point of a quadrilateral element thus depends on the common interface solutions at four flux points. The correction functions $g_L(\eta)$, $g_R(\eta)$, $g_B(\xi)$, and $g_T(\xi)$ are defined in a completely analogous manner as for the one-dimensional case. The corrected gradient $\hat{\mathbf{q}}^{\delta}$ within the entire element is then written as:
\begin{equation}
    \hat{\mathbf{q}}^{\delta} = (\hat{q}^{\delta}_{\xi},\hat{q}^{\delta}_{\eta}) = \sum_{i=1}^{(P+1)} \sum_{i=j}^{(P+1)} \hat{\mathbf{q}}_{i,j}^{\delta} \mathcal{l}_{i}(\xi) \mathcal{l}_{j}(\eta),
    \label{CorrectedGradient_Poly}
\end{equation}
where $\hat{\mathbf{q}}_{i,j}^{\delta}$ corresponds to the value of the corrected gradient at the solution point $(\xi_i,\eta_j)$. The discontinuous flux $\hat{f}^{\delta D}_{i,j}$ at each solution point $(\xi_i,\eta_i)$ is directly evaluated from the approximate solution $\hat{u}^{\delta}$ and the corrected gradient $\hat{\mathbf{q}}^{\delta}$. The values of the discontinuous flux at the $(P+1)^2$ solution points are then used as coefficients to construct the discontinuous flux within the entire element $\Omega_S$ as follows:
\begin{equation}
    \hat{\mathbf{f}}^{\delta D}(\xi,\eta) = \sum_{i=1}^{(P+1)}\sum_{j=1}^{(P+1)}
    \hat{\mathbf{f}}_{i,j}^{\delta D} \mathcal{l}_{i}(\xi) \mathcal{l}_{j}(\eta).
    \label{DiscFluxQuad_Poly}
\end{equation}

The continuous transformed approximated flux $\hat{\mathbf{f}}^{\delta} = (\hat{f}^{\delta},\hat{g}^{\delta})$ is obtained by adding a correction flux $\hat{\mathbf{f}}^{\delta C}$ to the discontinuous flux: $ \hat{\mathbf{f}}^{\delta} = \hat{\mathbf{f}}^{\delta D} + \hat{\mathbf{f}}^{\delta C}$.

The correction flux $\hat{\mathbf{f}}^{\delta C}$ is constructed using the correction functions $h_L$, $h_R$, $h_B$, and $h_T$, and the common interface fluxes $\hat{f}^{\delta I}_{L}$, $\hat{f}^{\delta I}_{R}$, $\hat{g}^{\delta I}_{B}$, and $\hat{g}^{\delta I}_{T}$ at the flux points located along the lines $\xi = \xi_i$ and $\eta = \eta_i$. These numerical interface fluxes are calculated following the same methodology as in the one-dimensional approach. The $\xi$- and the $\eta$-components of the correction flux $\hat{\mathbf{f}}^{\delta C}$ are computed independently as:
\begin{align}
    \hat{f}^{\delta C} &= ( \hat{f}^{\delta I}_{L} - \hat{f}^{\delta D}_{L}) h_L + ( \hat{f}^{\delta I}_{R} - \hat{f}^{\delta D}_{R}) h_R,\\
    \hat{g}^{\delta C} &= ( \hat{g}^{\delta I}_{B} - \hat{g}^{\delta D}_{B}) h_B + ( \hat{g}^{\delta I}_{T} - \hat{f}^{\delta D}_{T}) h_T,
\end{align}
where the discontinuous fluxes $\hat{f}^{\delta D}_{L}$, $\hat{f}^{\delta D}_{R}$, $\hat{g}^{\delta D}_{B}$, and $\hat{g}^{\delta B}_{T}$ are computed directly from equation \ref{DiscFluxQuad_Poly}.\\

Next, the divergence of the total approximate flux $\mathbf{\nabla_{\boldsymbol{\xi}}} \cdot \hat{\mathbf{f}}^{\delta}(\xi_i,\eta_i)$ is calculated at the solution points as the sum of the divergence of the discontinuous and the correction flux components. The divergence of the discontinuous flux at the solution point ($\xi_i,\eta_i$) is obtained from:
\begin{equation}
    \boldsymbol{\nabla_{\xi}} \cdot \hat{\mathbf{f}}^{\delta D} (\xi_i,\eta_i) = \sum_{i=1}^{(P+1)}\sum_{j=1}^{(P+1)} \hat{f}^{\delta D}_{i,j}\frac{d \mathcal{l}_i}{d \xi}(\xi_i)\mathcal{l}_j(\eta_j) + \sum_{i=1}^{(P+1)} \sum_{i=1}^{(P+1)} \hat{f}^{\delta D}_{i,j}\mathcal{l}_i(\xi_i)\frac{d \mathcal{l}_j}{d \eta}(\eta_j).
\end{equation}
Similarly, the divergence of the correction flux at the solution point ($\xi_i,\eta_i$) is computed from:
\begin{equation}
    \boldsymbol{\nabla_{\xi}} \cdot \hat{\mathbf{f}}^{\delta C} (\xi_i,\eta_i) = \frac{\partial \hat{f}^{\delta C}}{\partial \xi}(\xi_i,\eta_i) + \frac{\partial \hat{g}^{\delta C}}{\partial \eta}(\xi_i,\eta_i),
\end{equation}
where
\begin{align}
    \frac{\partial \hat{f}^{\delta C}}{\partial \xi}(\xi_i,\eta_j) &= ( \hat{f}^{\delta I}_{L} - \hat{f}^{\delta D}_{L}) \frac{d h_L}{d \xi}(\xi_i) + ( \hat{f}^{\delta I}_{R} - \hat{f}^{\delta D}_{R}) \frac{d h_R}{d \xi}(\xi_i), \\
    \frac{\partial \hat{g}^{\delta C}}{\partial \xi}(\xi_i,\eta_j) &= ( \hat{g}^{\delta I}_{B} - \hat{g}^{\delta D}_{B}) \frac{d h_B}{d \eta}(\eta_j) + ( \hat{g}^{\delta I}_{T} - \hat{g}^{\delta D}_{T}) \frac{d h_T}{d \eta}(\eta_j).
\end{align}
Finally, a time marching method is applied in order to advance the approximate solution at the solution points $\hat{u}^{\delta}_{i,j}$ in time via the following expression:
\begin{equation}
    \frac{d\hat{u}^{\delta}_{i,j}}{dt} = - \mathbf{\nabla_{\boldsymbol{\xi}}} \cdot \hat{\mathbf{f}}^{\delta} = - \frac{\partial \hat{f}^{\delta}}{\partial \xi}(\xi_i,\eta_j) - \frac{\partial \hat{g}^{\delta}}{\partial \eta} (\xi_i,\eta_j).
\end{equation}

\section{Shock Capturing}\label{sec:SC}

Robust shock capturing is the main pacing item for high-order FE-type CFD methods. In the vicinity of discontinuities within the flow field, spurious oscillations appear due to the Gibbs phenomenon. This effect is more severe for higher orders and for stronger shocks. The oscillations near the shock generally cause numerical instabilities and negative pressures or densities. Several schemes have been investigated in \cite{sheshadri2014shock,park2014higher,du2015simple,yu2015localized,li2017convergent} to avoid spurious oscillations around shock waves. However these schemes have not been successfully applied to viscous hypersonic flows and are generally only applicable to explicit time stepping. The present paper proposes a modified Localized Laplacian Artificial Viscosity (LLAV) scheme combined with a positivity preservation method in order to alleviate oscillations caused by the Gibbs phenomenon as well as harsh transient effects characterizing hypersonic flow calculations. 

\subsection{Localized Laplacian Artificial Viscosity}\label{sec:LLAV}

The concept of localized artificial viscosity is to add an artificial diffusive flux dependent on the local properties of the flow field in order to spread out the shock over one or more elements. In this way spurious oscillations around shocks are attenuated. Within the LLAV formulation, the artificial viscosity is of a Laplacian type, as proposed in \cite{yu2015localized,persson2006sub}. The local artificial viscosity is only activated in areas of the flow field where discontinuities are present. To this end, a smoothness detector is used, of which several are proposed in \cite{yu2015localized}.

LLAV shows better convergence characteristics compared to traditional limiters developed for FR, such as Multi-Dimensional Limiting Process (MLP) \cite{park2014higher}, especially for steady-state cases. Furthermore LLAV is readily adaptable to implicit time-marching schemes. However, LLAV cannot guarantee positivity of the solution for hypersonic test cases and requires fine-tuning of the maximum amount of artificial viscosity added, to work properly. A modified LLAV scheme is proposed to alleviate these difficulties.

Within the LLAV framework, an extra flux is added to the system of equations:
\begin{equation}
  \frac{\partial \mathbf{u}}{\partial t} = -\boldsymbol{\nabla} \cdot \mathbf{f}_C + \boldsymbol{\nabla} \cdot \mathbf{f}_D + \boldsymbol{\nabla} \cdot \mathbf{f}_{AV},
\end{equation}
where $\mathbf{f}_C$ is the convective flux, $\mathbf{f}_D$ the diffusive flux and $\mathbf{f}_{AV}$ the flux due to the artificial viscosity. $\mathbf{f}_{AV}$ is constructed as follows:
\begin{equation}
\mathbf{f}_{AV} = \varepsilon \boldsymbol{\nabla} \mathbf{u},
\end{equation}
where $\varepsilon$ is the artificial viscosity. Within the FR framework, this means that the space discretization procedure, as presented in section \ref{sec:FR}, is also applied to a third flux. The gradients $\nabla \mathbf{u}$ are approximated using the corrected solution gradients $\mathbf{q}$ as computed in equation \ref{CorrectedGrad}. As such, the FR procedure is applied to $\mathbf{f}_{AV} = \varepsilon  \mathbf{q}$.

The value of the artificial viscosity $\varepsilon$ is computed in such a way that it is only active in discontinuous regions:
\begin{equation}
\varepsilon = 
\begin{cases}
    0       & \quad \text{if } S < S_0 - \kappa,\\
    \frac{\varepsilon_0}{2}\left( 1+\sin\dfrac{\pi(S-S_0)}{2\kappa}\right)  & \quad \text{if } S_0-\kappa\leq S \leq S_0 + \kappa,\\
    \varepsilon_0 & \quad \text{if } S > S_0 + \kappa,
  \end{cases}
\end{equation}
where $S$ is the smoothness in the current element, $S_0$ the reference smoothness, $\kappa$ a parameter to control the working spectrum of the artificial viscosity and $\varepsilon_0$ the maximum artificial viscosity.

The smoothness for a $P$-th order FR method can be computed as follows, using the smoothness detector $S$ found in \cite{yu2015localized}:
\begin{equation}
S = \log_{10} \dfrac{\langle u_m-u_m^{P-1},u_m-u_m^{P-1}\rangle}{\langle u_m,u_m\rangle},
\end{equation}
where $u_m$ is the state variable that is monitored in order to detect discontinuities. For the Navier-Stokes equations, typically the density $\rho$ or pressure $p$ is used. $u_m^{P-1}$ represents the monitored state variable projected on a $P-1$-th order polynomial. The operator $\langle \cdot\ ,\cdot\rangle$ is the element-wise scalar product over the solution points. The reference smoothness is set equal to $S_0 = -3\log_{10} P$ as suggested in \cite{park2014comparative}. The $(P-1)$-projected state variable can be computed by projecting the solution polynomial, which is expression in a Lagrange polynomial basis, on a modal basis using the Vandermonde matrix. Next, the sum over the first $P-1$ polynomials is made to determine $u_m^{P-1}$.

A novel scheme to compute the maximum artificial viscosity is proposed. In order to match the flux dimensions, $\varepsilon$ must be expressed in $\left[ \frac{m^2}{s}\right]$, or a length-scale times a velocity-scale. Since a higher artificial viscosity is needed at higher speeds and in bigger elements (in order to spread out the shock over the element), $\varepsilon_0$ is set to be:
\begin{equation}\label{artEps}
\varepsilon_0 = f(P)h|\lambda|_{max},
\end{equation}
where $h$ is the characteristic length of the current element and $|\lambda|_{max}$ the maximum eigenvalue of the inviscid part of the system of equations. In order to take into account the order $P$ of the FR method, following the approach of \cite{yu2015localized}, $f(P)$ is set to:
\begin{equation}
f(P) = c(2-\Delta \xi_{max}(P)),
\end{equation}
where $\Delta \xi_{max}(P)$ is the subcell resolution, i.e. the largest distance between two solution points within the reference domain and $c$ a constant. In order to avoid the need for fine-tuning the constant $c$, an approximate relation between this parameter and the number of cells over which a shock is spread is defined here after. 

For strong shocks where viscosity is present, the following shock-related Reynolds number is of order of magnitude 1:
\begin{equation}
\text{Re}_s = \dfrac{\delta_s \Delta u \rho^*}{\mu^*} \approx 1,
\end{equation}
where $\delta_s$ is the shock thickness, $\Delta u$ the velocity difference over the shock, $\rho^*$ and $\mu^* $ respectively the density and dynamic viscosity in the shock at $M=1$. Consider the following assumptions:
\begin{equation}
\rho^* \sim \rho_{\infty},
\end{equation}
\begin{equation}
\mu^* \sim \varepsilon \rho_{\infty}.
\end{equation}
Combining these two relations, the following scaling relation is found:
\begin{equation}
\varepsilon \rho_\infty \sim \delta_s \Delta u \rho_\infty.
\end{equation}
Using the expression for the velocity drop $\Delta u$ over a shock:
\begin{equation}
\varepsilon \sim \delta_s \dfrac{2}{\gamma+1} \dfrac{M_\infty^2-1}{M_\infty^2},
\end{equation}
combining with expression \ref{artEps}:
\begin{equation}
c(2-\Delta\xi_{max})h|\lambda|_{max} \sim \delta_s \dfrac{M_\infty^2-1}{M_\infty^2},
\end{equation}
rearranging and setting $h^* = (2-\Delta \xi_{max})h$ the characteristic subcell length:
\begin{equation}
c \sim \dfrac{\delta_s}{h(u_\infty+\frac{u_\infty}{M_\infty})} \dfrac{M_\infty^2-1}{M_\infty^2},
\end{equation}
\begin{equation}
c \sim \dfrac{\delta_s}{h^*} \dfrac{M_\infty-1}{M_\infty},
\end{equation}
introducing the characteristic shock thickness $\delta_s^* = \dfrac{\delta_s}{h^*}$. This variable is a measure for the number of characteristic element lengths over which the shock will span. Using this variable, the following relation is found:
\begin{equation}\label{eq:delta}
c \sim \delta_s^* \dfrac{M_\infty-1}{M_\infty}.
\end{equation}
Using expression \ref{eq:delta}, the LLAV scheme can be characterized by a characteristic shock thickness $\delta_s^*$ which is a measure for the number of elements over which the shock is spread out. By calibrating $c$ as:
\begin{equation}\label{deltaCal}
c = \alpha \delta_s^* \dfrac{M_\infty-1}{M_\infty} + \beta.
\end{equation}
If there is no viscosity present, the characteristic shock thickness is zero. As a consequence, $\beta$ can be set to zero. For the value of $\alpha$, a calibrating test case can be used. As described in the section \ref{verification}, the following value is found:
\begin{equation}
\alpha \simeq 0.027\ .
\end{equation}
By using this scheme, only the parameters $\kappa$ and $\delta_s^*$ need to be specified for the LLAV method instead of $\kappa$ and $c$. The spectral sensitivity $\kappa$ is only related to the smoothness detector and is independent of the specific conditions of the test case. As proposed in \cite{yu2015localized}, a value of $\kappa$ ranging from 1 to 5  generally produces robust results. The characteristic shock thickness is a measure for the amount of elements over which the shock is spread, as a consequence, if $\delta_s^* < 1$ there is a risk of spurious oscillations, while if $\delta_s^* > 2 $, the shock will be spread over more than two elements, and accuracy may be deteriorated due to superfluous artificial viscosity. The LLAV scheme based on these two parameters requires little fine-tuning and generally produces robust results, independent of the characteristic length $h$ of the elements near the shock, of the Mach number of the far field, etc. In this way the fine-tuning of the novel LLAV scheme defined by $\delta_s^*$ is much more straightforward than that of the LLAV scheme defined by the scaling parameter $c$.


\subsection{Positivity Preservation Scheme}

For the Navier-Stokes equations the conservative state variables are $\mathbf{u} = (\rho, \rho\mathbf{v}, \rho e_t)$, where $\mathbf{v}$ is the velocity and $e_t$ the specific total energy. However not all states are admissible: density and pressure must remain positive:
\begin{equation}\label{eq:posConstraints}
\rho > 0 \quad \text{ and } \quad p = (\gamma - 1)\left( \rho e_t - \frac{1}{2} \dfrac{||\rho\mathbf{v}||^2}{\rho} \right) > 0.
\end{equation}
If the above conditions are not satisfied, a numerical blow-up of the solution will occur. A shock capturing method based on artificial viscosity cannot guarantee that equation \ref{eq:posConstraints} is satisfied. Especially in hypersonic test cases where very strong shocks and severe transients are present, artificial viscosity is inadequate in preserving positivity. In order to alleviate this, a positivity preservation scheme is proposed for FR. This scheme limits the state at each iteration such that each violation of equation \ref{eq:posConstraints} is corrected. The proposed scheme is similar to the one presented in \cite{zhang2017positivity} that was developed for the DG method, with the difference that now the limiting procedure must limit the states based on the evaluation of the approximate solution polynomials in the flux points. 

The first step is to limit the density. To this end the average state $\bar{\mathbf{u}}$ on the each element is computed. Based on $\bar{\mathbf{u}}$, the average density $\bar{\mathbf{\rho}}$ and pressure $p(\bar{\mathbf{u}})$ are known. A small number $\epsilon$ is defined to which $\rho$ will be limited:
\begin{equation}
\epsilon = \min(10^{-13},\bar{\mathbf{\rho}},p(\bar{\mathbf{u}})),
\end{equation}
where $p(\bar{\mathbf{u}})$ denotes the pressure computed using the element average state. Next, $\rho$ is recomputed in each solution point $i$, in order to obtain a new positive density $\tilde{\rho}_i$:
\begin{equation}
\tilde{\rho}_i = t_{1}(\rho_i - \bar{\rho}) + \bar{\rho} \quad \text{ with } \quad t_{1} = \min\left(\dfrac{\bar{\rho} - \epsilon}{\bar{\rho} - \rho_{min}},1\right).
\end{equation}
$\rho_{min}$ is computed in the following way:
\begin{equation}
\rho_{min} = \min_j \left( \rho_j \right),
\end{equation}
where the subscript $j$ refers to the $j$-th flux point. In this way, the density in each solution point of an element is limited such that all the densities in the flux points remain larger than or equal to $\epsilon$.

The state $\tilde{\mathbf{u}} = (\tilde{\rho}, \rho\mathbf{v}, \rho e_t)$ obtained after limiting the density, is now used to limit the pressure: if in the $j$-th flux point $p_j < \epsilon$, a coefficient $t_j$ is computed such that:
\begin{equation}\label{pLim}
p(t_j(\tilde{\mathbf{u}}_j - \bar{\mathbf{u}}) + \bar{\mathbf{u}}) = \epsilon.
\end{equation}
$\tilde{\mathbf{u}}_j$ represents the state after limiting the density in the $j$-th flux point. This can be computed by evaluating the approximate solution polynomial, as defined in equation \ref{ApproxSolution}, in the $j$-th flux point. The value of $t_j$ can be obtained from equation \ref{pLim} by solving a second-order polynomial equation. Note that the expression $0 < t_j < 1$ is used to select the correct solution of the second-order equation. Next, the whole state in each solution point $i$ is limited using the minimum value of the $t_j$ coefficients computed in the current element, obtaining $\tilde{\tilde{\mathbf{u}}}_i$:
\begin{equation}
\tilde{\tilde{\mathbf{u}}}_i = t_{2}(\tilde{\mathbf{u}}_i - \bar{\mathbf{u}}) + \bar{\mathbf{u}} \quad \text{ with } \quad t_{2} = \min_j (t_j).
\end{equation}
This scheme guarantees that the density and pressure remain positive in the flux points. For a $P1$ extrapolation, the most negative value will always be attained inside a flux point rather than a solution point, since they are situated on the boundary of the element. For higher order it is possible that a negative state appears first in a solution point. As such, the same procedure as outlined above is also applied to the solution points. 

\subsection{Interaction between LLAV and positivity preservation}

Generally for a high maximum artificial viscosity, which is accomplished by specifying a large $\delta_s^*$, positivity is only enforced during the transient of the simulation. However if $\delta_s^*$ is chosen lower, typically $\delta_s^*\simeq 1$, the positivity preservation scheme remains active until convergence. Since the positivity preservation scheme is applied outside of the residual computation, it causes the residuals to stall at convergence: the positivity preservation scheme exactly matches the contribution of the non-zero residuals to the solution at the next time step. As a consequence the solution remains unchanged, despite non-zero residuals. As proposed in \cite{li2017convergent}, a solution for this is to monitor $\frac{u(t+\Delta t) - u(t)}{\Delta t}$ instead of the residual. This is a behavior typically found in limiter-type shock capturing methods for high-order FE schemes, such as Multi-Dimensional Limiting Process, presented in \cite{li2017convergent}. However, since the positivity preservation scheme aims at keeping the solver stable during the transient, it should be avoided that it remains active until convergence. $\delta_s^*$ can simply be increased in order to avoid this behavior. In the present solver implementation it is possible to do this ``on the fly'' during the simulation.

Furthermore, the positivity preservation scheme is applied outside of the residual computation process and imposes a severe limit to the CFL number in implicit time marching. This is due to the fact that the contribution of the positivity preservation scheme in the Jacobian of the implicit scheme cannot readily be evaluated. However, since the positivity preservation scheme should only be active during the transient, the severe CFL limit is only imposed as long as it remains active.

\section{Implementation of the Solver}\label{sec:impl}

This section first presents an overview of the capabilities of the developed solver. Second, the general structure is described as well as the main interactions between various parts of the solver and the existing COOLFluiD framework. The present FR solver consists of two main segments: the data structure and the FR algorithm itself. Both parts are successively discussed in detail.

\begin{figure}[H]
  \centering
  \includegraphics[width=0.2\textwidth]{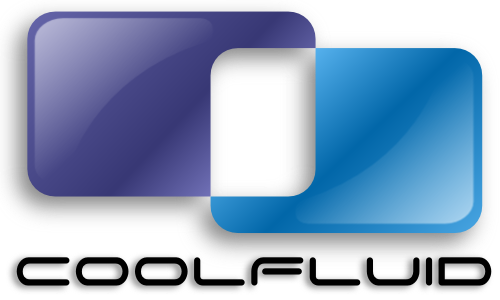}
  \caption{Logo of COOLFluiD}
  \label{fig:logoCOOLFluiD}
\end{figure}

The present FR solver is implemented as a plug-in of COOLFluiD, a platform written in C++ that offers a component-based framework oriented towards complex multi-physics simulations. Herein, each numerical method or physical model is encapsulated into an independent dynamic module (or {\it plug-in} library) that can be loaded on demand by user-defined applications. Some of the main features of COOLFluiD include: parallel solvers for compressible and incompressible flows based on multiple discretization techniques (e.g. FV, Residual Distribution schemes, Finite Element, Spectral Finite Differences) for unstructured meshes, interfaces to different linear systems solvers (e.g. PETSc, Trilinos, Pardiso), aerothermochemical models for flows and plasma \cite{knight2017assessment,panesi07, degrez09, knight12, munafo13, panesi13}, MHD, coupling algorithms for multi-physics and multi-domain simulations, Arbitrary Lagrangian Eulerian methods and radiation transport algorithms based on Monte Carlo \cite{lani13, lani13b}. More information on the development of this platform can be found in \cite{lani2008object,quintino2008component}. The logo of COOLFluiD is shown in figure \ref{fig:logoCOOLFluiD}. 

COOLFluiD takes as inputs unstructured meshes with arbitrary element types. The mesh of a test case for the FR solver can either be a 2D grid comprised of quadrilaterals or a 3D grid consisting of hexahedra. Most existing mesh generators can exclusively create low-order meshes. As pointed out by Wang et al. in \cite{wang2013high}, this is one of the most important challenges high-order CFD methods are confronted with nowadays. As a consequence, the ability to construct a high-order mesh based on an existing low-order mesh is highly desirable. The FR solver can upgrade a grid to up to ninth-order.

The present solver has the ability to increase the amount of elements in a mesh by dividing each element in a number of equally sized elements specified by the user. In order to determine the order of convergence of a high-order solver, a grid convergence study is performed. This is done by comparing the norm of a variable that represents the solution accuracy for a range of meshes consisting of a different number of elements. However, such meshes are often not readily available. The capability of increasing the grade of an existing mesh is thus regarded as a highly convenient feature.

The FR solver has been developed to solve the compressible Euler and Navier-Stokes equations in 2D and 3D. Section \ref{GoverningEqs} describes these equations in more detail. In COOLFluiD, the algorithms that depend on the physics of a problem, such as the evaluation of fluxes, are decoupled from the numerical algorithms. This means that the FR solver has been developed to solve advection and advection-diffusion problems in general. In addition, all necessary physics-dependent algorithms for the Euler and Navier-Stokes equations were implemented. These are the boundary conditions that compute the ghost states and gradients as well as the algorithm that computes the characteristic speeds for the evaluation of the CFL number. Extending the functionality of the FR solver to different types of advection-diffusion problems only requires the addition of the above mentioned physics-dependent functions.

The FR method is fully determined by the solution and flux point distributions, the type of correction function and the interface flux scheme. The present FR solver can employ Gauss-Legendre, Gauss-Lobatto or equidistant points for both solution and flux points. These point distributions are defined in separate C++ classes within the solver. The user can specify which distribution must be used. As a consequence, it suffices to add an extra class without changing the existing solver, to add new distributions. The VCJH scheme for the correction functions has been implemented in the present solver for quadrilaterals and hexahedra. The following interface flux schemes have been implemented: Centered flux \cite{van1987comparison}, Lax-Friedrichs flux \cite{rusanov1961calculation}, Roe flux  \cite{roe1981approximate}, AUSM$^+$ \cite{liou1996sequel} and AUSM$^+$-up flux \cite{liou2006sequel}. These methods are used to compute the common interface fluxes. 

In the COOLFluiD framework, the time and space discretizations are decoupled. This means that the FR solver computes the residuals and, for implicit cases, the numerical flux Jacobian. Next, the COOLFluiD framework uses a time marching procedure, such as the backward Euler method, to compute the states at the next discrete time step. The present FR solver is able to use both an explicit and an implicit time-marching scheme. Forward Euler can be used as an explicit scheme, whereas backward Euler and the Newton method can be used as implicit time marching schemes. These schemes are discussed in more detail in section \ref{time marching}.

Finally, the solver can either be executed serially or in parallel. For the serial execution, one CPU computes the update values of the whole mesh at each iteration. For the parallel execution, the spatial domain is partitioned between a number of CPUs specified by the user. Details of the partitioning procedure, which are out of the scope of the present paper, are a high-level feature of COOLFluiD and are explained in \cite{lani2008object,laniPart}.

The developed FR solver consists of two major components: the FR data structure and the FR algorithm. The main function of the data structure is to create the data that defines the high-order unstructured mesh. This is comprised of geometric entities, namely elements and faces, how these elements are connected to each other, and their properties: solution points, flux points, base functions, etc. The FR algorithm subsequently uses this data to compute the numerical solution of a given test case in an iterative way.

\begin{figure}[ht]
  \centering
  \includegraphics[width=0.75\textwidth]{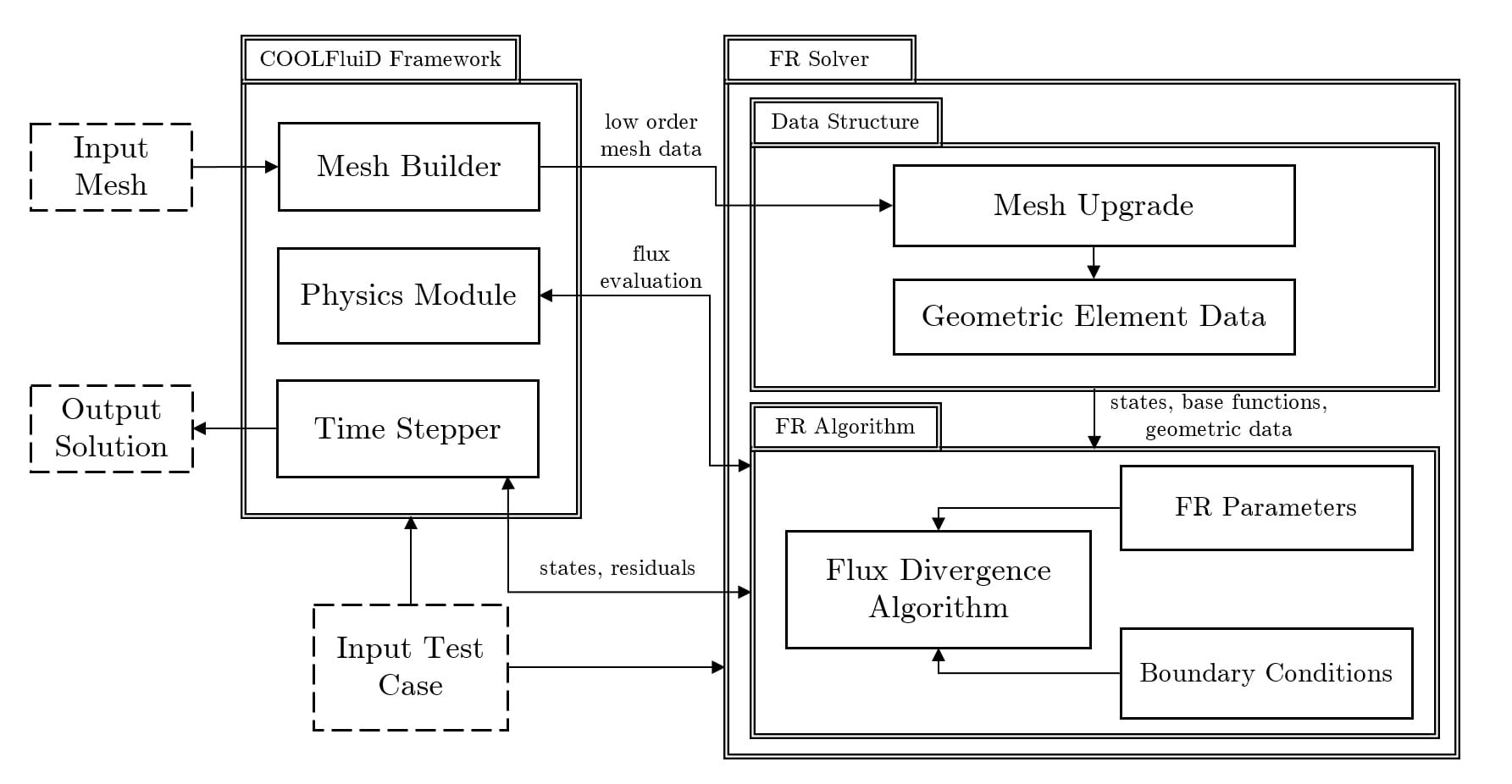}
  \caption{Overview of the major parts of the FR solver}
  \label{fig: overviewSolver}
\end{figure}

Both parts of the FR solver interact with the existing COOLFluiD framework, as schematically shown in figure \ref{fig: overviewSolver}. In the next sections, the two major parts of the solver are discussed in detail.

\subsection{Implementation of the Data Structure}\label{data structure}

The FR data structure allows the solver to loop over the geometric entities of the mesh. Geometric entities are either elements or faces. In addition, a mesh is specified by a list of nodes and states. For unstructured meshes there is no trivial way to access neighboring geometric entities. As a consequence, the creation of look-up tables, called connectivity tables, is an important function of the data structure. These tables store how geometric entities are linked to each other. The collection of a certain type of geometric entities and their relevant connectivity tables is referred to as a Topological Region Set  (TRS). TRSs include the boundary as well as the internal part of the computational domain. The first function of the FR data structure  is the creation of TRSs. Since the present FR solver builds the high-order TRSs by upgrading the order of the mesh data generated in the COOLFluiD framework, the block ``Mesh Upgrade'' in figure \ref{fig: overviewSolver} represents this functionality. The second function of the data structure is to store the geometric data and properties of the different element types that are needed for the FR algorithm.

\subsubsection{Building the Topological Region Sets}\label{TRS}

\begin{figure}[H]
  \centering
  \includegraphics[width=0.5\textwidth]{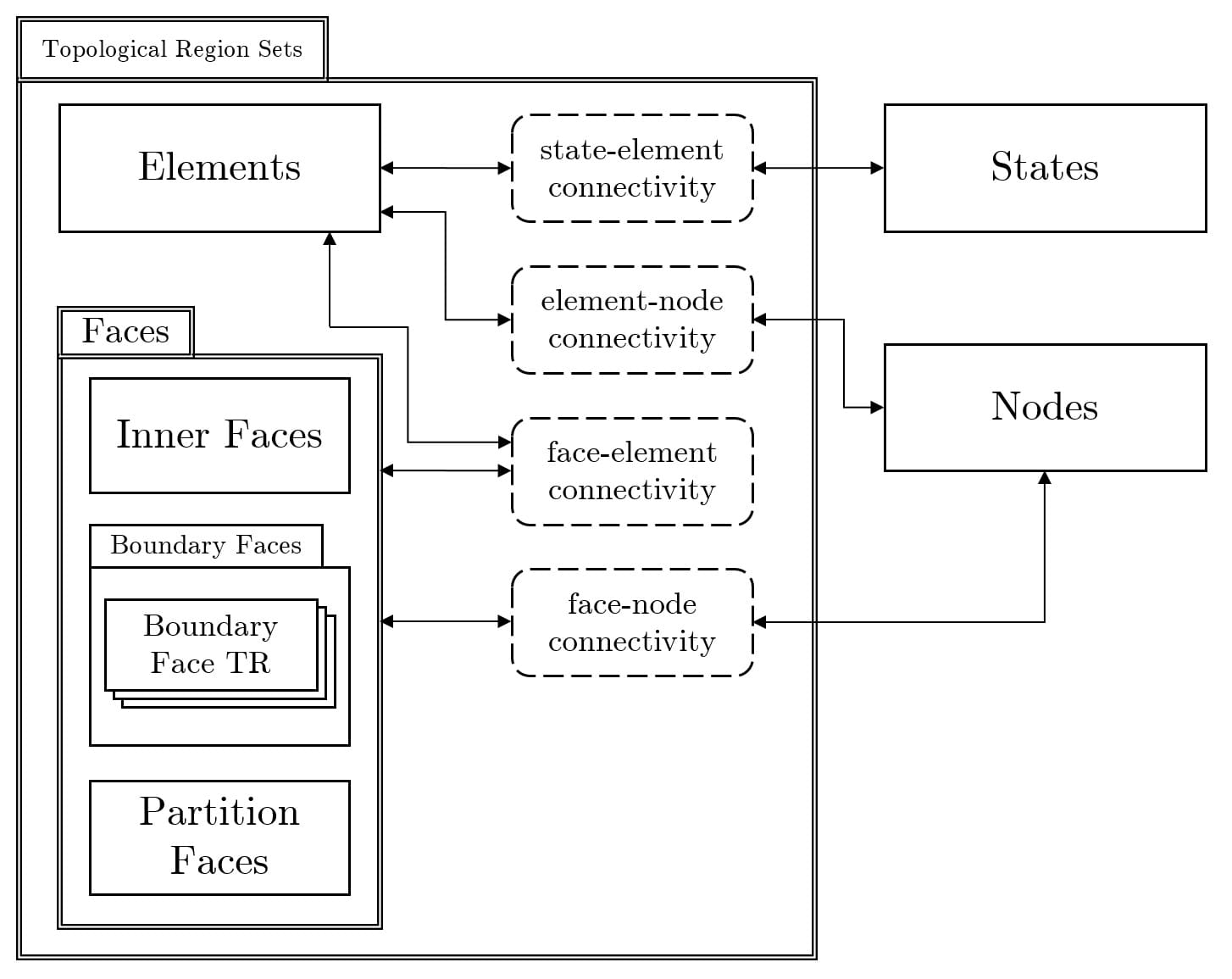}
  \caption{Overview of the TRSs that are created by the FR data structure}
  \label{fig: TRSoverview}
\end{figure}

The FR data structure creates the following topological region sets: a set of all elements, inner faces, boundary faces and partition faces. Figure \ref{fig: TRSoverview} gives an overview of these TRSs.

\paragraph{Element TRS}
The TRS of the elements consists of all the element (a.k.a. cells) geometric entities in the mesh. The position of a certain element in the mesh is defined by the coordinates of its nodes. As a consequence, a connectivity table that links elements to their nodes is necessary. The approximate solution polynomial on an element is determined by the states in its solution points. Consequently, a connectivity table linking the elements to states is present. Finally, a connectivity table stores indices of the neighboring faces of each element. This is necessary for the evaluation of the state and fluxes in the flux points on the faces. These connectivity tables together with the element geometric entities form the element TRS.

If the amount of elements has to be increased, the FR solver creates a completely new element TRS. For each element of the old TRS, a number of new elements is created as specified by the user. Since the nodes that define these new elements are not created directly by the core COOLFluiD infrastructure but by a local mesh upgrader, the FR data structure builds a new list of nodes and a new element-node connectivity table. Finally, the correct amount of states is created based on the order of the FR method and the state-element connectivity table is updated for each new element.

\paragraph{Inner and Boundary Face TRS}
After the creation of the element TRS, the inner and boundary face TRSs are built. The position of a face in the mesh is defined by its corner nodes. A 2D face has two corner nodes, while a 3D face of a hexahedral element has four corner nodes. As a consequence, a face-node connectivity table is made. Since there are solution points only within an element and not on a face, there are no states linked to faces, only nodes. Using the face-element connectivity table, the number of neighboring elements of a face is determined. If a face has only one neighboring element, it is added to the boundary faces TRS. If it has two neighboring elements, it is added to the inner faces TRS. Since a number of different boundary conditions can apply to the boundary faces, the boundary face TRS is split up in at least the same amount of Topological Regions (TR). In this manner, a boundary condition can be applied to a subset of the boundary faces.

The core COOLFluiD infrastructure creates face TRSs based on the low-order input mesh. If the amount of elements needs to be increased, the FR solver creates a new set of faces. By looping over the old elements of the mesh, the neighboring faces of the new elements are created. However, a face should only be constructed once, so a search in the updated element-face connectivity table is performed to check whether the face has already been created. If this is the case, only the face-element connectivity table is updated and no new face is made.

Finally, an additional TRS of the partition faces must be created if the solver runs in parallel. Partition faces are inner faces of the global mesh, but make up the edge of the domain of a certain CPU. Since these faces have only one neighboring element, but have no boundary condition attached to them, they must be treated separately by the FR algorithm.

\subsubsection{Geometrical Data of the Elements}\label{elementData}

During the execution of the numerical computation of the update values for the states, the FR algorithm needs to access the following geometrical data of the elements:
\begin{itemize}
    \item type of the element: shape, geometric order, solution order and dimension;
    \item coordinates of the solution and flux points in the standard domain;
    \item local numbering of faces, flux points and solution points;
    \item approximate solution polynomial and its derivative.
\end{itemize}
For this purpose, a generic C++ class is defined for each type of element that provides the above information to the solver. When the data structure of the test case is built, an instance of the class is created. Since all coordinates in this class are defined in the reference domain $\Omega_S$, as defined in section \ref{sec:FR}, it is sufficient to create only one instance of an element data class for each type of element. When the global coordinates are needed, the transformation between the reference and global domain is performed, as in equation \ref{TransformationQuad}.

The distribution of the solution and flux points is a parameter of the FR method. As a consequence, the point distributions are decoupled from the generic element data classes. Point distributions are defined by the 1D coordinates of the points in the domain $[-1,1]$. Based on these 1D coordinates, the element data classes construct the 2D or 3D coordinates of the flux and solution points depending on the type of element.

The approximate solution is only stored in the solution points. This means that in order to evaluate the approximate solution polynomial in the flux points, the Lagrange interpolation of equation \ref{ApproxSolution_Quad} is used. Since the Lagrange base functions belonging to the $i$-th solution point $\mathcal{l}_i$ are defined in the standard domain, it is sufficient to compute them once for each element type and subsequently use the polynomials for all elements of this type. In addition, the derivative of $\mathcal{l}_i$ is computed in order to be able to determine the derivative of the discontinuous fluxes as follows:
\begin{equation}
    \frac{\partial \hat{\mathbf{f}}^{\delta D}}{\partial \xi_j}(\boldsymbol{\xi}) = \sum_{i=1}^{N_s} \hat{\mathbf{f}}^{\delta D}_{i} \frac{\partial \mathcal{l}_i}{\partial\xi_j}(\boldsymbol{\xi}).
\end{equation}

\subsection{Implementation of the FR Algorithm}\label{impl algo}

\begin{figure}[ht]
  \centering
  \includegraphics[width=0.9\textwidth]{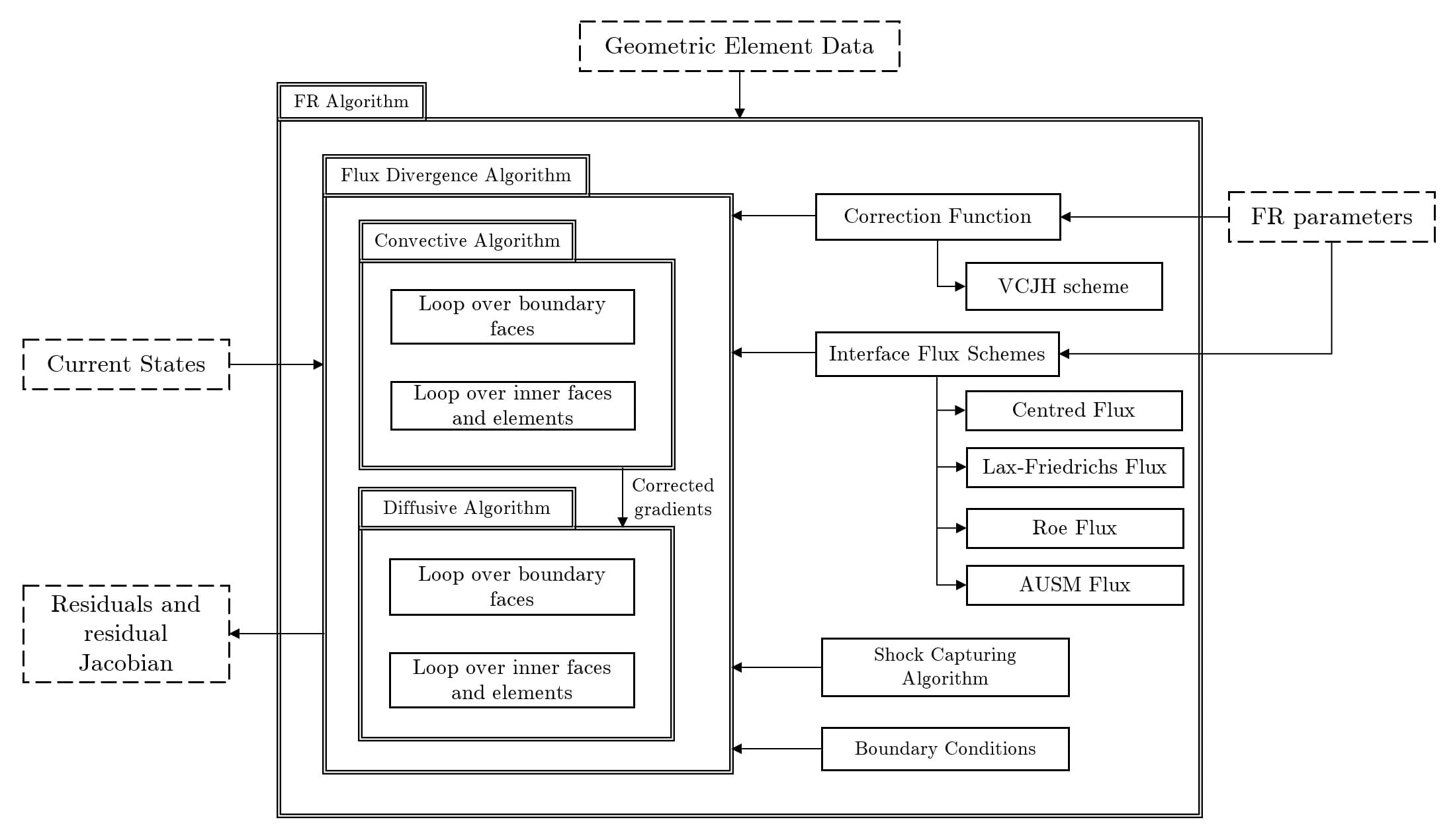}
  \caption{Overview of the structure of the FR algorithm}
  \label{fig: algorithm}
\end{figure}

The compressible Navier-Stokes and Euler equations can be written as a system of PDEs in the conservative form as presented in equation \ref{cons law}. The goal of the FR solver is to compute the divergence of the reconstructed flux in each element at every time step. For this purpose, the FR algorithm is divided into different parts as shown in figure \ref{fig: algorithm}. As mentioned in section \ref{GoverningEqs}, the flux is split up in a convective and diffusive part:
\begin{equation}
    \frac{\partial \mathbf{u}}{\partial t} = -\div \mathbf{f}_{C}(\mathbf{u}) + \div \mathbf{f}_{D}(\mathbf{u}, \grad\mathbf{u}).
    \label{consVgl}
\end{equation}
The evaluation of the physical fluxes based on $\mathbf{u}$ and $\grad\mathbf{u}$ is performed in the physics module of the COOLFluiD framework, as shown in figure \ref{fig: overviewSolver}. The convective and diffusive flux can be computed independently. As a consequence, the FR algorithm consists of two major parts: the convective algorithm and the diffusive algorithm. These subdivisions both interact with classes that define the parameters of the FR method that are specified by the user. These parameters are: the flux and solution point distribution, the interface flux scheme and the correction function type. However, the first of these parameters is used in the FR data structure as explained in the previous section. Additionally the boundary conditions as well as the shock capturing method are implemented in the FR algorithm.

\subsection{Computation of the Convective Flux Divergence}

The reconstructed convective flux in the reference domain consists of a discontinuous flux $\hat{\mathbf{f}}^{\delta D}_C$ and a correction flux $\hat{\mathbf{f}}^{\delta C}_C$ which in general are expressed as follows:
\begin{equation}
    \hat{\mathbf{f}}^\delta_{C} = \hat{\mathbf{f}}^{\delta D}_C(\hat{\mathbf{u}}) + \hat{\mathbf{f}}^{\delta D}_C(\hat{\mathbf{u}}) = \hat{\mathbf{f}}^{\delta D}_C(\hat{\mathbf{u}}) + \sum_{i = 1}^{N_f} \mathbf{h}_i(\boldsymbol{\xi})\left(\left(\hat{\mathbf{f}}_{C,i}^{\delta I} - \hat{\mathbf{f}}_{C}^{\delta D}(\hat{\mathbf{u}}(\boldsymbol{\xi}_i))\right) \cdot \hat{\mathbf{n}}_i\right),
\end{equation}
where $\mathbf{h}_i$ is the correction field associated with the $i$-th flux point, $N_f$ the total number of flux points and $\hat{\mathbf{n}}_i$ the unit normal vector to the face in flux point $i$. The divergence of this expression is:
\begin{equation}
    \boldsymbol{\nabla_{\xi}}\cdot\hat{\mathbf{f}}^\delta_{C} = \boldsymbol{\nabla_{\xi}}\cdot\hat{\mathbf{f}}^{\delta D}_C(\hat{\mathbf{u}}) + \sum_{i = 1}^{N_f} \boldsymbol{\nabla_{\xi}}\cdot\mathbf{h}_i(\boldsymbol{\xi})\left(\left(\hat{\mathbf{f}}_{C,i}^{\delta I} - \hat{\mathbf{f}}_{C}^{\delta D}(\hat{\mathbf{u}}(\boldsymbol{\xi}_i))\right) \cdot \hat{\mathbf{n}}_i\right).
    \label{convNumFlux}
\end{equation}
The computation of the discontinuous and correction flux divergence can be done separately. For the divergence of the discontinuous flux, the states in the solution points must be accessed, as shown by its dependence on $\hat{\mathbf{u}}$. Since the solution points are part of the element geometrical entities, this term is computed by performing a loop over the elements.

For the computation of the correction flux divergence, the states in the flux points $\boldsymbol{\xi}_i$ for both the left and right neighboring elements are needed, as can be seen in the second term in the RHS of equation \ref{convNumFlux}: this contains the interface flux $\hat{\mathbf{f}}_{C,i}^{\delta I}$ which depends on the left and right states on a face. Since the flux points are part of the face geometric entities, this term is computed by looping over the faces. When looping over the faces, each flux point state only needs to be computed once, since each flux point is only part of one iteration of the loop. However, if a loop over the elements would be performed instead of over the faces, each face would be considered in two iterations as it belongs to two neighboring elements. This means that each flux point state would be computed twice which is inefficient. This is an additional reason why the computation of the correction flux divergence is performed by looping over the face geometric entities.

For the computation of $\hat{\mathbf{f}}_{C}^{\delta D}(\hat{\mathbf{u}}(\boldsymbol{\xi}_i))$ in the correction flux divergence however, the discontinuous fluxes in the solution points of an element need to be accessed. This is necessary for the extrapolation of the discontinuous fluxes to the flux points. As such, it is more efficient to compute this term in the loop over elements together with the discontinuous flux divergence. In this manner, the discontinuous fluxes in the solution points only need to be evaluated once for each element. Equation \ref{differentTerms} summarizes which terms are computed in which loops in each solution point:
\begin{subequations}
\begin{align}
    \text{loop over elements: } & \boldsymbol{\nabla_{\xi}}\cdot\hat{\mathbf{f}}^{\delta D}_C(\hat{\mathbf{u}}) - \sum_{i = 1}^{N_f}\boldsymbol{\nabla_{\xi}}\cdot\mathbf{h}_i(\boldsymbol{\xi})\left( \hat{\mathbf{f}}_{C}^{\delta D}(\hat{\mathbf{u}}(\boldsymbol{\xi}_i)) \cdot \hat{\mathbf{n}}_i\right),
    \label{loopElemsEq}
    \\
    \text{loop over faces: } & \sum_{i = 1}^{N_f}\boldsymbol{\nabla_{\xi}}\cdot\mathbf{h}_i(\boldsymbol{\xi})\left( \hat{\mathbf{f}}_{C,i}^{\delta I}(\hat{\mathbf{u}}(\boldsymbol{\xi}_i)) \cdot \hat{\mathbf{n}}_i\right).
    \label{loopFacesEq}
\end{align}
\label{differentTerms}
\end{subequations}
Note that the divergence of the correction field $\boldsymbol{\nabla_{\xi}}\cdot\mathbf{h}(\boldsymbol{\xi}_i)$ only has to be computed once, since it is defined within the reference domain. Therefore its appearance in both loops simply means that its value needs to be accessed twice, not that the same computation needs to be performed twice.

\subsubsection{Loop over Elements}

The first term in equation \ref{loopElemsEq} is the discontinuous flux divergence, which can be computed as follows in the $i$-th solution point:
\begin{equation}\label{divDiscFlux}
    \boldsymbol{\nabla_{\xi}}\cdot\hat{\mathbf{f}}^{\delta D}_C(\hat{\mathbf{u}}_i) = \sum_{j = 1}^{N_{dim}} \frac{\partial\hat{\mathbf{f}}^{\delta D}_C}{\partial \xi_j} (\hat{\mathbf{u}}_i) \cdot \hat{\mathbf{n}}_{j} = \sum_{j = 1}^{N_{dim}}\sum_{k = 1}^{N_s} \left(\hat{\mathbf{f}}^{\delta D}_C(\hat{\mathbf{u}}_k) \cdot \hat{\mathbf{n}}_{j}\right)\frac{\partial \mathcal{l}_k}{\partial \xi_j} (\boldsymbol{\xi}_i),
\end{equation}
where $N_{dim}$ is the dimensionality of the problem, $N_s$ the number of solution points and $\hat{\mathbf{n}}_{j}$ the unit vector in the direction of $\xi_j$ in the reference coordinate frame. From this equation, it is clear that the discontinuous flux divergence is computed through summation over all solution points of the projected flux multiplied by the derivative of the Lagrange base function of the corresponding solution point. The base function derivatives are defined on the reference domain and are stored in the FR data structure, as explained in section \ref{elementData}. The states in the solution points can easily be accessed when building an element geometric entity. The same applies to the unit axes $\hat{\mathbf{n}}_j$. As a consequence, all information required to evaluate equation \ref{divDiscFlux} is available when looping over the elements.

The second term in equation \ref{loopElemsEq} depends on the extrapolation of the discontinuous fluxes to the flux points. For the $j$-th flux point, this is equal to:
\begin{equation}\label{fD}
    \hat{\mathbf{f}}^{\delta D}_C(\hat{\mathbf{u}}(\boldsymbol{\xi}_j)) = \sum_{k=1}^{N_{s}} \hat{\mathbf{f}}^{\delta D}_C (\hat{\mathbf{u}}_{k}) \mathcal{l}_{k}(\boldsymbol{\xi}_j).
\end{equation}
In this expression, every variable is now known, since the discontinuous fluxes in the solution points were evaluated to compute expression \ref{divDiscFlux}.

Finally, the correction field of quadrilaterals and hexahedra associated with the $j$-th flux point, can be expressed in terms of a scalar correction function: $\mathbf{h}_j = h_j\ \hat{\mathbf{n}}_j$. The scalar correction function in 2D or 3D is constructed using the tensor product of 1D correction functions, as explained in section \ref{sec:FR2D}. The divergence of the correction field in the $i$-th solution point due to the correction associated with the $j$-th flux point for quadrilaterals and hexahedra can simply be evaluated as follows:
\begin{equation}\label{divh}
    \boldsymbol{\nabla}_{\boldsymbol{\xi}}\cdot\mathbf{h}_j (\boldsymbol{\xi}_i) =  \boldsymbol{\nabla}_{\boldsymbol{\xi}}\cdot \left(h_j  (\boldsymbol{\xi}_i)\ \hat{\mathbf{n}}_j\right) = \frac{\partial h_j}{\partial \xi_j}(\boldsymbol{\xi}_i),
\end{equation}
where $\xi_j$ is the direction along the unit normal in the $j$-th flux point $\hat{\mathbf{n}}_j$.

In summary, the computation that is performed for the $i$-th solution point in the loop over elements is given in expression \ref{loopElems}. This is the implementation of the evaluation of expression \ref{loopElemsEq}:
\begin{equation}\label{loopElems}
\sum_{j = 1}^{N_{dim}}\sum_{k = 1}^{N_s} \left(\hat{\mathbf{f}}^{\delta D}_C(\hat{\mathbf{u}}_k) \cdot \hat{\mathbf{n}}_{j}\right)\frac{\partial \mathcal{l}_k}{\partial \xi_j} (\boldsymbol{\xi}_i) - \sum_{f=1}^{N_f}\frac{\partial h_f}{\partial \xi_f}(\boldsymbol{\xi}_i)\sum_{n=1}^{N_s}\left(\hat{\mathbf{f}}^{\delta D}_C (\hat{\mathbf{u}}_{n})\cdot \hat{\mathbf{n}}_f \right)\mathcal{l}_{n}(\boldsymbol{\xi}_f),
\end{equation}
considering once again that $\xi_f$ denotes the direction along the unit vector $\hat{\mathbf{n}}_f$. Note that in this expression $N_f$ is equal to the total amount of flux points of an element. The pseudo-code of this computation is presented in algorithm \ref{algo: divDiscFlux}.
\begin{algorithm}[H]
\caption{Element loop contribution to the computation of the flux divergence}\label{algo: divDiscFlux}
\begin{algorithmic}[1]
\Procedure{ElementLoop}{}
\For{$iElement \gets 0,\, N$}
\State $states \gets$ get states of $iElement$
\For{$iState \gets 0,\, N_{s}$}
\For{$iDim \gets 0,\, N_{dim}$}
\State $fluxes(iState,iDim) \gets$ get flux of $iState$ along axis $iDim$
\EndFor
\EndFor
\For{$iFlux\gets 0,\, N_f$}
\State $dim \gets $ get the normal direction in $iFlux$
\For{$iState\gets 0,\, N_s$}
\State $discFlux(iFlux) \gets 0$
\Comment{discontinuous flux}
\State $baseFct \gets$ get $\mathcal{l}_{iState}$ in $iFlux$
\State $discFlux(iFlux)  \mathrel{{+}{=}} fluxes(iState,dim) \cdot baseFct$
\EndFor
\EndFor
\For{$iState \gets 0,\, N_{s}$}
\State $result(iElement,iState) \gets 0$
\For{$iDim \gets 0,\, N_{dim}$}
\For{$jState \gets 0,\, N_{s}$}
\State $dBaseFct \gets$ get $\dfrac{\partial\mathcal{l}_{jState}}{\partial \xi_{iDim}}$ in $iState$
\State $result(iElement,iState) \mathrel{{+}{=}} fluxes(jState,iDim)\cdot dBaseFct$
\EndFor
\EndFor
\EndFor
\For{$iState \gets 0,\, N_{s}$}
\For{$iFlux\gets 0,\, N_f$}
\State $dim \gets $ get the normal direction in $iFlux$
\State $dh \gets$ get $\dfrac{\partial h_{iFlux}}{\partial \xi_{dim}}$ in $iState$
\State $result(iElement,iState)  \mathrel{{-}{=}} discFlux(iFlux) \cdot dh$
\EndFor
\EndFor
\EndFor
\State \textbf{return} $result$
\EndProcedure
\end{algorithmic}
\end{algorithm}

\subsubsection{Loop over Faces}
When computing equation \ref{loopFacesEq} in a loop over faces, the common interface flux needs to be evaluated. The common interface flux $\hat{\mathbf{f}}_{j}^{\delta I}$ can be computed if the states in the flux point of the left and right neighboring elements are known: a single flux point always belongs to two elements, as such, a left and right state $\hat{\mathbf{u}}^-$ and $\hat{\mathbf{u}}^+$ can be computed in each flux point. Therefore, the neighboring element geometric entities need to be accessed. A flux scheme such as Roe, Lax-Friedrichs or centered flux then provides the interface flux value $\hat{\mathbf{f}}_{j}^{\delta I}$. The terms $\hat{\mathbf{u}}^-$ and $\hat{\mathbf{u}}^+$ are evaluated using the Lagrange interpolation, as in equation \ref{ApproxSolution}. The divergence of the correction field can be computed in the same way as in equation \ref{divh}. Using equations \ref{fD} and \ref{divh}, the final computation performed in the loop over faces for the $i$-th solution point of the neighboring elements becomes the following:
\begin{equation}
    \sum_{j = 1}^{N_f} \frac{\partial h_j}{\partial \xi_j}(\boldsymbol{\xi}_i) \left(\left(\hat{\mathbf{f}}_{C,j}^{\delta I}(\hat{\mathbf{u}}^-(\boldsymbol{\xi}_{j}),\hat{\mathbf{u}}^+(\boldsymbol{\xi}_{j})) \right) \cdot \hat{\mathbf{n}}_j\right),
\end{equation}
where $\xi_j$ once again denotes the direction along the unit normal vector $\hat{\mathbf{n}}_j$. Note that here $N_f$ denotes the number of flux points on a face. Algorithm \ref{algo: divCorrFlux} presents how this is executed in pseudo-code.
\begin{algorithm}[H]
\caption{Face loop contribution to the computation of the flux divergence}\label{algo: divCorrFlux}
\begin{algorithmic}[1]
\Procedure{FaceLoop}{}
\For{$iOrientation \gets 0,\, $ maximum number of orientations}
\For{$iFace \gets 0,\, $ number of faces with $iOrientation$}
\State $leftNeighbor \gets$ get left neighbor element of $iFace$
\State $rightNeighbor \gets$ get right neighbor element of $iFace$
\For{$iFlux \gets 0,\, N_{f}$}
\State $leftLocalIndex(iFlux) \gets$ left index of $iFlux$ for $iOrientation$
\State $rightLocalIndex(iFlux) \gets$ right index of $iFlux$ for $iOrientation$
\State $leftState \gets 0$
\State $rightState \gets 0$
\For{$iState \gets 0,\, N_{s}$}
\State $iLeftState \gets$ get state of left neighbor at $iState$ 
\State $iRightState \gets$ get state of right neighbor at $iState$
\State $leftBaseFct \gets \mathcal{l}_{iState}$ in $leftLocalIndex(iFlux)$
\State $rightBaseFct \gets \mathcal{l}_{iState}$ in $rightLocalIndex(iFlux)$
\State $leftState \mathrel{{+}{=}} iLeftState\cdot leftBaseFct$
\State $rightState \mathrel{{+}{=}} iRightState\cdot rightBaseFct$
\EndFor
\State $interfaceFlux(iFlux) \gets \hat{\mathbf{f}}^{\delta I}_{C,iFlux}(leftState, rightState)\cdot \hat{\mathbf{n}}_{iFlux}$
\EndFor
\For{$iState \gets 0,\, N_{s}$}
\State $leftResult(leftNeighbor,iState) \gets 0$
\State $rightResult(rightNeighbor,iState) \gets 0$
\For{$iFlux \gets 0,\, N_{f}$}
\State $leftDivh \gets$ get divergence of $h_{leftLocalIndex(iFlux)}$ in $iState$
\State $rightDivh \gets$ get divergence of $h_{rightLocalIndex(iFlux)}$ in $iState$
\State $leftResult(leftNeighbor,iState) \mathrel{{+}{=}}$ 
\State $\qquad \qquad interfaceFlux(iFlux) \cdot leftDivh$
\State $rightResult(rightNeighbor,iState) \mathrel{{+}{=}}$
\State $\qquad \qquad interfaceFlux(iFlux) \cdot rightDivh$
\EndFor
\EndFor
\EndFor
\EndFor
\State \textbf{return} $leftResult$, $rightResult$
\EndProcedure
\end{algorithmic}
\end{algorithm}

\subsubsection{Treatment of the Boundary Faces}

A boundary face has only one neighboring element. As a consequence, only one state (``internal'' to the domain) can be evaluated in the flux points. However, there needs to be a left and right state in order to determine the interface flux. This is solved by imposing a ghost state in each flux point. Ghost states depend on the boundary condition attached to the boundary face. The way these states are computed is discussed in section \ref{ghost states}.

Algorithm \ref{algo: divCorrFlux} cannot be used directly for the boundary faces. The algorithm is executed on the inner faces TRS and a separate loop over the boundary faces TRS is performed. This procedure is very similar to algorithm \ref{algo: divCorrFlux}. The only differences being that the right state is now a ghost state that is computed by means of a separate boundary condition implementation and that the divergence of the correction flux is only computed for one neighbor element, i.e. the internal element attached to the boundary face.

\subsubsection{Computation of the Diffusive Flux Divergence}

As section \ref{GoverningEqs} shows, the diffusive flux depends on both the state $\hat{\mathbf{u}}$ and the corrected gradient of the state $\hat{\mathbf{q}} = \boldsymbol{\nabla} \hat{\mathbf{u}}$. Once $\hat{\mathbf{q}}$ is known, the algorithms for the computation of the convective flux divergence can be used (algorithms \ref{algo: divDiscFlux} and \ref{algo: divCorrFlux}). The approximate corrected gradient in the $\xi_i$-direction, in the reference domain, $\hat{\mathbf{q}}^{\delta}_{\xi_i}$ is computed as follows:
\begin{equation}\label{corrGrad}
    \hat{\mathbf{q}}^{\delta}_{\xi_i} = \frac{\partial \hat{\mathbf{u}}^\delta}{\partial \xi_i} + \sum_{j=1}^{N_f} (\hat{\mathbf{u}}^{\delta I}_j - \hat{\mathbf{u}}^{\delta}(\boldsymbol{\xi}_j))\frac{\partial \mathbf{h}}{\partial \xi_i}\cdot\hat{\mathbf{n}}_i,
\end{equation}
where $\hat{\mathbf{n}}_i$ is the unit normal vector along direction $\xi_i$.

The way the corrected gradient is constructed is analogous to the corrected flux. As a consequence, there is a term depending on the approximate state polynomial within an element and a term depending on the left and right state in a flux point. Following the same approach as the computation of the divergence of the convective flux, a loop over the elements and the inner and boundary faces is performed. 
The computation of the convective flux divergence is performed before the diffusive computation, such that the algorithm to determine the corrected gradient can be done in the loops over elements and faces of the convective algorithm. In this manner, no separate loops are needed to calculate the corrected gradients. After the execution of the convective algorithm, $\hat{\mathbf{q}}^{\delta}$ can be accessed to compute the diffusive flux divergence.

On the boundary, the ghost state can be used to determine the corrected gradient similarly to the approach used for the convective flux divergence. However, the method to evaluate the interface flux on the boundary must be modified. The diffusive flux depends on the state and the state gradient in the flux point. Consequently it does not suffice to compute a ghost state, but a ghost gradient has to be computed as well. Section \ref{ghost states} describes how these are determined.

Since the correction procedure of the gradients is analogous to the correction of the fluxes, an algorithm analogous to algorithms \ref{algo: divDiscFlux} and \ref{algo: divCorrFlux} is used. The main difference being that for the correction of the fluxes the variable extrapolated to the flux points is the discontinuous flux, while for the correction of the gradients the states are extrapolated to the flux points. Now only these extrapolated states are used to compute an interface value: this interface value is in this case not dependent on the extrapolated discontinuous fluxes. 

\subsubsection{Correction Functions}\label{implCorrFct}

The present solver uses VCJH correction functions for quadrilaterals and hexahedra. For these types of elements, the correction field can be constructed based on a 1D function defined in the 1D standard domain $[-1,1]$. This function is:
\begin{equation}\label{1DCorrFct}
    h = \frac{(-1)^P}{2}\left( \Upsilon_P - \left(\frac{\eta_P \Upsilon_{P-1} + \Upsilon_{P+1}}{1+\eta_P}\right)\right),
\end{equation}
where $\Upsilon_P$ is the $P$-th order Legendre polynomial. $\eta_P$ is equal to $\dfrac{c(2P+1)(a_P P!)^2}{2}$ and $a_P = \dfrac{(2P)!}{2^P (P!)^2}$. The scalar $c$ must be within the range $c_{\_} < c < \infty$, where
\begin{equation}\label{cMin}
    c_{\_} = \frac{-2}{(2P+1)(a_P P!)^2}.
\end{equation}
$c_{\_}$ corresponds to the smallest value of $c$ for which the FR schemes have been proven to be stable in 1D. Many values of $c$ are proposed in the literature: the value $c_{DG}$ for which the FR method reduces to a DG method, the value $c_{SD}$ for which it reduces to an SD method and the value proposed by Huynh $c_{g2}$: \cite{huynh2007flux,williams2011extension}
\begin{equation}
    c_{DG} = 0, \quad c_{SD} = \frac{2P}{(2P+1)(P+1)(a_P P!)^2}, \quad c_{g2} = \frac{2(P+1)}{(2P+1)P(a_P P!)^2}.
\end{equation}

Let $h_f (\boldsymbol{\xi}_i)$ denote the correction function associated with the $f$-th flux point, evaluated in the $i$-th solution point. Then $h_f(\boldsymbol{\xi}_i)$ is not zero if $\boldsymbol{\xi}_i$ and $\boldsymbol{\xi}_f$ have $N_{dim}-1$ coordinates in common. Here, $N_{dim}$ is the spatial dimensionality of the problem. Let $\xi_f$ be the coordinate of the flux point that it does not have in common with the solution point, and let $\xi_s$ be the corresponding solution point coordinate. Since flux points are on faces, $\xi_f$ is always equal to $\pm1$ in the reference domain. In the case where $h_f (\boldsymbol{\xi}_i)$ is not equal to zero, it can be evaluated as follows:
\begin{equation}
    h_f (\boldsymbol{\xi}_i) = \begin{cases}
    h(\xi_s), & \text{if $\xi_f=-1$,}\\
    h(-\xi_s), & \text{if $\xi_f=1$.}
  \end{cases}
\end{equation}
The gradient of the correction function along the $j$-th axis is denoted as $\frac{\partial h_f }{\partial \xi_j}(\boldsymbol{\xi}_i)$. The gradient of the correction function can only be non-zero if $h_f (\boldsymbol{\xi}_i)$ is non-zero. In addition, the derivative direction $\xi_j$ must be in the same coordinate direction as the axis defined by the solution and flux point. For example, if the solution point and flux point have the same $\xi$- and $\eta$-coordinate, but different $\zeta$-coordinates, only the derivative with respect to $\zeta$ is non-zero. If the above mentioned conditions are verified, the gradient of the correction function can be computed as follows:
\begin{equation}\label{derivh}
    \frac{\partial h_f}{\partial \xi_j}(\boldsymbol{\xi}_i) = \begin{cases}
    \frac{\partial h}{\partial \xi}(\xi_s), & \text{if $\xi_f=-1$,}\\
    -\frac{\partial h}{\partial \xi}(-\xi_s), & \text{if $\xi_f=1$.}
  \end{cases}
\end{equation}

\subsection{Boundary Conditions}\label{ghost states}

The boundary of a mesh consists of faces that hold flux points. As such, it is precisely in these points that boundary conditions are introduced in the FR formulation. In a flux point of an inner face, a left and right state and gradient can be defined. However, there can only be one state extrapolated to a flux point that lies on a boundary face. This is called the internal state $\mathbf{u}_{inner}$. In order to introduce the boundary condition that applies to a face, the second state and gradient in the flux point are computed based on this boundary condition. They are called ghost states $\mathbf{u}_{ghost}$ and ghost gradients $\mathbf{q}_{ghost}$.

In general, a state value in the boundary $u_{bnd}$ can be imposed by choosing the following ghost state: (Dirichlet condition)
\begin{equation}\label{dirichlet}
    u_{ghost} = 2u_{bnd} - u_{inner}.
\end{equation}
If a boundary condition does not impose a certain state value, the ghost state value can be set equal to the inner one, corresponding to applying a homogeneous Neumann condition for that specific value.
For the gradients, the same approach is used with the exception that a boundary condition will typically impose the gradient projected on the boundary normal:
\begin{equation}
    \mathbf{q}_{ghost} = \mathbf{q}_{inner} + 2(q_{bnd} - \hat{\mathbf{n}}\cdot \mathbf{q}_{inner})\ \hat{\mathbf{n}}.
\end{equation}
Once again, if no gradient needs to be imposed, $\mathbf{q}_{ghost}$ can simply be set to $\mathbf{q}_{inner}$.

\subsection{Time Marching Procedure}\label{time marching}

The RHS of the set of equations to be solved (equation \ref{consVgl}) is determined by discretizing the spatial domain of the problem in elements. The FR method is then used to evaluate the RHS in every solution point in each element. This discretized form of the RHS is called the residual $\mathbcal{R}$:
\begin{equation}\label{semi disc prob}
    \frac{\partial \mathbf{u}_{ce}}{\partial t} = \mathbcal{R}(\mathbf{u}_{ce},\mathbf{u}_{ne}),
\end{equation}
where the subscript $ce$ refers to the current element, while $ne$ represents the neighboring elements. This results in a semi-discretized equation. In order to fully solve the problem, the LHS of \ref{semi disc prob} also needs to be discretized. This is done by dividing the time domain in discrete time steps $\Delta t$. Subsequently, a time marching scheme can determine $\mathbf{u}(t+\Delta t)$.

For implicit time stepping in addition to the residual, the Jacobian matrix of the residuals is needed. Consider as an example the backward Euler implicit time marching scheme that can be written as follows:
\begin{equation}\label{bwd euler}
    \mathbf{u}_{ce}(t+\Delta t) = \mathbf{u}_{ce}(t) + \Delta t\ \mathbcal{R}(\mathbf{u}_{ce}(t+\Delta t),\mathbf{u}_{ne}(t+\Delta t)),
\end{equation}
The value of $\Delta t$ is determined by a $CFL$ number specified by the user. At time $t$, the value of $\mathbcal{R}(\mathbf{u}_{ce}(t+\Delta t),\mathbf{u}_{ne}(t+\Delta t))$ is not readily available since $\mathbf{u}(t+\Delta t)$ is unknown. Because $\mathbcal{R}$ is non-linear for the Navier-Stokes equations, expression \ref{bwd euler} is a non-linear equation that has to be solved during each iteration. This system of equations is linearized resulting in the following:
\begin{equation}\label{timemarch}
    \mathbf{u}(t+\Delta t) = \mathbf{u}(t) + \Delta t\ \left(\mathbcal{R}(\mathbf{u}(t)) + \left. \dfrac{\partial \mathbcal{R}}{\partial \mathbf{u}}\right|_t (\mathbf{u}(t+\Delta t) - \mathbf{u}(t)) \right),
\end{equation}
where $\left. \dfrac{\partial \mathbcal{R}}{\partial \mathbf{u}}\right|_t$ is the Jacobian matrix of the residuals with respect to the state vectors, evaluated at time $t$. Here the subscripts $ce$ and $ne$ are omitted since here $\mathbf{u}$ represents the vector of all state vectors in the spatial domain. The resulting large linear system is solved with for example a Generalized Minimal Residual approach (GMRES) \cite{saad1986gmres}.

The residual Jacobian matrix is determined by numerically approximating the partial derivatives with a simple first-order finite difference scheme in the $i$-th element:
\begin{equation}\label{implEval}
    \dfrac{\partial\mathbcal{R}}{\partial\mathbf{u}_i} = \dfrac{\mathbcal{R}(\mathbf{u}_i+\varepsilon_u) - \mathbcal{R}(\mathbf{u}_i)}{\varepsilon_u}.
\end{equation}


This philosophy of computing the residual Jacobian matrix is applied to the proposed LLAV scheme. In expression \ref{implEval}, the value of $\mathbcal{R}(\mathbf{u}_i)$ is known, since this is simply what is computed by the FR method. For the LLAV scheme, this is the artificial flux divergence. However $\mathbcal{R}(\mathbf{u}_i+\varepsilon_u)$ also needs to be computed. This can be written as:
\begin{equation}\label{jacobAV}
\mathbcal{R}(\mathbf{u}_i+\varepsilon_u) = -\boldsymbol{\nabla}\cdot \mathbf{f}_C(\mathbf{u}_i+\varepsilon_u) + \boldsymbol{\nabla}\cdot \mathbf{f}_D(\mathbf{u}_i+\varepsilon_u, \mathbf{q}(\mathbf{u}_i+\varepsilon_u)) + \boldsymbol{\nabla}\cdot \mathbf{f}_{AV}(\mathbf{q}(\mathbf{u}_i+\varepsilon_u)).
\end{equation}
It is clear that for the evaluation of the artificial viscosity part of expression \ref{jacobAV}, only the corrected gradients $\mathbf{q}$ computed based on the perturbed state $\mathbf{u}_i+\varepsilon_u$ is needed. This is however already computed for the diffusive part of equation \ref{jacobAV}. As such these variables are not computed again. Once $\boldsymbol{\nabla}\cdot \mathbf{f}_{AV}(\mathbf{q}(\mathbf{u}_i+\varepsilon_u))$ is calculated, the residual Jacobian matrix, including the contribution of the LLAV scheme, is known. After solving the time marching equation \ref{timemarch}, the positivity preservation scheme is applied to all elements on the newly computed state $\mathbf{u}(t+\Delta t)$.

\section{Verification of the FR Solver}\label{verification}
In this section, the developed FR solver is investigated and verified by means of various test cases dealing with both the compressible Euler and Navier-Stokes equations. First of all, the inviscid subsonic flow through a channel with a smooth sinusoid bump is studied in 2D and 3D. As a second test case governed by the Euler equations, the 2D inviscid flow around a cylinder is studied. This test case is also extended to viscous flow governed by the Navier-Stokes equations. More specifically, a laminar flow around the cylinder with stable vortices is studied in both 2D and 3D. Finally, the ability of the FR solver to accurately capture shocks is discussed based on the hypersonic flow over a cylinder and cone. All test cases that are presented in this section are solved using an implicit backward Euler time marching procedure, unless mentioned otherwise. Gauss-Legendre solution and flux points are used.

\subsection{Inviscid Subsonic Flow Through a Channel with a Sinusoid Bump in 2D}\label{2Dsinebump}
\begin{figure}[H]
  \centering
  \includegraphics[width=0.6\textwidth]{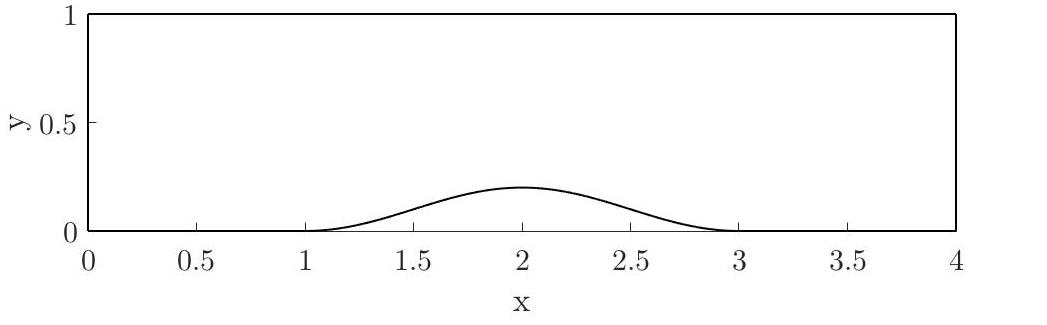}
  \caption{Channel with a smooth sinusoid bump between $x=1$ and $x=3$}
  \label{fig:SineBumpGeometry}
\end{figure}

\begin{figure}[H]
    \centering
    \begin{subfigure}[b]{0.49\textwidth}
        \includegraphics[width=\textwidth]{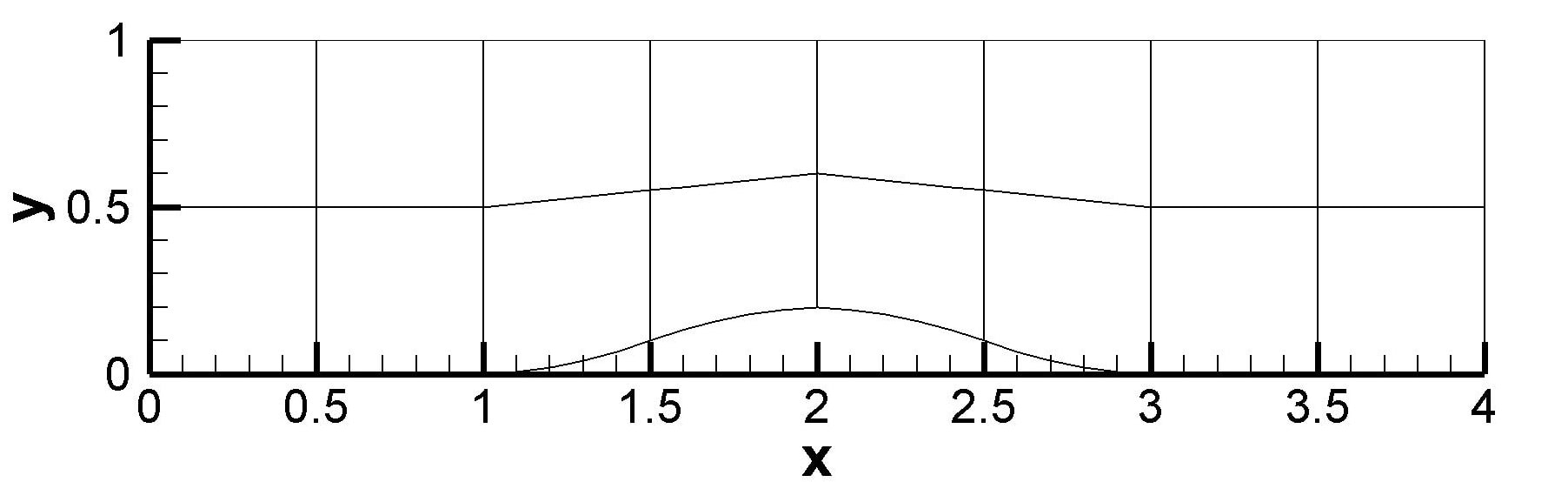}
        \caption{Grid 1, 3 $\times$ 9 nodes}
        \label{fig:MeshBump1}
    \end{subfigure}
    \begin{subfigure}[b]{0.49\textwidth}
        \includegraphics[width=\textwidth]{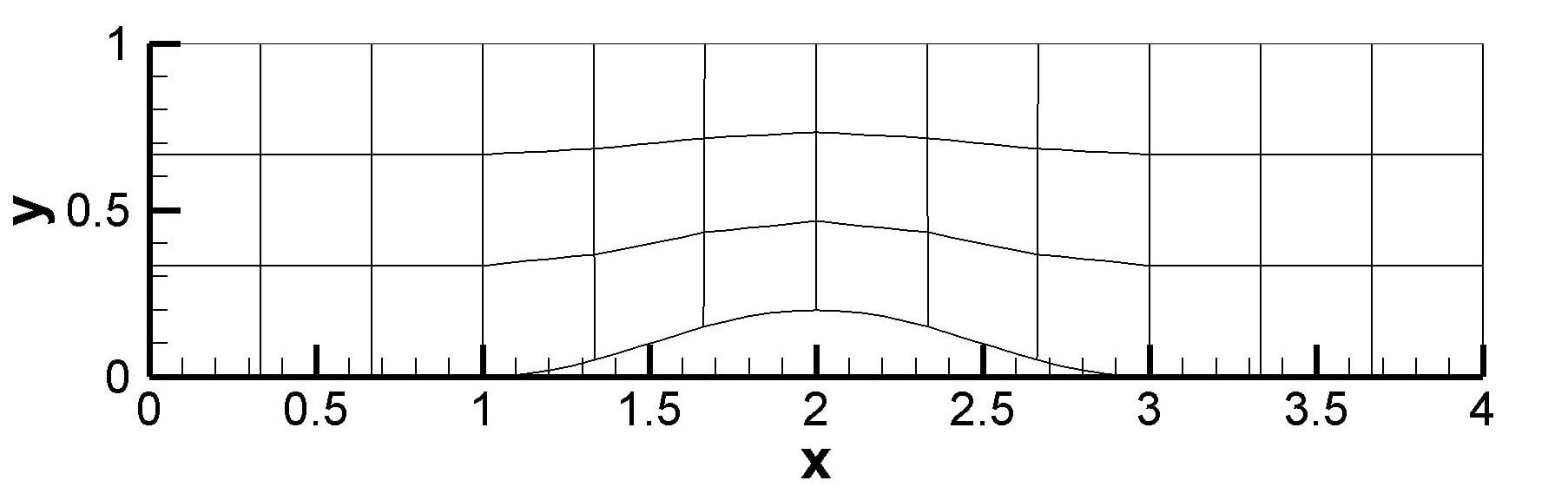}
        \caption{Grid 2, 4 $\times$ 13 nodes}
        \label{fig:MeshBump2}
    \end{subfigure}
    \par\bigskip
    \begin{subfigure}[b]{0.49\textwidth}
        \includegraphics[width=\textwidth]{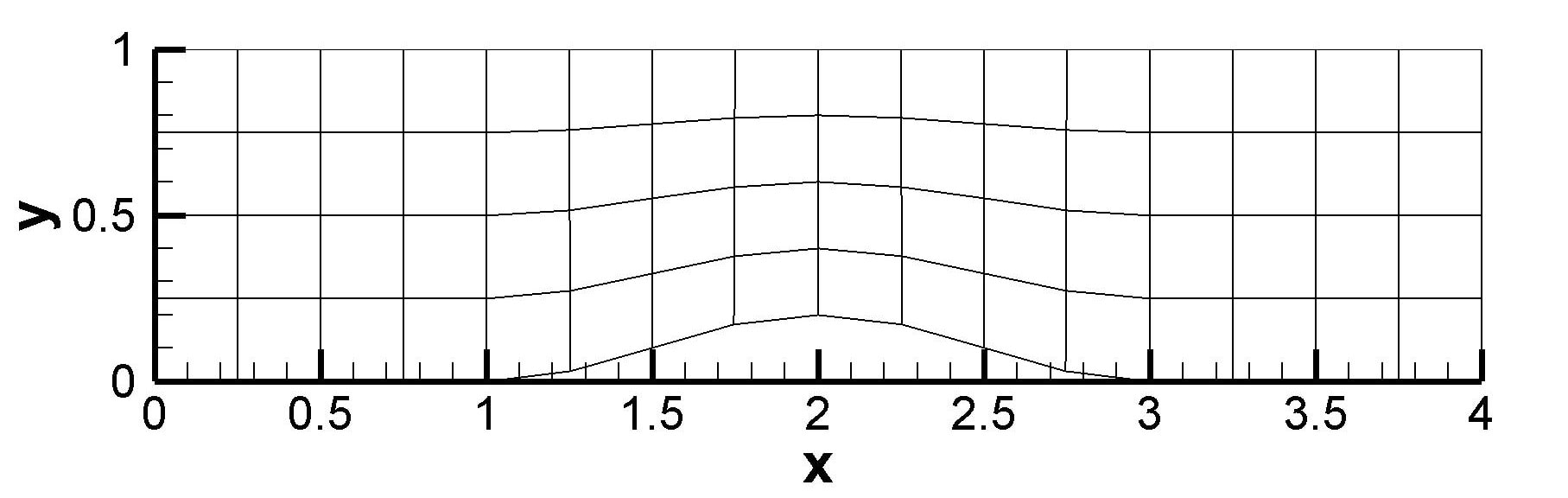}
        \caption{Grid 3, 5 $\times$ 17 nodes}
        \label{fig:MeshBump3}
    \end{subfigure}
    \begin{subfigure}[b]{0.49\textwidth}
        \includegraphics[width=\textwidth]{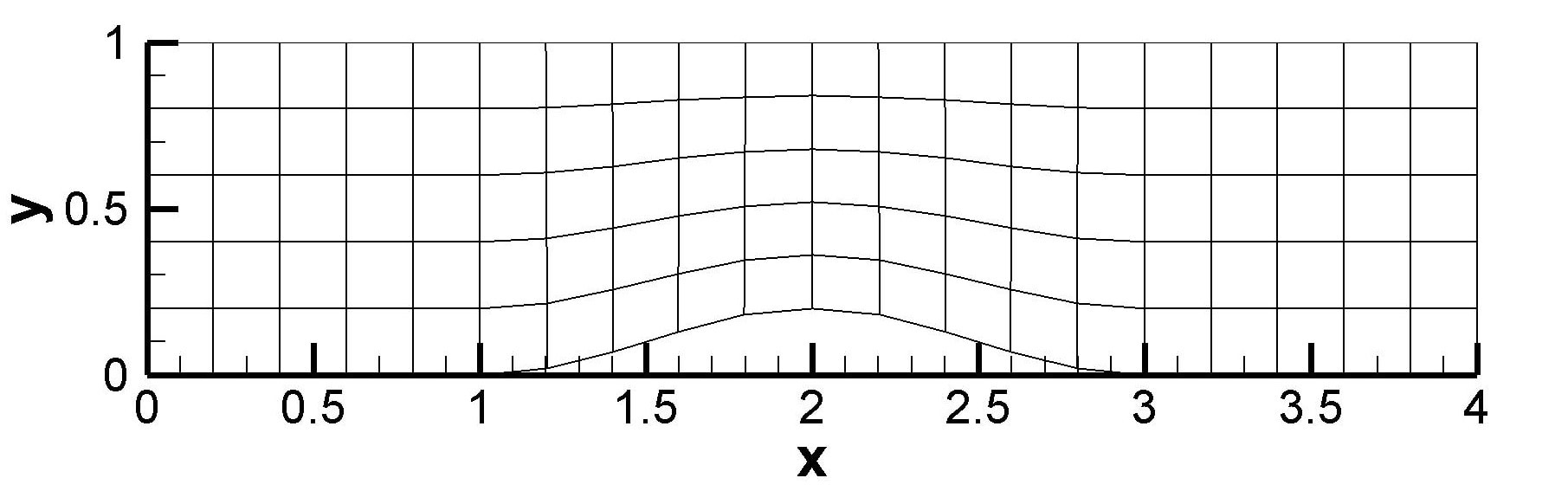}
        \caption{Grid 4, 6 $\times$ 21 nodes}
        \label{fig:MeshBump4}
    \end{subfigure}
    \caption{Discretization of the channel with a bump with quadrilateral cells}
    \label{fig:MeshesSineBump}
\end{figure}

\begin{figure}[H]
  \centering
  \includegraphics[width=0.5\textwidth]{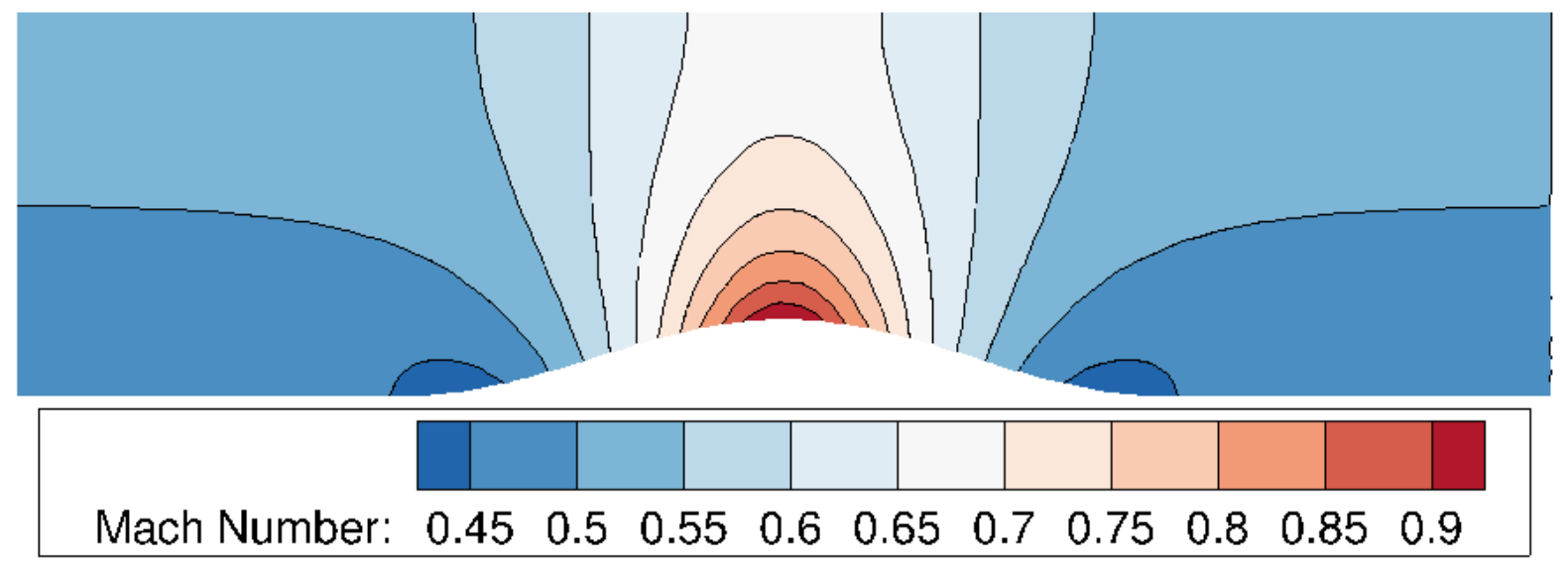}
  \caption{Mach contours for the inviscid flow through a channel with a bump obtained with P3Q2 elements on grid 4}
  \label{fig:MachSineBump}
\end{figure}

\begin{figure}[H]
  \centering
  \includegraphics[width=0.35\textwidth]{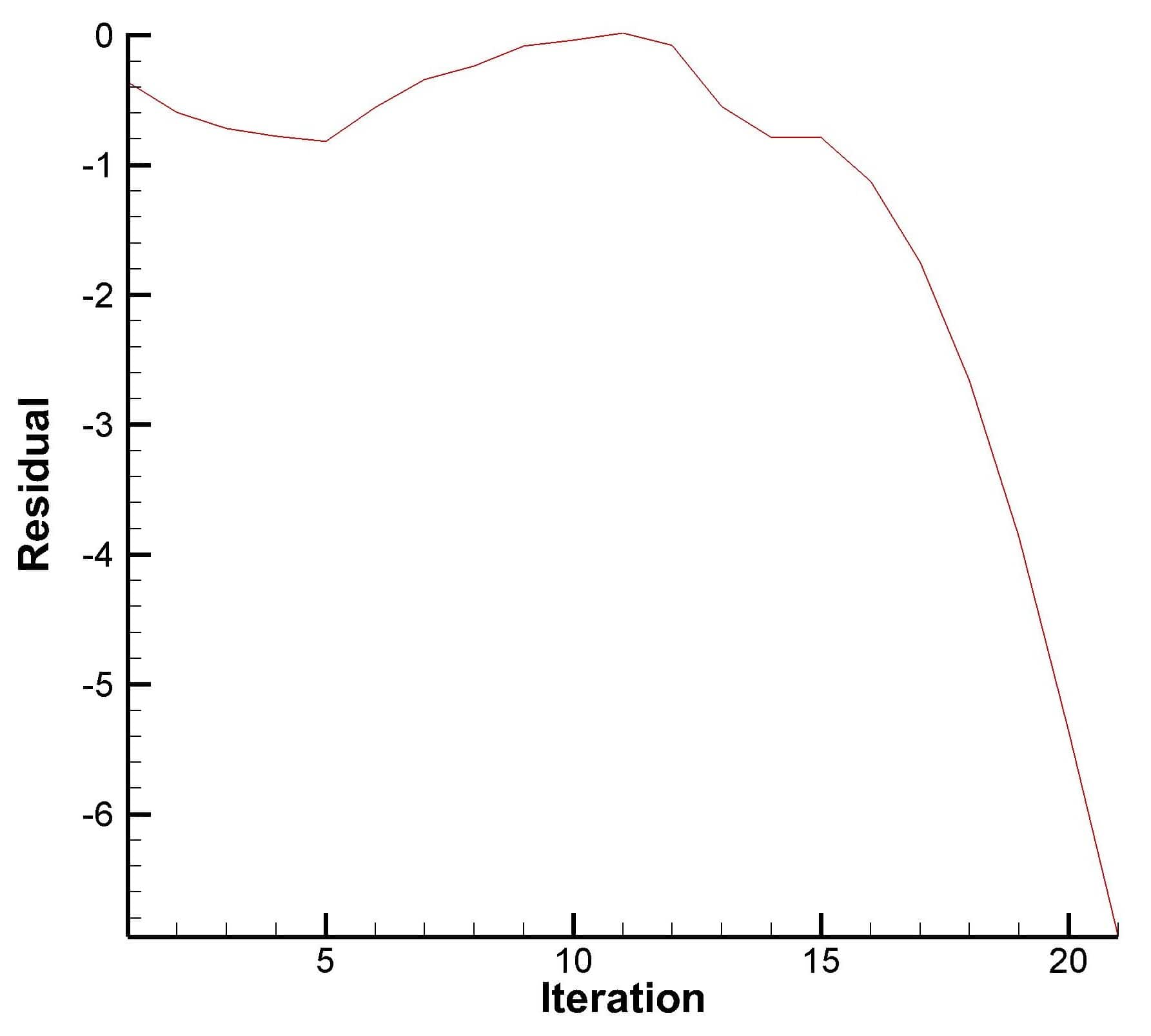}
  \caption{Logarithm of the residual of the density for the inviscid flow through a channel with a bump obtained with P3Q2 elements on grid 4}
  \label{fig:bumpRes}
\end{figure}

In this section, the inviscid flow at $M = 0.5$ through a channel with a smooth sinusoid bump is investigated. The physical domain is bounded between the inlet ($x=0$) and the outlet ($x=4$), and between the bottom ($y=0$) and the top ($y=1$) edges. The top is straight, while the bottom consists of a smooth bump symmetrically positioned between the in- and outlet, as shown in figure \ref{fig:SineBumpGeometry}. The bump is defined by the following expression:
\begin{equation}
    f(x) = 0.1 + 0.1  \cos(\pi(x-2)), \quad \text{with} \quad x \in [1;3].
\end{equation}
A slip-wall boundary condition is imposed to both the bottom and the top edge. Subsonic inlet (imposing total inlet pressure and temperature) and subsonic outlet conditions (imposing outlet pressure) are imposed such that the resulting outflow Mach number is 0.5 at zero angle of attack as proposed by \cite{wang2013high}. The free-stream flow values are summarized in table \ref{tab:freeStreamBump}.
\begin{table}[H]
\caption{\label{tab:freeStreamBump} Free-stream values}
\centering
\begin{tabular}{ccccc}
\hline
$M_\infty$ & $\rho_\infty\ [-]$ & $p_\infty\ [-]$ & $v_{x,\infty}\ [-]$ & $v_{y,\infty}\ [-]$\\\hline
0.5 & 1 & 1 & $0.5\sqrt{1.4}$ & 0\\
\hline
\end{tabular}
\end{table}
The physical domain is discretized by means of four different grids, consisting of quadrilateral elements. These grids are composed of $3 \times 9$, $4 \times 13$, $5 \times 17$, and $6 \times 21$ nodes, respectively. The first value denotes the number of nodes on the inlet and the outlet, while the second refers to the number of nodes on the top and bottom edges of the channel. The four meshes are constructed by equally distributing the nodes along the boundaries as represented in figure \ref{fig:MeshesSineBump}. Both straight edged elements with linear geometric mapping (Q1), as well as curvilinear elements with quadratic geometric mapping (Q2) are used. However, large errors may pollute the solution in the vicinity of the boundary if the physical boundary does not coincide with the computational boundary. Therefore, special treatment is required for the straight edged elements in order to correctly take into account the curvature of the bump as demonstrated by Krivodonova et al. \cite{krivodonova2006high}. As a consequence, all the simulations are run with curvilinear quadrilateral elements in order to avoid the generation of an unphysical ``boundary layer'' as discussed further on in this section. In this work, the elements are denoted by P$p$Q$q$ following the convention of Bassi and Rebay \cite{bassi1997euler}. Here, $p$ refers to the order of the polynomials used to approximate the solution, and $q$ refers to the order of the polynomials used for the geometric mapping.
The Mach contour plot obtained with P3Q2 elements on grid 4 (6 $\times$ 21 nodes) is shown in figure \ref{fig:MachSineBump}. For an inviscid flow, no entropy should be produced within the flow field. In addition, the geometry of the sinusoid bump channel is completely symmetrical with respect to the vertical plane at the bump ($x=2$). Consequently, the solution for the Euler equations of this test case should also be symmetrical. From figure \ref{fig:MachSineBump} it is clear that the flow field is smooth and is almost exactly symmetrical with respect to the bump. Figure \ref{fig:bumpRes} shows the logarithm of the density residual as a function of the simulation iterations. The CFL law is the following:
\begin{equation}
    CFL(n) = \begin{cases}
    0.5, & \text{if $n\leq 5$,}\\
    \min(0.5\cdot 2^{n-5},10^4), & \text{otherwise},
    \end{cases}
\end{equation}
where $n$ is the iteration. Using implicit time stepping methods, there is no constraint on the CFL number, which greatly speeds up convergence: by being able to use large CFL numbers, convergence is reached in 21 iterations for this test case. If an explicit time stepping method is used, the CFL limit for FR is:
\begin{equation}
    CFL_{max} = \frac{1}{P+1}.
\end{equation}
For the P3Q2 test case, this means that $CFL_{max} = 0.25$.


Since the entropy should remain constant through the entire domain, the entropy error $\varepsilon_s$ is used as an indication for the solution accuracy as proposed in \cite{wang2013high}. The entropy error $\varepsilon_s$ in the $i$-th state is defined as:
\begin{equation}
    \varepsilon_{s,i} = \frac{s - s_i}{s}.
\end{equation}


The exact value of the specific entropy $s$ can be calculated using the inlet variables. Based on the total pressure $p_t$ and the total temperature $T_t$ imposed by the subsonic inlet boundary condition, the specific entropy $s$ of the flow field is calculated as $s = p \rho^{-\gamma}$ where $\rho$ and $p$ are the inlet density and pressure. The $L_1$- and the $L_2$-norm of the entropy error $\varepsilon_s$ are defined as follows:
\begin{equation}\label{L1norm}
    L_1(\varepsilon_s) = \sum_{n=1}^{N} \int_{\Omega_n} |\varepsilon_s| d\Omega_n \approx \sum_{n=1}^{N} \sum_{j=1}^{N_q} |\hat{\varepsilon}_s| J_n \omega_{j} \quad \text{ and } \quad L_2(\varepsilon_s) = \sqrt{\sum_{n=1}^{N} \int_{\Omega_n}  {\varepsilon_s}^2 d\Omega_n}  \approx \sqrt{\sum_{n=1}^{N} \sum_{j=1}^{N_q}  {\hat{\varepsilon}_s}^2 J_n \omega_j },
\end{equation}
where $\Omega_{S,n}$ is the standard domain associated with the $n$-th element, $J_n$ the Jacobian determinant of the $n$-th element, $N_q$ the number of quadrature points and $\omega_j$ the $j$-quadrature weight. As equation \ref{L1norm} shows, Gauss quadrature is used to approximate the integral over the element domain as in \cite{golub1969calculation}.

A grid convergence study is carried out by computing the $L_1$- and the $L_2$-norm of the entropy error $\varepsilon_s$ on the four successively refined grids for the P1Q2, P2Q2, P3Q2, P4Q2 and P5Q2 FR schemes. The calculation is considered to be converged when the density residual $\mathcal{R} \leq$ 1e-6. In all simulations, Roe flux was used to compute the inviscid fluxes and the VCJH parameter $c$ was set equal to the $g_{2}$ value as described in section \ref{implCorrFct}. The order of accuracy of the FR solver is then computed as the slope of the linear least squares fit of 
\begin{equation}
    \log \Delta x \propto \log L(\varepsilon_s),
\end{equation}
 where $L$ is either the $L_1$- or the $L_2$-norm. The characteristic spacing of the grid $\Delta x$ is approximated as $N^{-1/2}$ where $N$ is the total number of elements in the grid, following the procedure of Witherden \cite{witherden2015heterogeneous}. The results of the grid convergence study are summarized in table \ref{ConvergenceStudyTable}. The obtained order of accuracy is approximately P+1 for most cases, as is expected. In the P2Q2 and P4Q2 cases a significantly higher order of accuracy than $P+1$ is reached.
 
\begin{figure}[H]
    \centering
    \begin{subfigure}[b]{0.48\textwidth}
        \includegraphics[width=\textwidth]{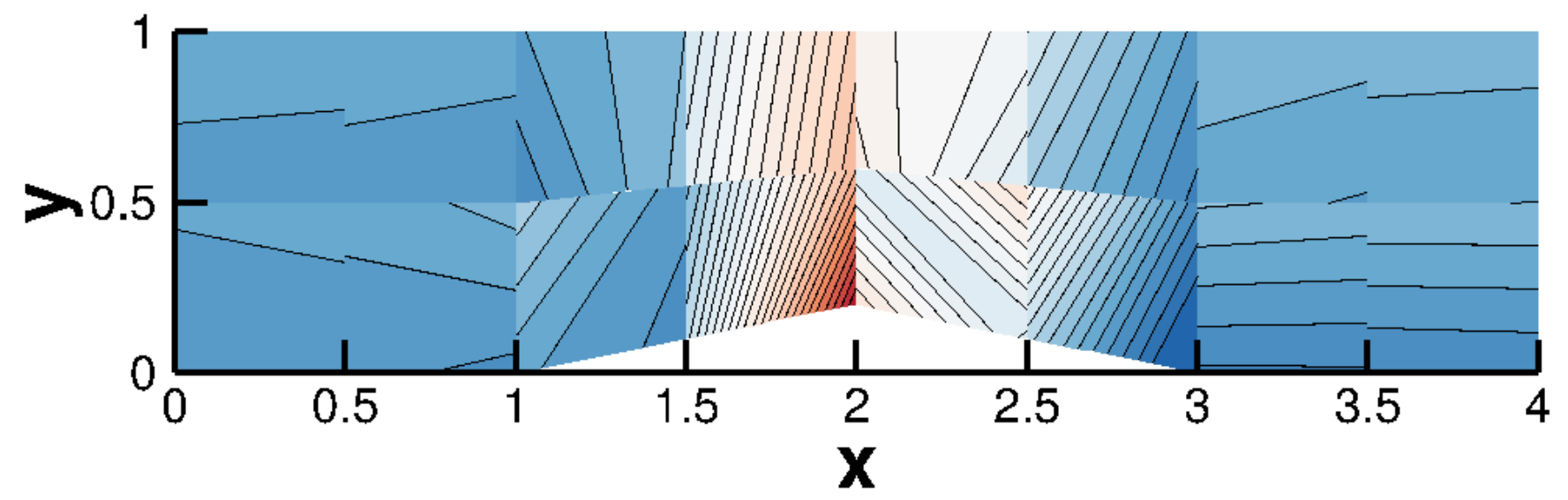}
        \caption{P1Q2}
        \label{fig:SineBump_P1Q2_2_8}
    \end{subfigure}
    \quad
    \begin{subfigure}[b]{0.48\textwidth}
        \includegraphics[width=\textwidth]{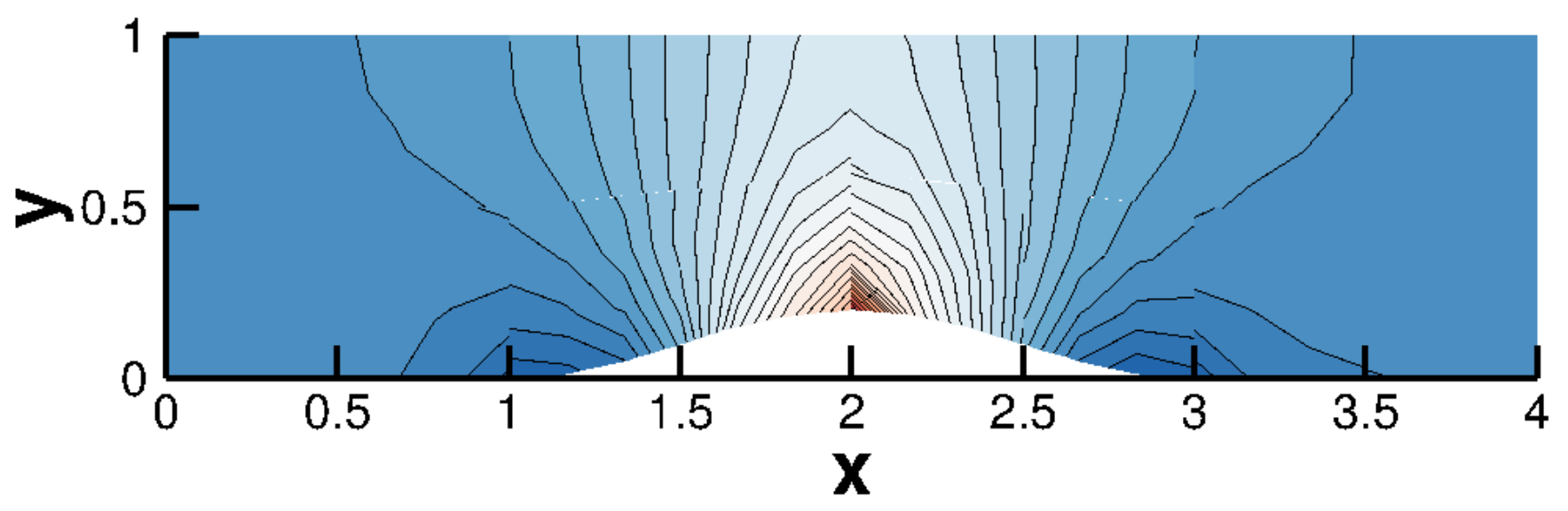}
        \caption{P3Q2}
        \label{fig:SineBump_P3Q2_2_8}
    \end{subfigure}
    \quad
    \begin{subfigure}[b]{0.48\textwidth}
        \includegraphics[width=\textwidth]{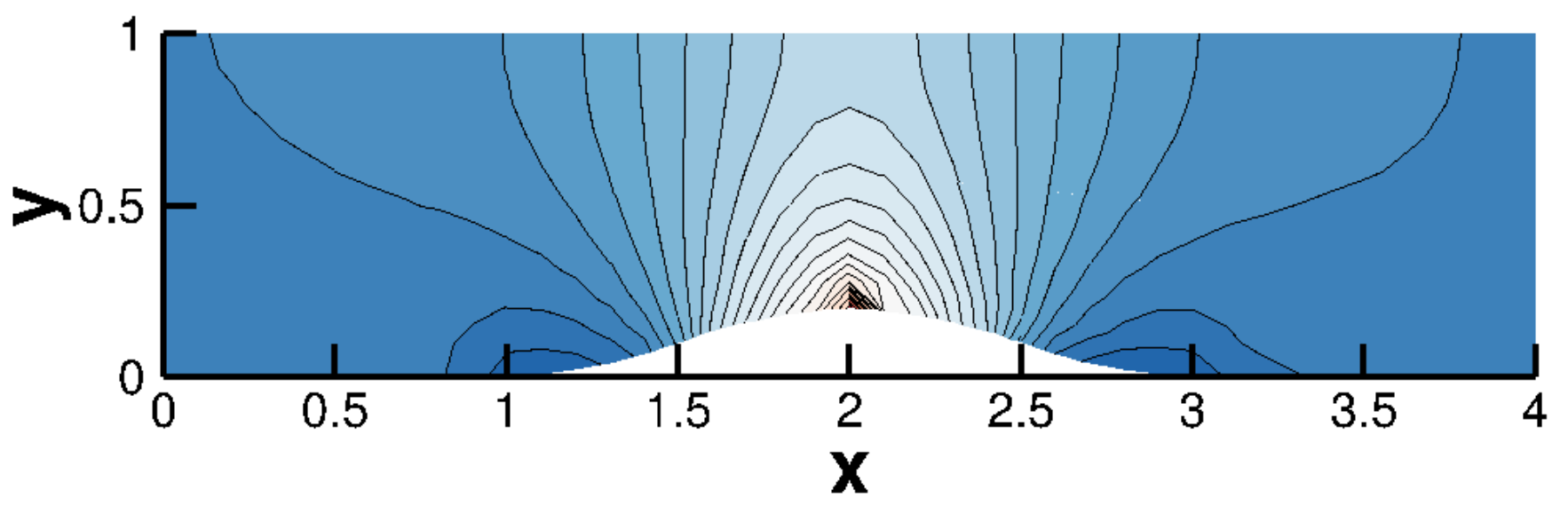}
        \caption{P5Q2}
        \label{fig:SineBump_P5Q2_2_8}
    \end{subfigure}
    \caption{Mach contours for the inviscid flow through a channel with a smooth bump on the coarsest grid (3$\times$9 nodes) under $p$-refinement}
    \label{fig:SineBump_2_8_Prefinement}
\end{figure}

It is clear that the accuracy is improved when either the polynomial order P to approximate the solution within an element is increased while maintaining the same element size, or when the mesh is refined for the same polynomial order P. The former method is commonly referred to as $p$-refinement, whereas the latter is known as $h$-refinement. Table \ref{ConvergenceStudyTable} clearly illustrates the advantage of $p$-refinement over $h$-refinement. Both the $L_1$- and the $L_2$-norm of the entropy error obtained with P1Q2 elements on the finest grid are approximately reduced by a factor of 10 when using P4Q2 elements on the coarsest grid, even though the same number of degrees of freedom is used, namely 400. The same conclusion can be drawn when comparing the error obtained with P2Q2 elements on the finest grid to those obtained with P4Q2 elements on grid 2, while the number of DOFs is 900 for both cases. The strength of $p-$refinement is clearly illustrated in figure \ref{fig:SineBump_2_8_Prefinement}, which shows the Mach contours obtained with second- (P1Q2), fourth- (P3Q2), and sixth-order (P5Q2) FR schemes on the coarsest grid. 

\begin{table}[ht!]
\centering
\begin{tabular}{crrrrr}
\hline
$P$ & \textbf{\#DOF} & $L_1$ \textbf{error} & $L_1$ \textbf{order} & $L_2$ \textbf{error} & $L_2$ \textbf{order}\\ \hline
1          & 64             &        3.16e-02         & -                 &        2.84e-02           & -       \\
           & 144            &        1.33e-02           &         2.13      &         1.58e-02          &     1.44  \\
           & 256           &         7.66e-03          &            1.50       &          1.03e-02         &    1.92   \\
           & 400           &          4.85e-03         &    2.05               &           7.35e-03        &   1.51      \\           &            &                   &                   &                   &             \\
2          & 144            &        5.19e-03           & -                 &      5.53e-02             & -           \\
           & 324            &         1.46e-03          &    3.13               &        1.84e-03           &     2.72   \\
           & 576           &        5.88e-04           &           3.16        &          7.60e-04         &     3.06  \\
           & 900           &        2.99e-04           &          3.04         &          3.83e-04         &     3.07   \\           &            &                   &                   &                   &         \\
3          & 256            &           1.05e-03        & -                 &          1.43e-03         & -            \\
           & 576           &         2.00e-04          &    4.08               &          2.85e-04         &     3.98     \\
           & 1024           &        6.98e-05           &          3.67         &         1.05e-04          &    3.49    \\
           & 1600          &          3.31e-05         &            3.34       &          5.12e-05         &     3.20   \\           &             &                   &                   &                   &             \\
4          & 400            &        3.26e-04           & -                 &         6.6e-04          & -          \\
           & 900           &         3.22e-05          &           5.70        &         7.16e-05          &     5.42    \\
           & 1600           &         6.11e-06          &          5.78         &           1.65e-05        &    5.11     \\
           & 2500          &         2.35e-06          &            4.28       &           6.15e-06        &     4.42  \\           &             &                   &                   &                   &           \\
5          & 576            &         2.38e-04          & -                 &          5.96e-04         & -              \\
           & 1296           &           2.15e-05        &            5.93       &          6.05e-05         &     5.64    \\
           & 2304           &        3.35e-06           &   6.46                &           1.07e-05        &    6.02     \\
           & 3600          &        1.06e-06           &    5.04               &          3.36e-06         &      5.19  \\ \hline
\end{tabular}
\caption{Results of the grid convergence study for the 2D inviscid flow through a channel with a bump (Q2 curvilinear elements) using the $L_1$- and $L_2$-norm of the entropy error $\varepsilon_s$ as an indication for the solution accuracy}
\label{ConvergenceStudyTable}
\end{table}

\subsubsection{Treatment of the Boundary}

\begin{figure}[H]
    \centering
    \begin{subfigure}[b]{0.49\textwidth}
        \includegraphics[width=\textwidth]{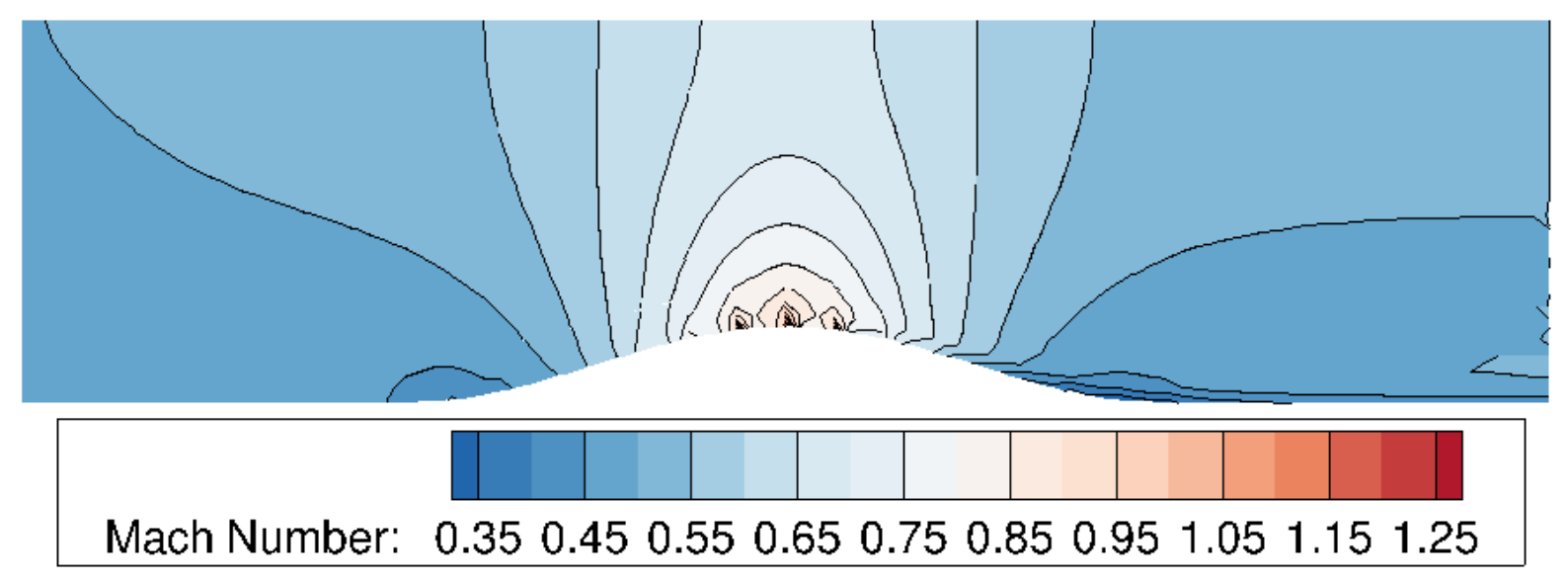}
        \caption{P3Q1}
    \end{subfigure} 
    \begin{subfigure}[b]{0.49\textwidth}
        \includegraphics[width=\textwidth]{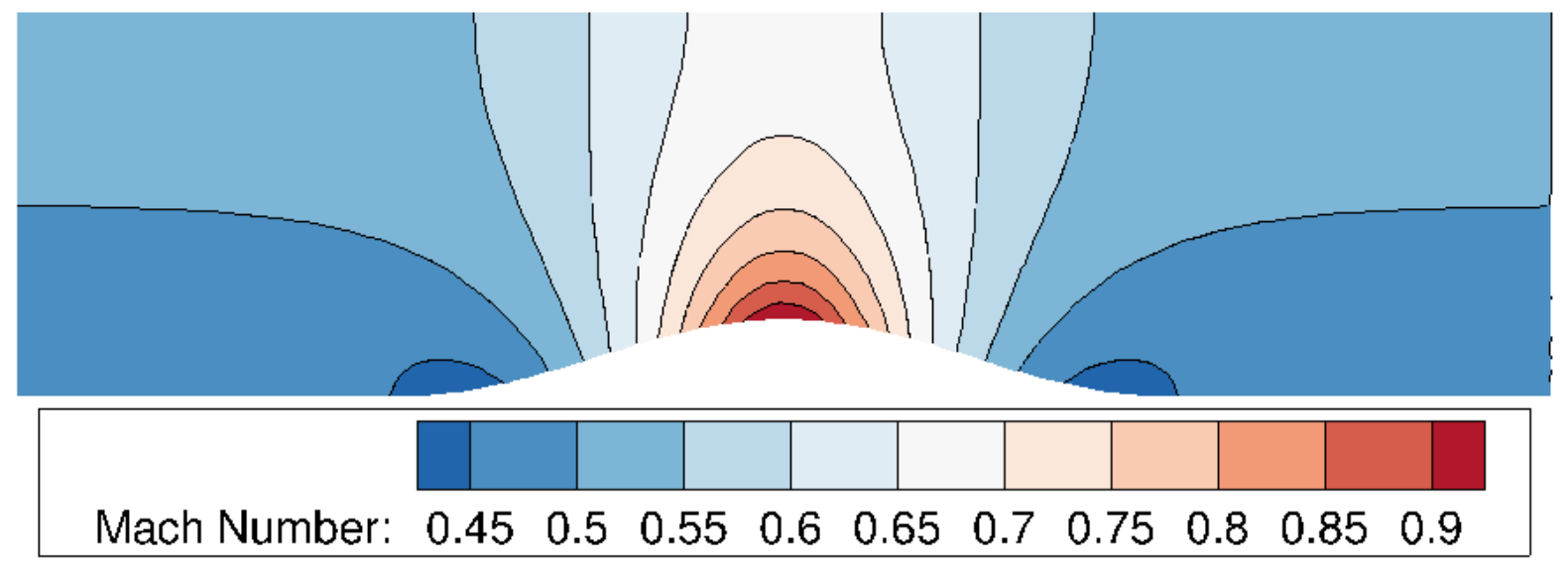}
        \caption{P3Q2}
    \end{subfigure}
    \caption{Mach contours obtained with P3Q1 and P3Q2 elements for grid 4}
    \label{fig:Q1wake}
\end{figure}

For high-order methods, the mesh approximation of the physical boundary has a large influence on the solution as discussed in  \cite{krivodonova2006high,luo2001influence,van2009development}. If the order of the geometric approximation of the boundary is too low in comparison to the solution polynomial order, the solution accuracy drops significantly and unphysical phenomena for Euler problems, such as spurious wakes and boundary layers, can form. To illustrate this, consider the P3Q2 and P3Q1 solutions on grid 4 using Lax-Friedrichs flux. For the solution where the mesh has a geometric order of Q1, it is clear that an unphysical ``entropy layer'' is generated behind the bump as shown in figure \ref{fig:Q1wake}. This layer however is not present in the Q2 solution. As a consequence, the accuracy obtained with curvilinear Q2 elements is significantly higher than the accuracy obtained with straight edged Q1 elements:
\begin{align}
    L_1(\varepsilon_s(t_{\infty}))_{Q1} &= \text{2.88e-03}, &L_2(\varepsilon_s(t_{\infty}))_{Q1} = \text{6.48e-03},\\
    L_1(\varepsilon_s(t_{\infty}))_{Q2} &= \text{1.33e-05}, &L_2(\varepsilon_s(t_{\infty}))_{Q2} = \text{2.12e-05}.
\end{align}

\begin{figure}[H]
  \centering
  \includegraphics[width=0.4\textwidth]{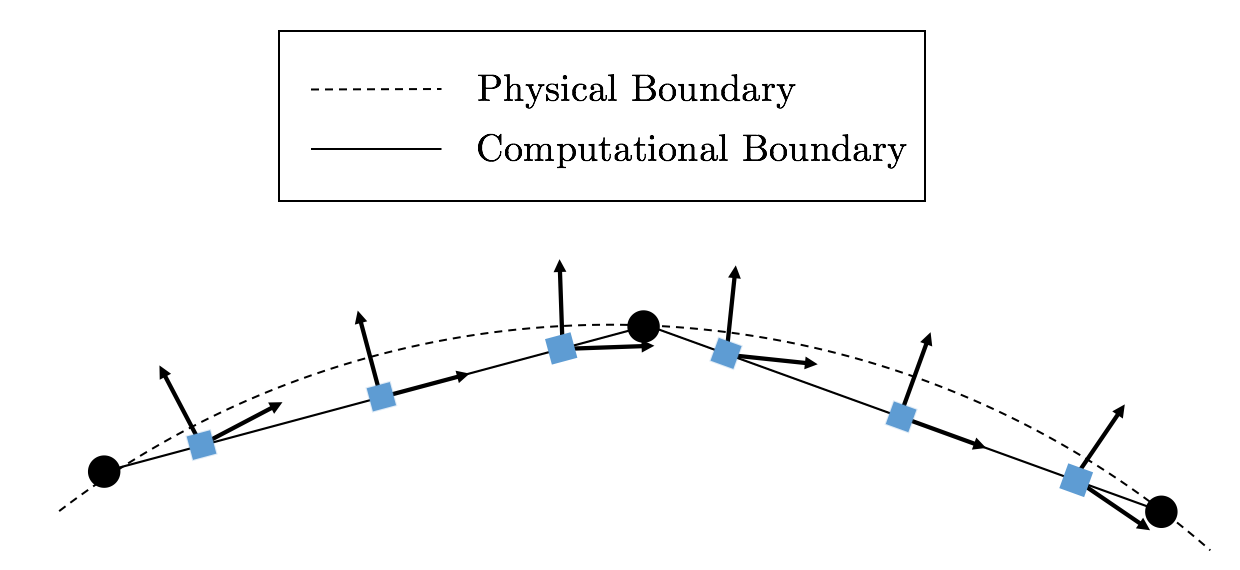}
  \caption{Schematic presentation of the use of analytical boundary normals}
  \label{fig:bndNormals}
\end{figure}

Therefore, the use of straight edged Q1 elements requires special treatment in order to take into account the curvature of the bump. A possible solution for this is the introduction of analytical boundary normals, as applied in \cite{krivodonova2006high,van2009development}. It consists of using the normal to the exact analytical form of the physical boundary for the flux projection on the boundary instead of the constant normal along the straight edged element. This is illustrated in figure \ref{fig:bndNormals}. This approach gives a similar result as the use of curvilinear elements.

\subsubsection{Parameter Study}

This section investigates the influence of the parameters of the VCJH $c$-parameter as well as the flux and solution point distribution on the accuracy of the obtained solution for the 2D channel with a sinusoid bump test case. In order to investigate the influence of a parameter, a reference test case is chosen. This case is the 2D channel with a sinusoid bump with geometric order Q2 on grid 3. Unless mentioned otherwise, the FR parameters are: Roe flux, the $c_{g_2}$ value of the $c$ parameter and Gauss-Legendre points for the solution and flux point distribution.

\begin{figure}[ht]
  \centering
  \includegraphics[width=0.4\textwidth]{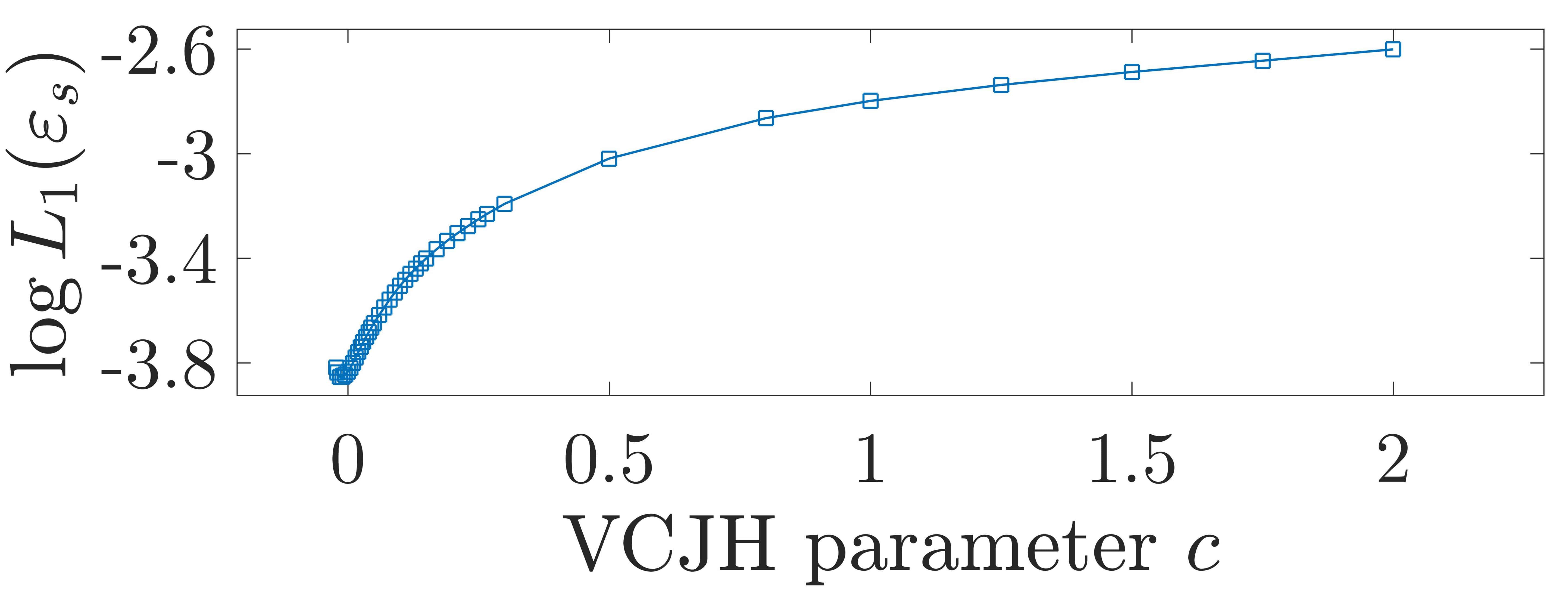}
  \caption{Stagnation of the $L_1$-norm of the entropy error $\varepsilon_s$ as $c$ approaches $c_{\text{large}}$}
  \label{fig:StagnationError}
\end{figure}

\begin{figure}[ht!]
    \centering
    \begin{subfigure}[b]{0.3\textwidth}
        \includegraphics[width=\textwidth]{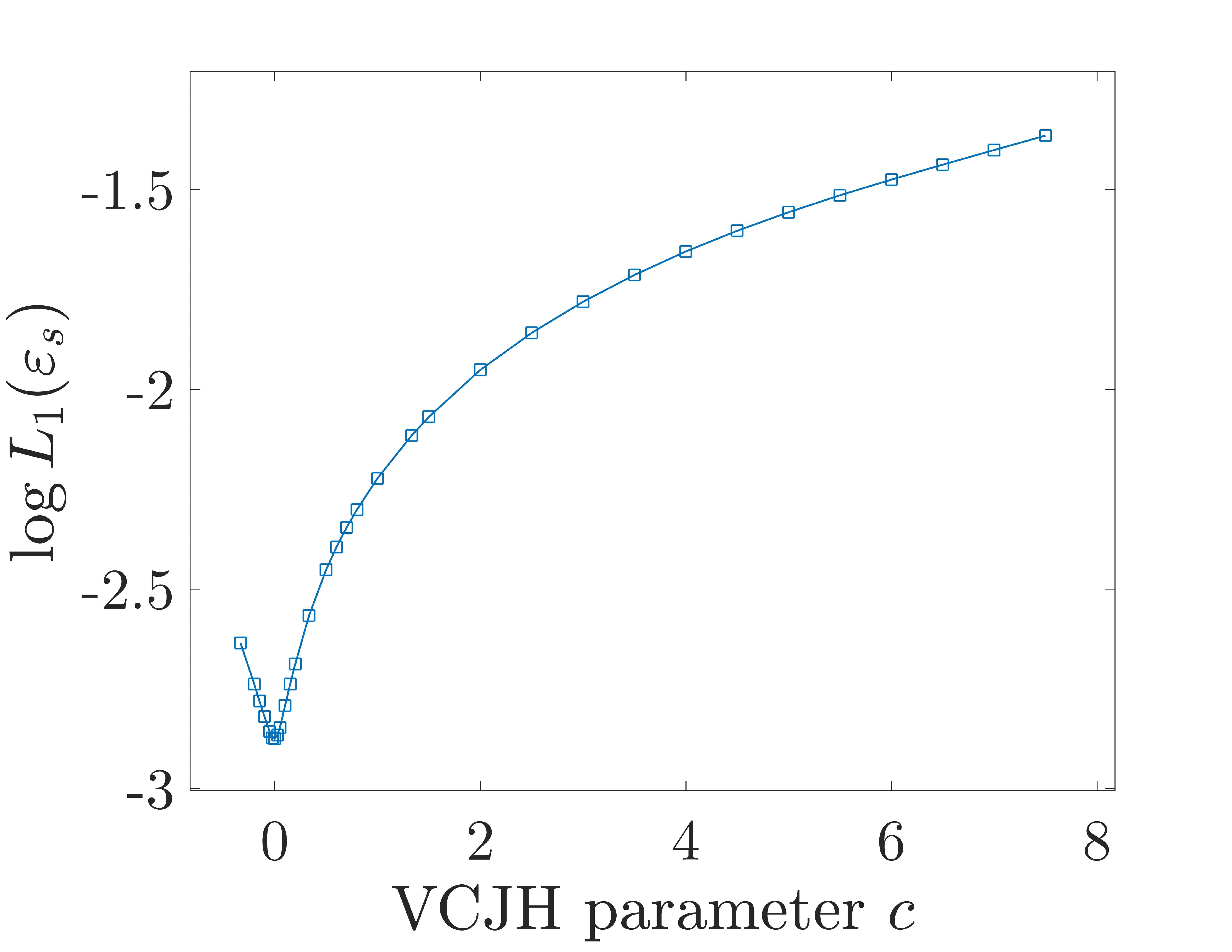}
        \caption{P1}
        \label{fig:cInfl_P1}
    \end{subfigure}
    \quad
    \begin{subfigure}[b]{0.3\textwidth}
        \includegraphics[width=\textwidth]{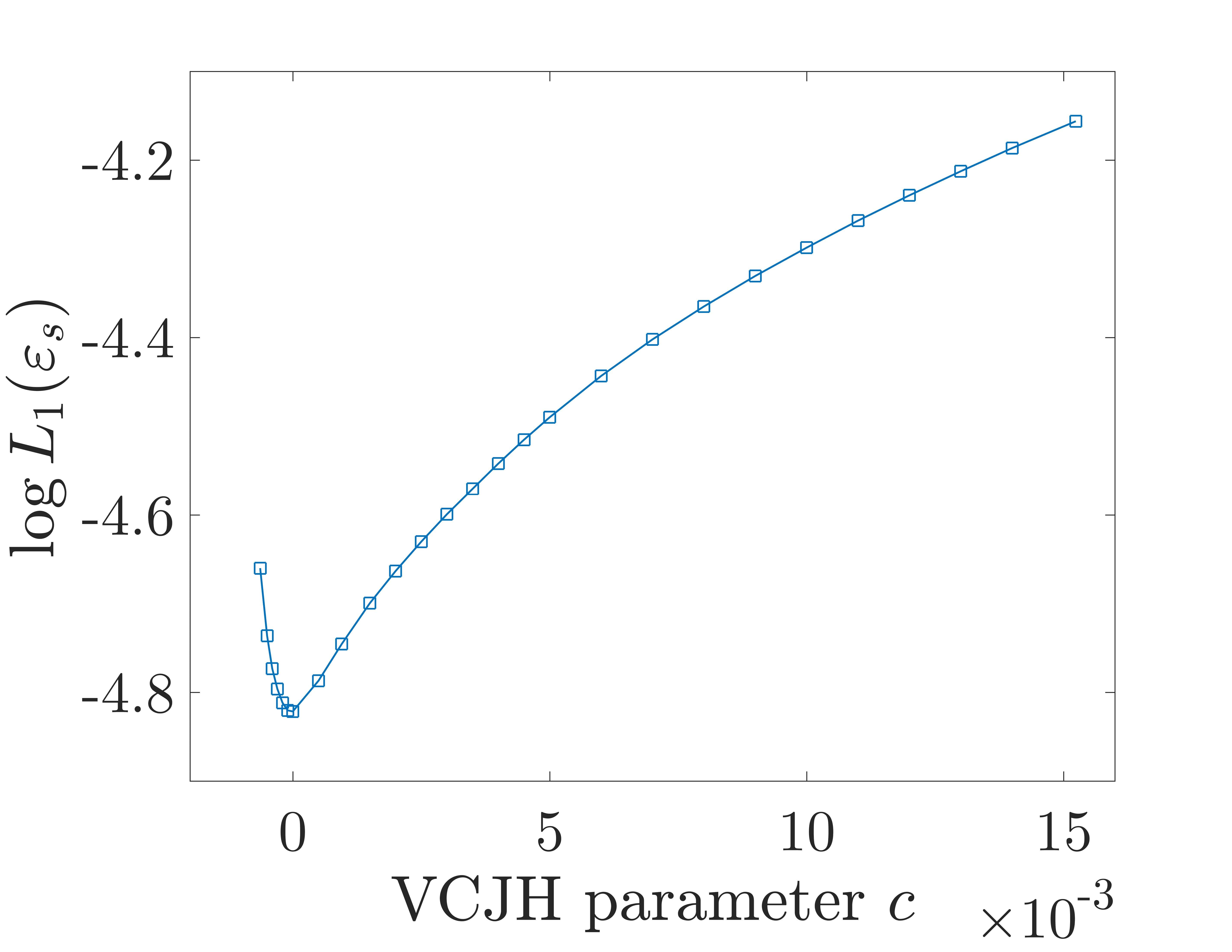}
        \caption{P3}
        \label{fig:cInfl_P3}
    \end{subfigure}
    \quad
    \begin{subfigure}[b]{0.3\textwidth}
        \includegraphics[width=\textwidth]{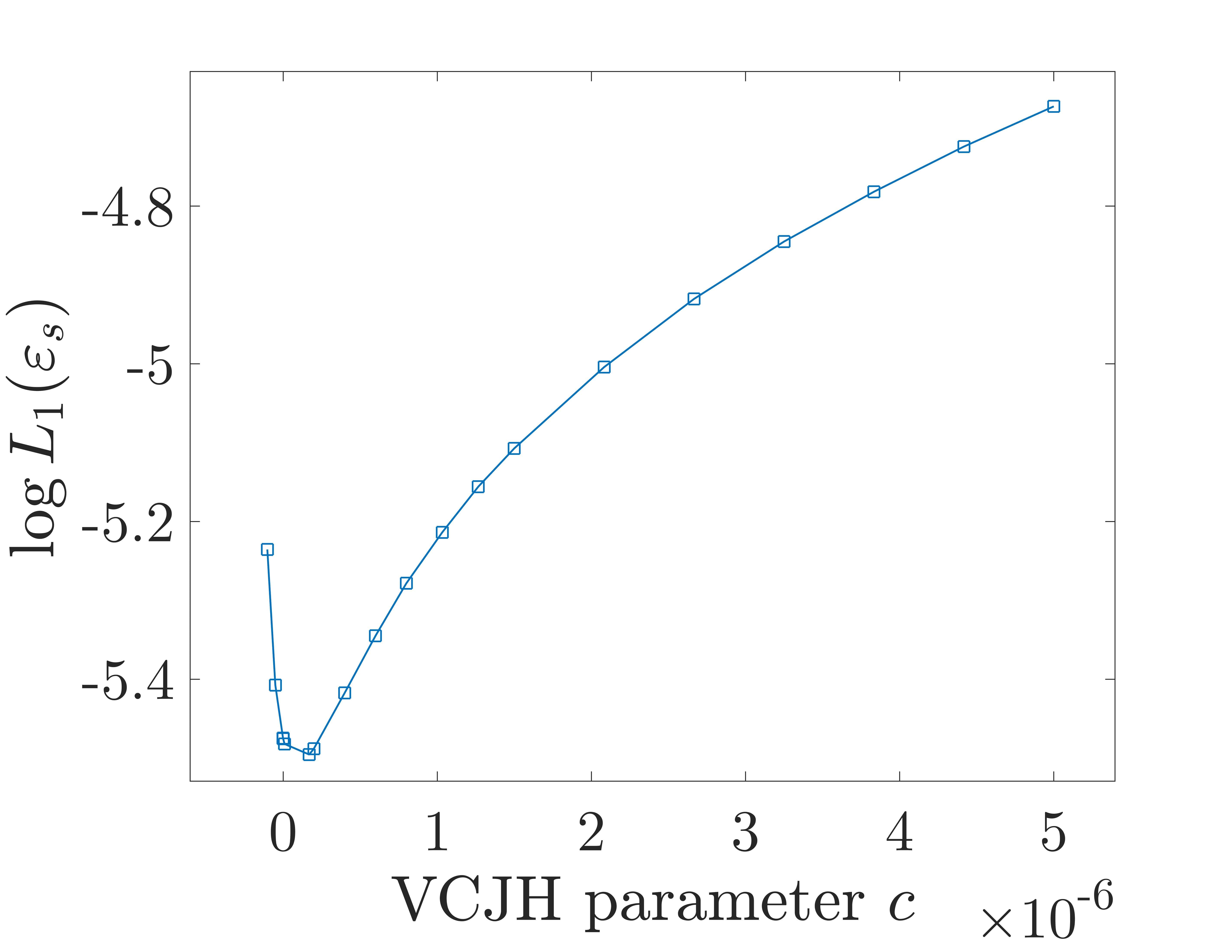}
        \caption{P5}
        \label{fig:cInfl_P5}
    \end{subfigure}
    \caption{Influence of the VCJH parameter $c$ on $L_1(\varepsilon_s)$}
    \label{fig:cInfl}
\end{figure}

As discussed in section \ref{implCorrFct}, the VCJH correction field for quadrilaterals and hexahedra is fully defined by a single parameter $c$. The FR method with a VCJH correction function is proven to be energy stable for $c_{\_} < c < \infty$ \cite{williams2011extension}. The value of $c_{\_}$ depends on the order of the FR method and is given by equation \ref{cMin}. Let $c_{min}$ be the minimum value for which the developed FR solver does not diverge for the flow through the 2D channel with a sinusoid bump. Consequently, the present FR solver converges for this test case if $c_{min} \leq c < \infty$. This value is determined experimentally by progressively lowering $c$ until no convergence is reached. Table \ref{tab:Cmin} shows the values of $c_{\_}$ and $c_{min}$ for orders P1 through P5. As expected, the values of $c_{min}$ are close to those of $c_{\_}$.
\begin{table}[H]
\centering
\begin{tabular}{ccc}
\hline
$\mathbf{P}$ & $c_{\_}$ & $c_{min}$\\ \hline
1          & -6.667e-1             &      -6.402e-1         \\
2          & -4.444e-2            &        -4.345e-2       \\
3          & -1.270e-3            &         -1.232e-3        \\
4          & -2.016e-5            &          -1.926e-5  \\
5          & -2.036e-7            &          -1.912e-7      \\
\hline
\end{tabular}
\caption{Values of $c_{\_}$ and $c_{min}$ for orders P1 through P5}
\label{tab:Cmin}
\end{table}
Once $c$ attains a certain value $c_{\text{large}}$, the form of the correction function does not significantly change anymore. When $c$ is further increased beyond $c_{\text{large}}$ the $L_1$-norm of $\varepsilon_s$ reaches a constant value and the accuracy stagnates as shown in figure \ref{fig:StagnationError}. Figure \ref{fig:cInfl} shows the variation of the $L_1$-norm of the entropy error $\varepsilon_s$ for different values of $c$ for orders P1, P3 and P5.  It is clear that the $L_1$-norm of the entropy error $\varepsilon_s$ reaches a minimum for a value of $c$ close to $c_{DG}$, which is in line with the results in \cite{castonguay2011application}.

As proposed by Jameson, Vincent and Castonguay in \cite{castonguay2011application,jameson2012non}, the coordinates of the solution and flux points have a large influence on non-linear aliasing driven instabilities. The use of quadrature points such as Gauss-Legendre is advised. In the present FR solver, the Gauss-Legendre points, Gauss-Lobatto points and an equidistant set of points have been implemented. The reference test case is executed with these three types of point distributions. The solution and flux point distributions are always chosen to be the same. The $L_1$-norm of the entropy error for these distributions is shown in table \ref{tab:inflPnt}.
\begin{table}[H]
\centering
\begin{tabular}{cccc}
\hline
$\mathbf{P}$ & \textbf{Equidistant} & \textbf{Gauss-Legendre} & \textbf{Gauss-Lobatto}\\ \hline
1          &  5.94e-03  &   7.66e-03  &  1.48e-02 \\
2          &   3.21e-04 &  5.88e-04   &  1.17e-03\\
3             &  6.27e-05  &   6.98e-05 &  1.11e-04 \\
4           &  7.88e-06  &  6.11e-06  &  1.25e-05 \\
5           & -  &  3.35e-06 &  7.14e-06    \\
\hline
\end{tabular}
\caption{$L_1$ norm of the entropy error for different convective interface flux schemes for orders P1 through P5}
\label{tab:inflPnt}
\end{table}
Although the performance of the equidistant set is similar to Gauss-Legendre for lower orders, for higher orders it is considerably worse. For order P5, the FR method with an equidistant set of solution and flux points was not able to converge. This is consistent with the observation that quadrature points should be used to minimize aliasing driven instabilities \cite{castonguay2011application,jameson2012non}.

Furthermore, it is clear that the Gauss-Legendre distribution is an order of magnitude more accurate than the Gauss-Lobatto points. Although the Gauss-Lobatto points are quadrature points as well, the solution points coincide with the flux points. This has the potential to reduce the computational cost since extrapolation of states and fluxes is avoided. However it is clear that its performance is considerably worse. Intuitively, this is logical, since for the case of Gauss-Lobatto points, the interface fluxes only depend on one solution point value of the approximate solution polynomial as opposed to all solution points of the element.

\subsection{Inviscid Subsonic Flow Through a Channel with a Sinusoid Bump in 3D}

In order to verify the FR solver in 3D, the 2D grids presented in the previous section are extruded to 3D while keeping the geometry of the $xy$-plane constant. The length of the domain in the $z$-direction equals one. The 3D meshes consist of $3 \times 9 \times 3$, $4 \times 13 \times 4$, $5 \times 17 \times 5$, and $6 \times 21 \times 6$ nodes where the last number refers to the number of nodes in the $z$-direction. The grids consist of curvilinear Q2 hexahedral elements.

\begin{figure}[H]
  \centering
  \includegraphics[width=0.5\textwidth]{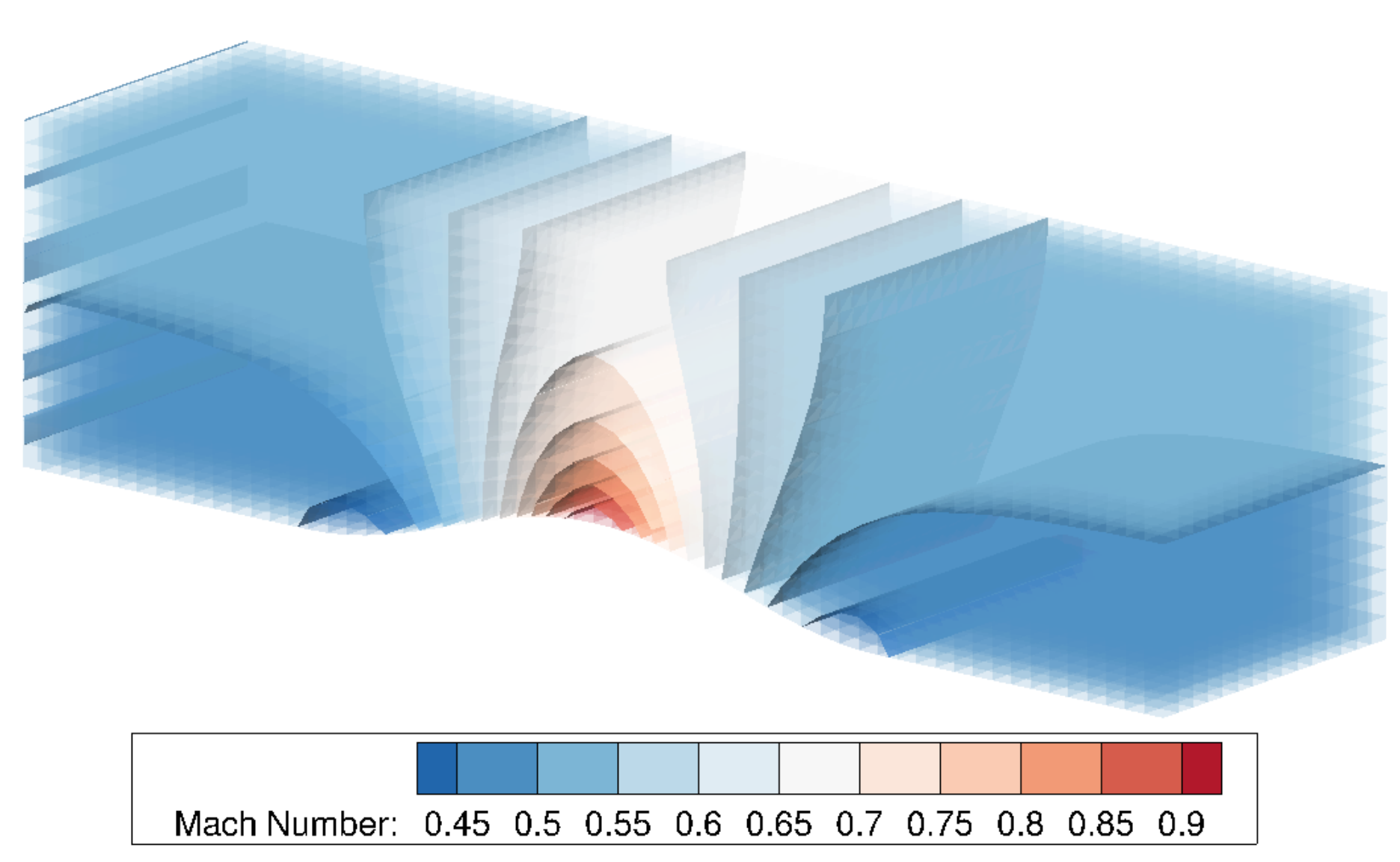}
  \caption{Mach contours obtained with P3Q2 elements on grid 4 in 3D}
  \label{fig:P3Q2G4}
\end{figure}

The boundary conditions used in this test case are similar to the 2D case: an inlet and outlet with the same parameters are used, however, the inlet and outlet are planes in the 3D case instead of edges. For the bottom and top wall, a slip-wall boundary condition is used. The same applies to the newly created walls by extruding the 2D mesh. The same FR parameters and CFL law as for the 2D case were used. Figure \ref{fig:P3Q2G4} presents the Mach contours on grid 4 for P3Q2 elements. As expected, this solution is equivalent to the 2D Mach contours shown in figure \ref{fig:MachSineBump} in the $xy$-plane, while the solution is constant in the $z$-direction.

A convergence study is performed using the same approach as presented in section \ref{2Dsinebump}. For the 3D case, the characteristic grid spacing is approximated by $\log \Delta x = \log \left(N^{-1/3}\right)$. Table \ref{tab:convStudy3D} presents the results of this convergence study. As expected, the results are similar to the 2D convergence study, which are shown in table \ref{ConvergenceStudyTable}.
\begin{table}[ht]
\centering
\begin{tabular}{crrrrr}
\hline
$P$ & \textbf{\#DOF} & $L_1$ \textbf{error} & $L_1$ \textbf{order} & $L_2$ \textbf{error} & $L_2$ \textbf{order}\\ \hline
1          & 256             &        3.16e-02         & -                 &        2.84e-02           & -       \\
           & 864            &        1.33e-02           &         2.13      &         1.58e-02          &     1.44  \\
           & 2048           &         7.49e-03          &            2.00       &          9.87e-03         &    1.64   \\
           & 4000           &          4.72e-03         &    2.07               &           7.11e-03        &   1.47      \\           &            &                   &                   &                   &             \\
2          & 864            &        5.19e-03           & -                 &      5.54e-03             & -           \\
           & 2916            &         1.46e-03          &    3.13               &        1.84e-03           &     2.72   \\
           & 6912           &        4.84e-04           &           3.84        &          6.60e-04         &     3.56  \\
           & 13500           &        2.53e-04           &          2.91        &          3.35e-04         &     3.04   \\           &            &                   &                   &                   &         \\
3          & 2048            &           1.06e-03        & -                 &          1.45e-03         & -            \\
           & 6912           &         2.02e-04          &    4.10               &          2.88e-04         &     3.99     \\
           & 16384           &        7.05e-05           &          3.65         &         1.03e-04          &    3.57    \\
           & 32000          &          3.31e-05         &            3.38       &          5.12e-05         &     3.14   \\           &             &                   &                   &                   &             \\
4          & 4000            &        5.34e-04           & -                 &         8.38e-04          & -          \\
           & 13500           &         6.11e-05          &           5.35        &         1.04e-04          &     5.14    \\
           & 32000           &         8.35e-06          &           6.92        &         2.03e-05          &     5.68    \\
           & 62500           &           2.51e-06        &           5.38        &          6.13e-06         &     5.38  \\\hline
\end{tabular}
\caption{Results of the grid convergence study for the 3D inviscid flow through a channel with a bump using the $L_1$- and $L_2$-norm of the entropy error $\varepsilon_s$ as an indication for the solution accuracy}
\label{tab:convStudy3D}
\end{table}

\subsection{Inviscid Subsonic Flow Around a Cylinder in 2D}
In this section, the inviscid flow at $M_{\infty} = 0.38$ around a circular cylinder with a diameter equal to one is studied. The obtained results are compared to reference solutions provided by Bassi and Rebay \cite{bassi1997euler}, Krivodonova and Berger \cite{krivodonova2006high}, and Van den Abeele \cite{van2009development}. The freestream values are given in table \ref{tab:freeStreamCyl}.
\begin{table}[H]
\caption{\label{tab:freeStreamCyl} Free-stream values}
\centering
\begin{tabular}{ccccc}
\hline
$M_\infty$ & $\rho_\infty\ [-]$ & $p_\infty\ [-]$ & $v_{x,\infty}\ [-]$ & $v_{y,\infty}\ [-]$\\\hline
0.38 & 1 & 1 & $0.38\sqrt{1.4}$ & 0\\
\hline
\end{tabular}
\end{table}

\begin{figure}[H]
    \centering
    \begin{subfigure}[b]{0.23\textwidth}
        \includegraphics[width=\textwidth]{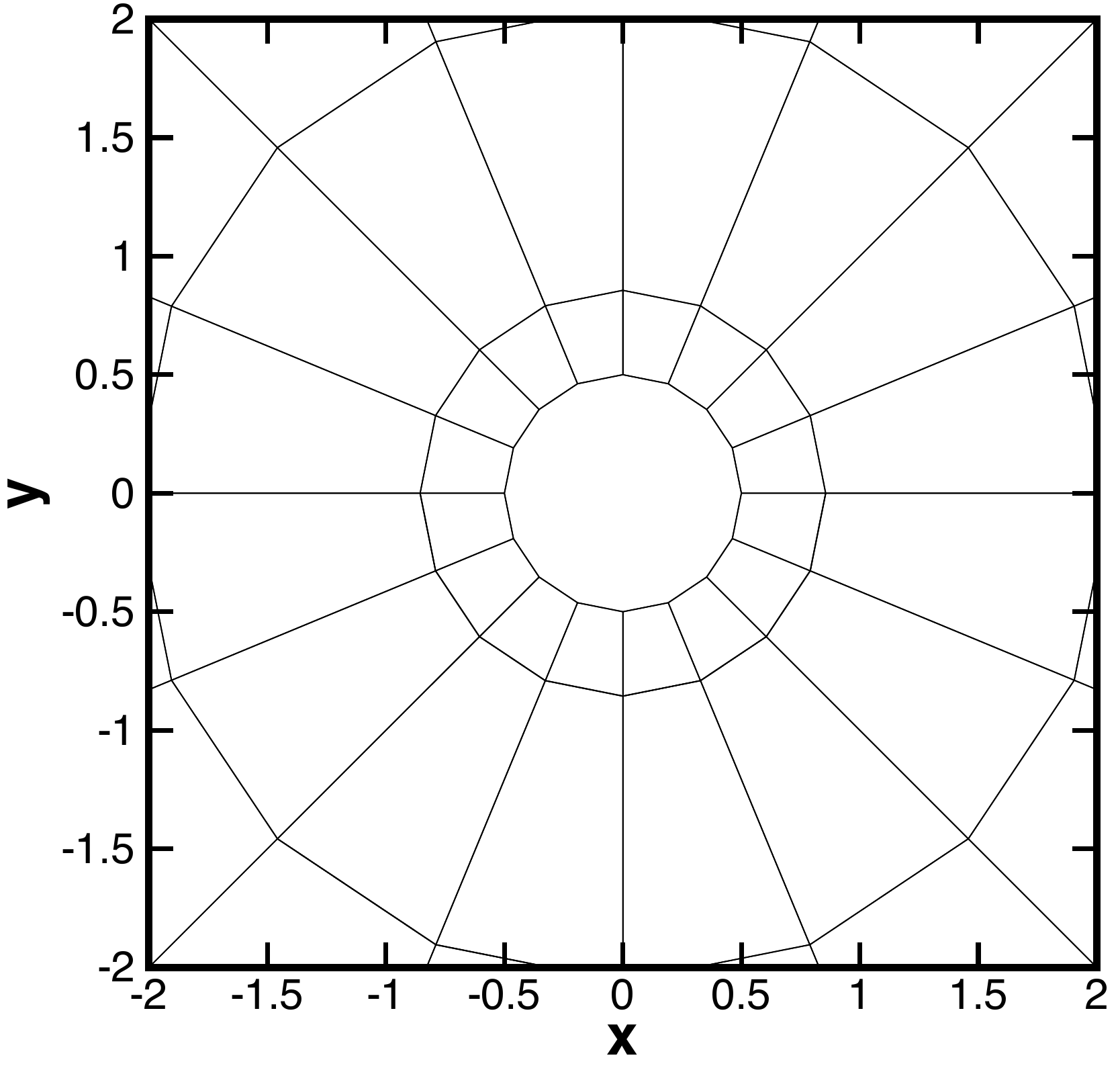}
        \caption{Grid 1, 16 $\times$ 5 nodes}
        \label{fig:CylinderMesh1}
    \end{subfigure}
    \quad
    \begin{subfigure}[b]{0.23\textwidth}
        \includegraphics[width=\textwidth]{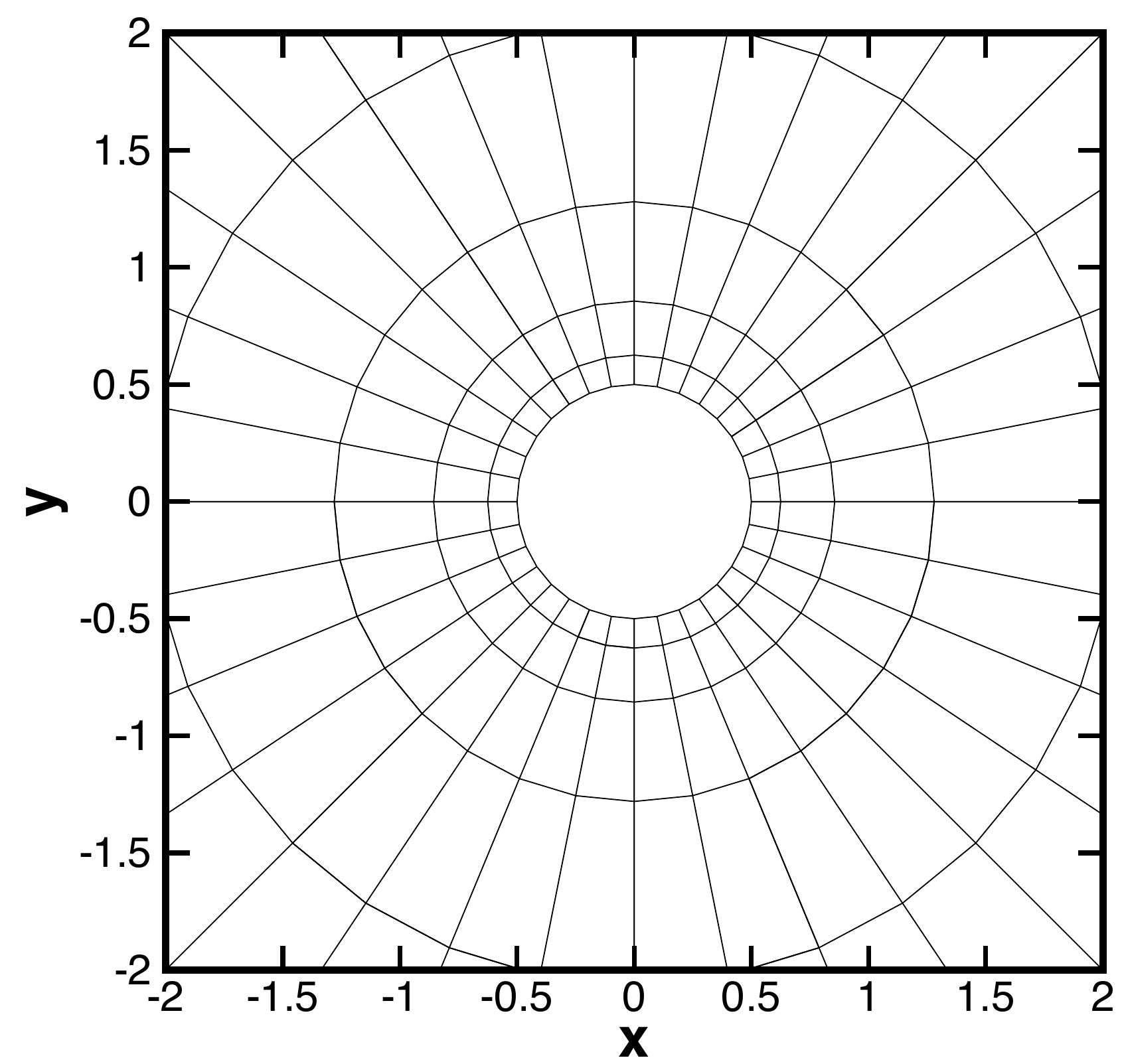}
        \caption{Grid 2, 32 $\times$ 9 nodes}
        \label{fig:CylinderMesh2}
    \end{subfigure}
    \quad
    \begin{subfigure}[b]{0.23\textwidth}
        \includegraphics[width=\textwidth]{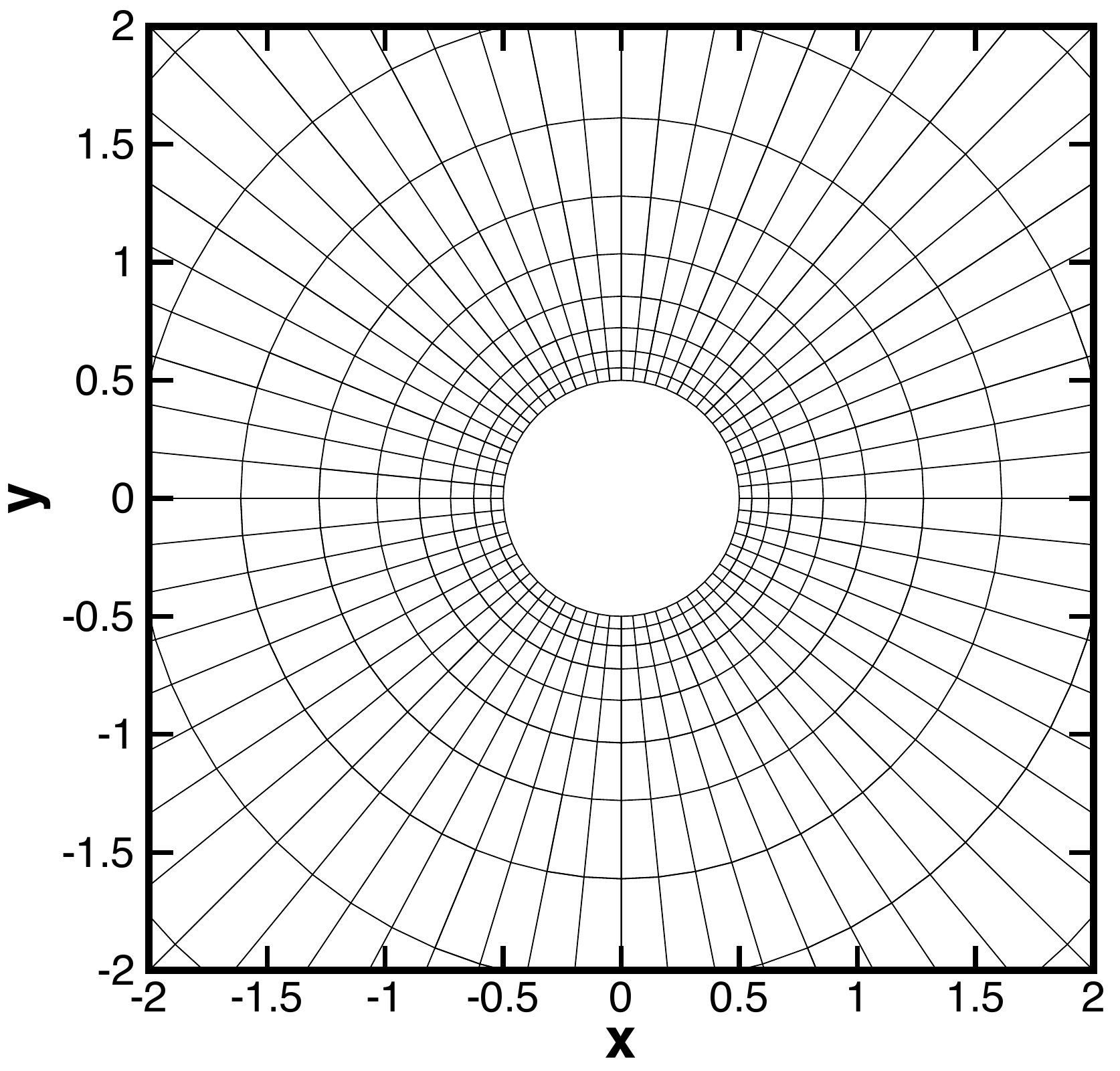}
        \caption{Grid 3, 64 $\times$ 17 nodes}
        \label{fig:CylinderMesh3}
    \end{subfigure}
    \quad
    \begin{subfigure}[b]{0.23\textwidth}
        \includegraphics[width=\textwidth]{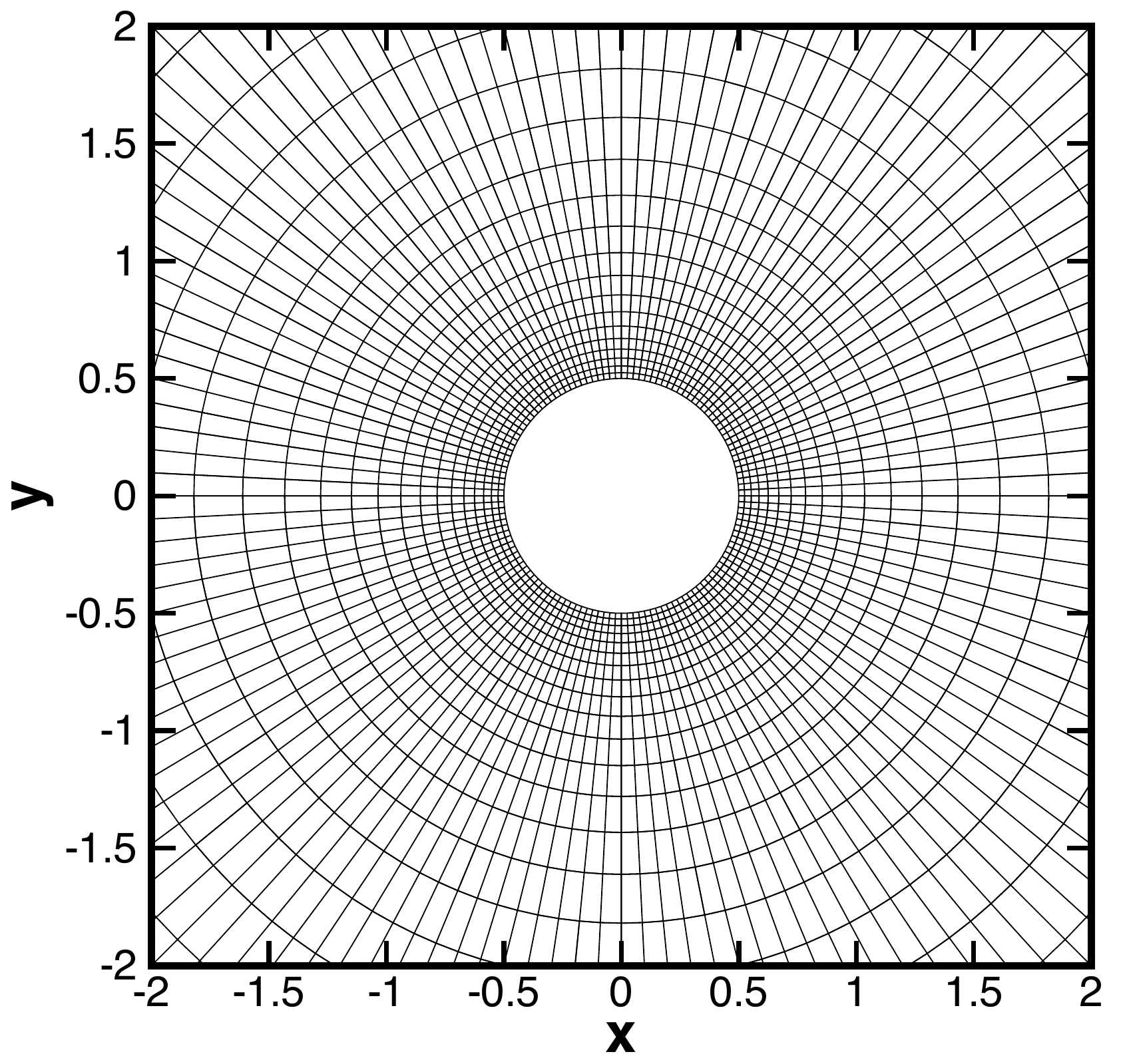}
        \caption{Grid 4, 128 $\times$ 33 nodes}
        \label{fig:CylinderMesh4}
    \end{subfigure}
    \caption{Meshes used for the inviscid flow around a cylinder test case, consisting of Q2 quadrilaterals, same meshes as used in \cite{van2009development}}
    \label{fig:MeshesCylinder}
\end{figure}

The physical domain is discretized by means of four O-grids consisting of quadrilateral elements, as shown in figure \ref{fig:MeshesCylinder}. These grids are equivalent to those used by Van den Abeele \cite{van2009development} and consist of $16 \times 5$, $32 \times 9$, $64 \times 17$, and $128 \times 33$ nodes, respectively. The first value denotes the number of nodes on the inner and outer circle, while the second number refers to the number of concentric circles. The nodes in the circumferential direction are equally distributed over the cylinder and the far field boundary, while those in the radial direction are distributed by means of a geometric progression using the Gmsh mesh generator \cite{geuzaine2009gmsh}. In this manner, the radii of the circles in the finest grid are given by the following expression:
\begin{equation}
    r_i = r_0 \left( 1 + \frac{2\pi}{128} \sum_{n=0}^{i-1} \zeta^n \right), \qquad i = 0, \dots, 32,
\end{equation}
where $r_0 = 0.5$ is the radius of the cylinder. The parameter $\zeta = 1.1648336$ is chosen such that the radius of the outer circle $r_{32}$, associated with the far field boundary, is 20. This grid is then successively coarsened in order to obtain the three other grids, as shown in figure \ref{fig:MeshesCylinder}. A slip-wall boundary condition is introduced at the cylinder wall ($r_{\text{cylinder}} = 0.5$). At the boundary $r_{\text{farfield}} = 20$, a far field boundary condition is imposed.

As demonstrated by Bassi and Rebay \cite{bassi1997euler}, curvilinear Q2 elements with quadratic geometric mapping are mandatory in order to take into the account the curvature of the cylinder. In the case of straight edged elements, the solution is heavily polluted in the vicinity of the cylinder wall due to the fact that the computational boundary substantially differs from the physical boundary. As a result, an unphysical ``wake'' is generated at the downstream side of the cylinder leading to inaccurate solutions. As outlined in the previous section, special measures need to be taken in order to avoid this spurious entropy production.

\begin{figure}[H]
    \centering
    \begin{subfigure}[b]{0.23\textwidth}
        \includegraphics[width=\textwidth]{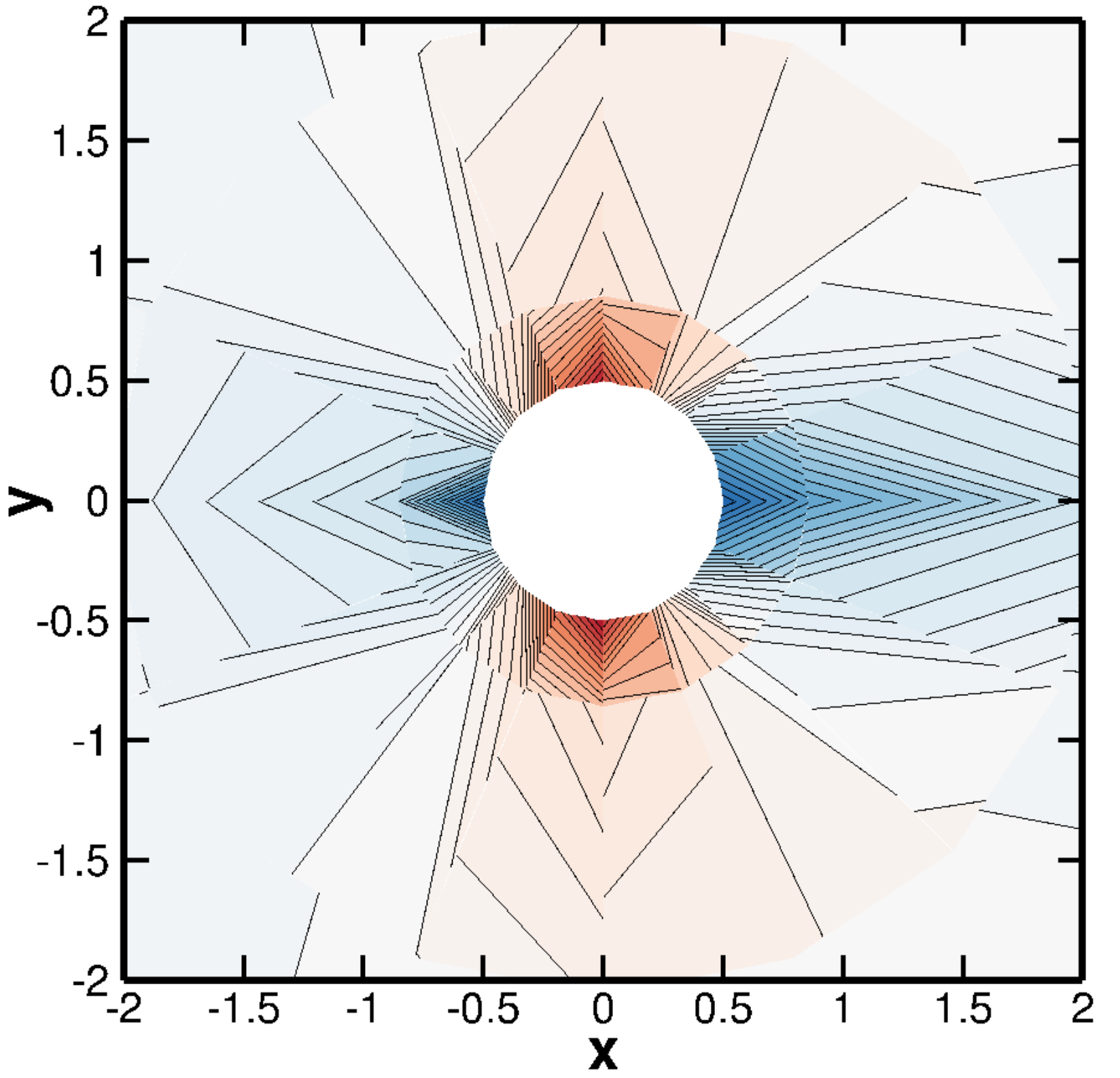}
        \caption{Grid 1}
        \label{fig:CylinderP1Q2_16_5b}
    \end{subfigure}
    \quad
    \begin{subfigure}[b]{0.23\textwidth}
        \includegraphics[width=\textwidth]{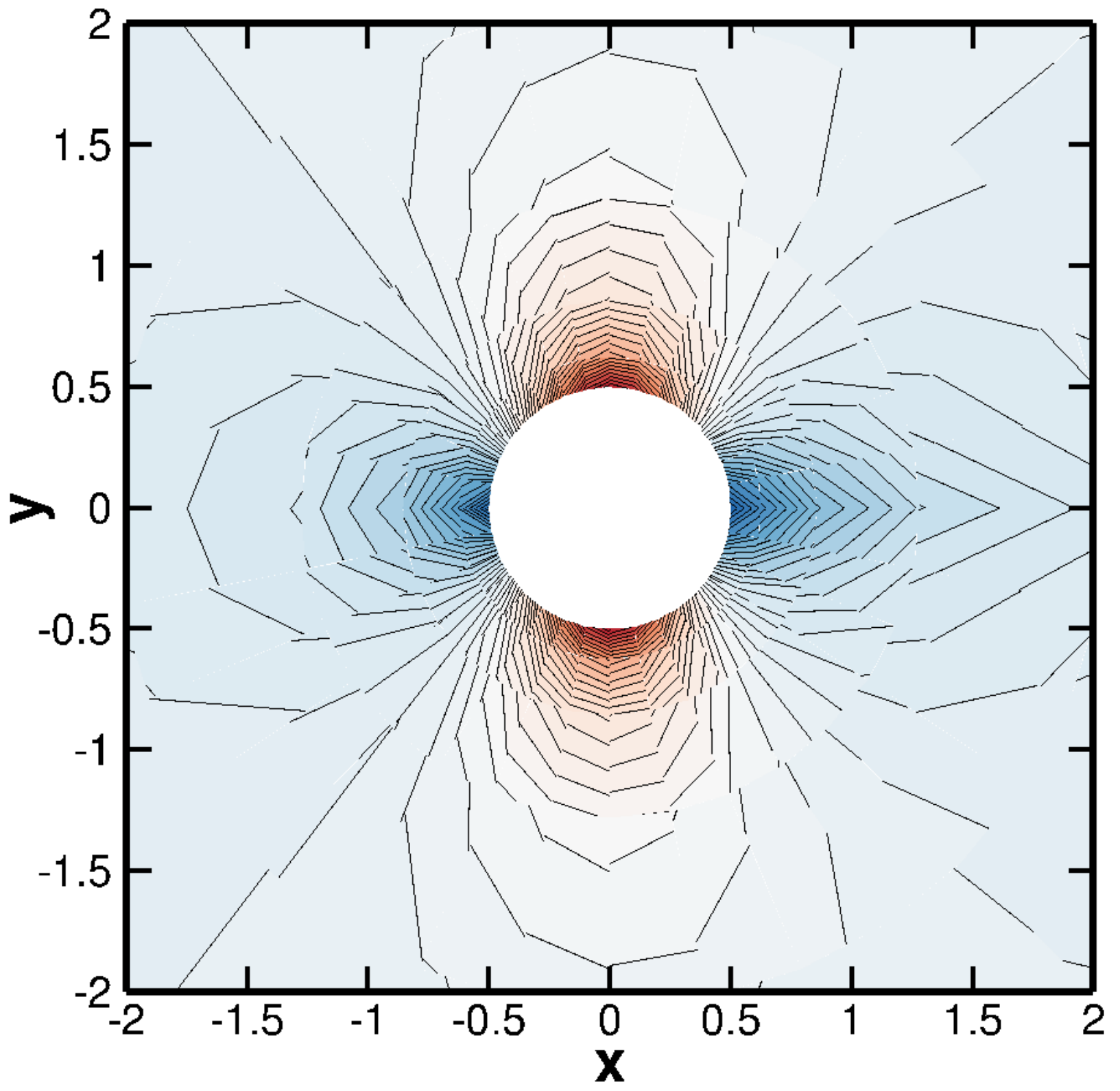}
        \caption{Grid 2}
        \label{fig:CylinderP1Q2_32_9}
    \end{subfigure}
    \quad
    \begin{subfigure}[b]{0.23\textwidth}
        \includegraphics[width=\textwidth]{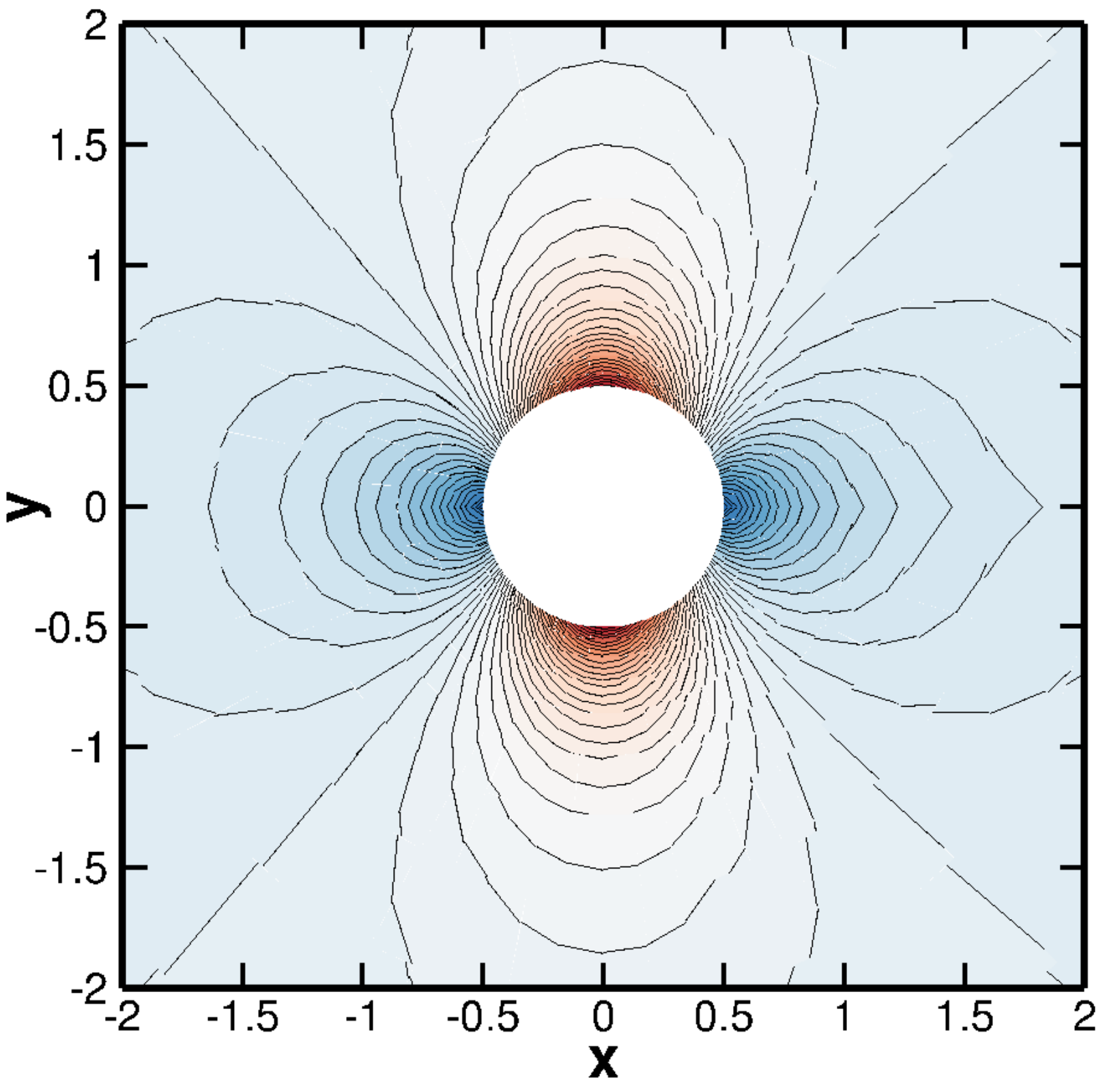}
        \caption{Grid 3}
        \label{fig:CylinderP1Q2_64_17}
    \end{subfigure}
    \quad
    \begin{subfigure}[b]{0.23\textwidth}
        \includegraphics[width=\textwidth]{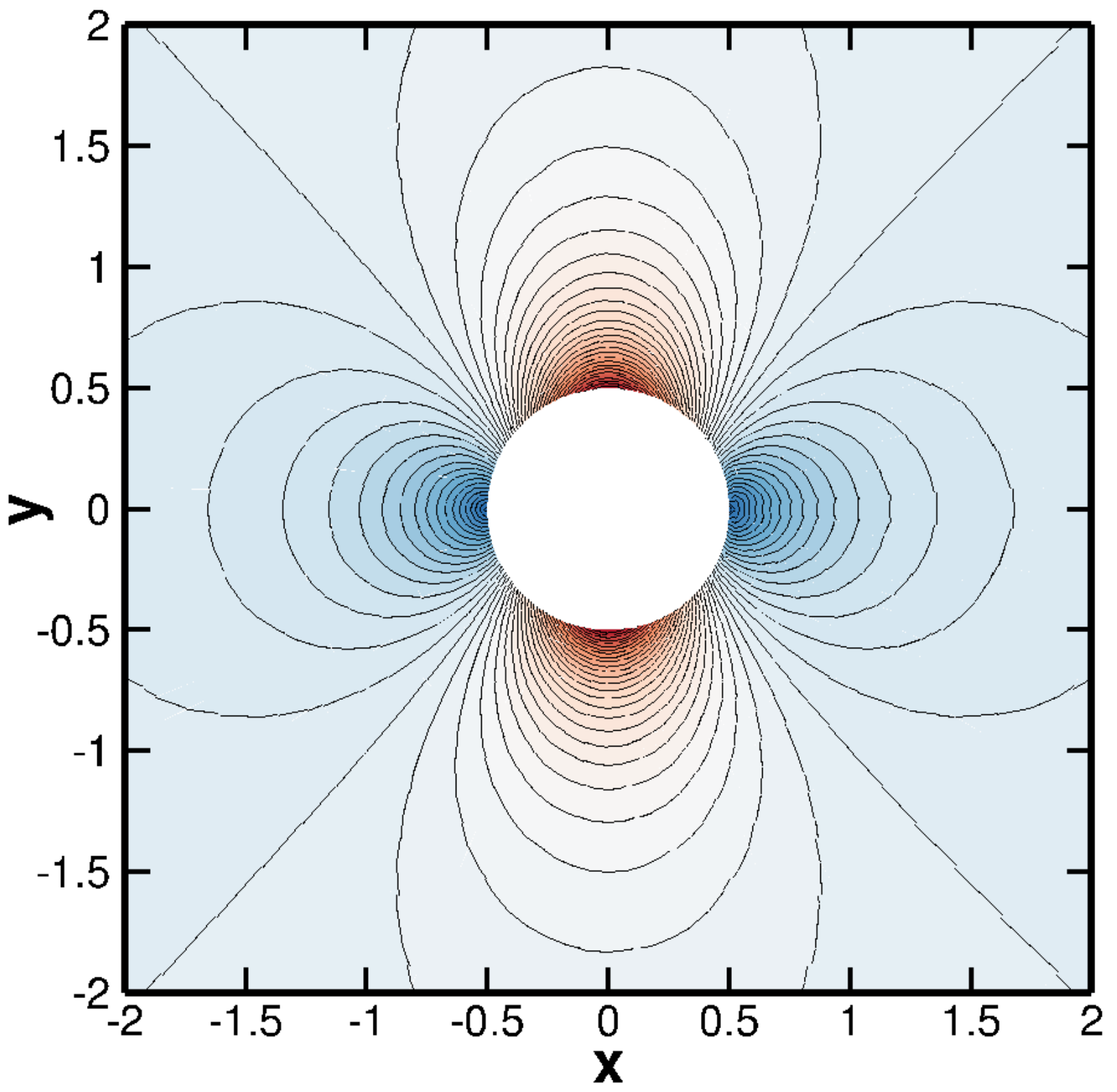}
        \caption{Grid 4}
        \label{fig:CylinderP1Q2_128_33}
    \end{subfigure}
    \caption{Mach contours for the inviscid flow around a cylinder obtained with P1Q2 elements under $h$-refinement}
    \label{fig:CylinderP1Q2}
\end{figure}

\begin{figure}[ht!]
    \centering
    \begin{subfigure}[b]{0.25\textwidth}
        \includegraphics[width=\textwidth]{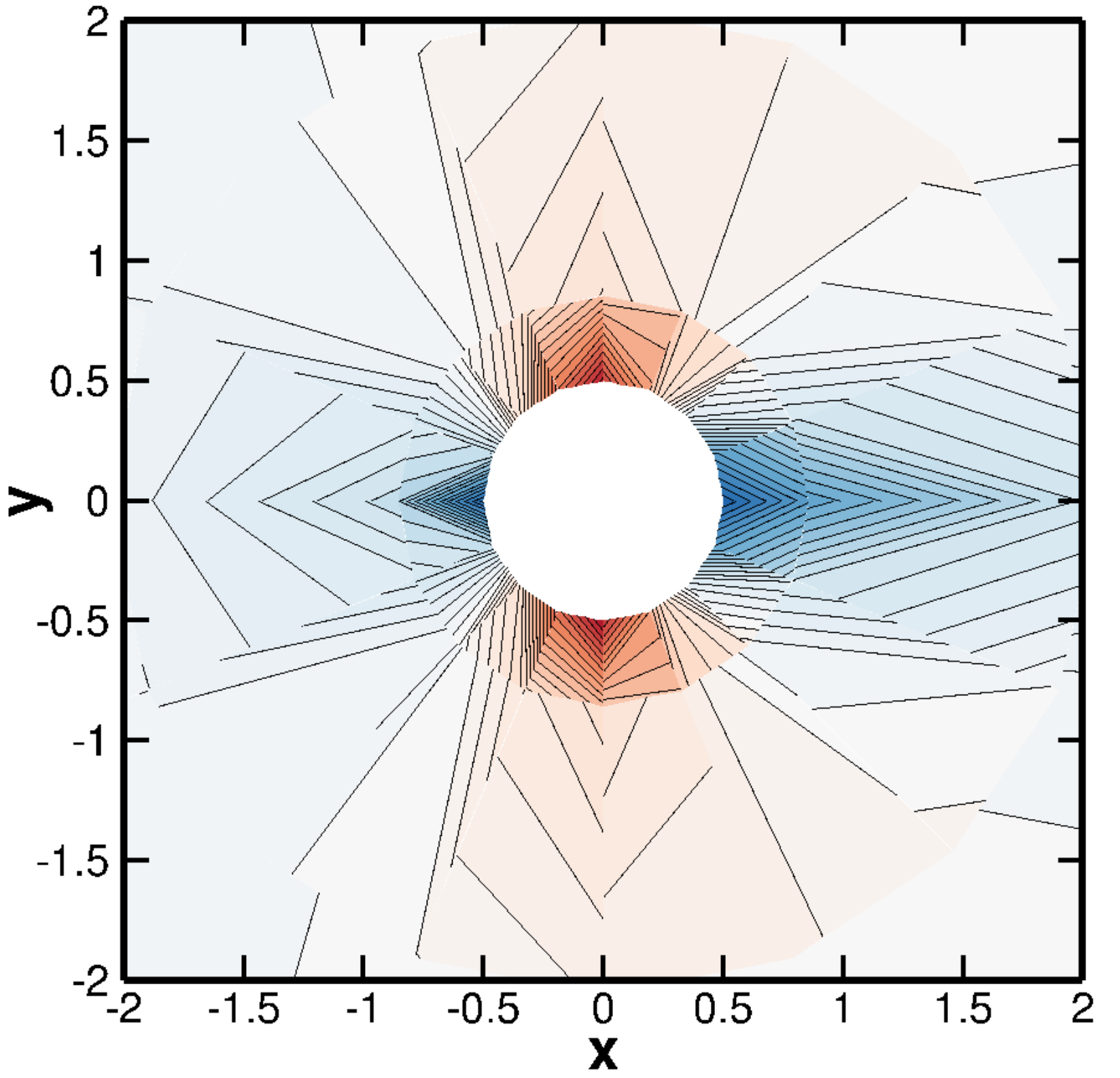}
        \caption{P1Q2}
        \label{fig:CylinderP1Q2_16_5a}
    \end{subfigure}
    \quad
    \begin{subfigure}[b]{0.25\textwidth}
        \includegraphics[width=\textwidth]{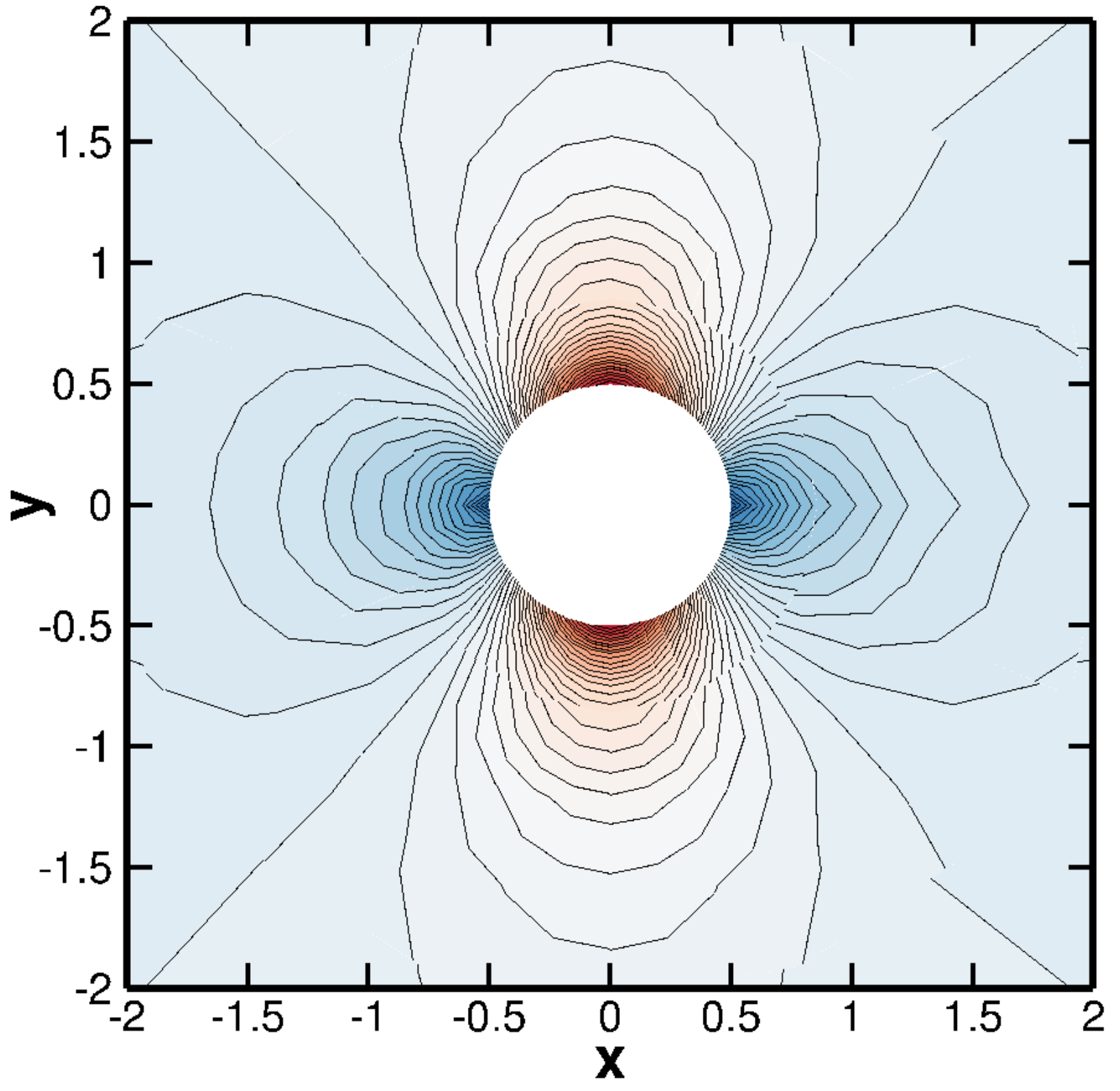}
        \caption{P3Q2}
        \label{fig:CylinderP3Q2_16_5}
    \end{subfigure}
    \quad
    \begin{subfigure}[b]{0.25\textwidth}
        \includegraphics[width=\textwidth]{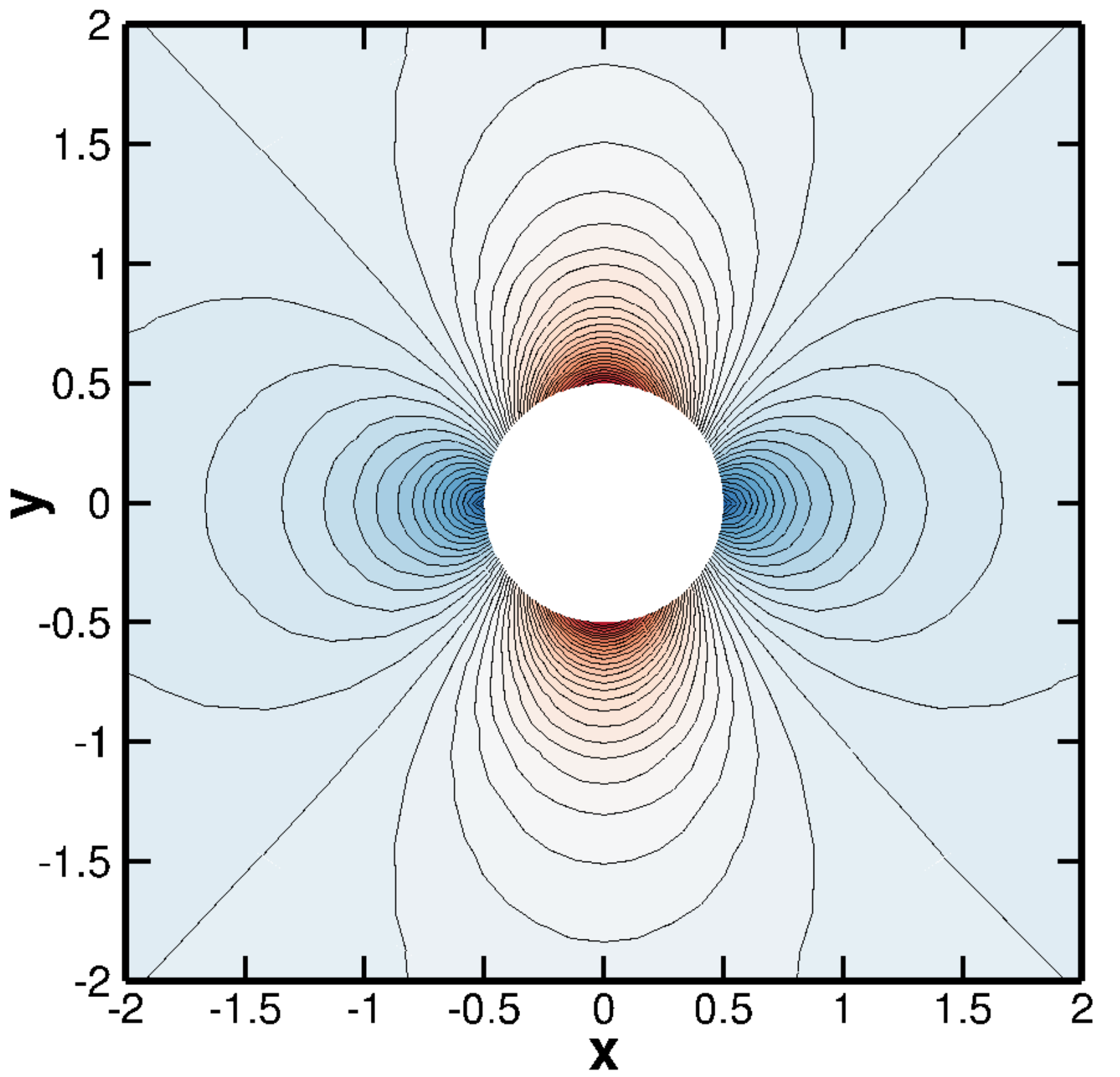}
        \caption{P5Q2}
        \label{fig:CylinderP5Q2_16_5}
    \end{subfigure}
    \caption{Mach contours for the inviscid flow around a cylinder obtained on the coarsest grid (16$\times$5 nodes) under $p$-refinement}
    \label{fig:Cylinder_16_5_Prefinement}
\end{figure}

The Mach isolines obtained with P1Q2 elements are plotted in figure \ref{fig:CylinderP1Q2} for the four successively refined grids. Figure \ref{fig:Cylinder_16_5_Prefinement} shows the Mach isolines resulting from second- (P1Q2), fourth- (P3Q2) and sixth-order (P5Q2) FR schemes on the coarsest grid. The VCJH parameter $c$ is chosen such that the SD method is retrieved, as used by Van den Abeele in \cite{van2009development}. The obtained solutions show great similarities to those in \cite{krivodonova2006high,van2009development,bassi1997euler}. The same CFL law as for the sinusoid bump test case is used. By comparing figure \ref{fig:CylinderP1Q2} to \ref{fig:Cylinder_16_5_Prefinement}, it is once again clear that $p$-refinement is more favorable than $h$-refinement. The Mach contours obtained with P5Q2 elements on the coarsest grid, which requires 1600 DOFs, compare well with those obtained with P1Q2 elements on the finest grid, with 16384 DOFs.


\subsection{Viscous Subsonic Flow Around a Cylinder in 2D}

The viscous flow around cylinders is greatly influenced by the Reynolds number:
\begin{equation}
    \text{Re} = \frac{\rho v_{\infty} D}{\mu},
\end{equation}
where $v_{\infty}$ is the flow speed far away from the cylinder, $D$ the diameter of the cylinder and $\mu$ the dynamic viscosity. Based on the value of Re, different flow regimes can be defined. The flow is characterized as laminar up to $\text{Re}=10^3$. In the vicinity of this Re-number, there is a transitional phase to a fully turbulent flow. For a very low Reynolds number, e.g. $\text{Re}<1$, the flow is symmetrical and no separation occurs. 

For $1<\text{Re}<40$ a stable re-circulation bubble is formed behind the cylinder. In this manner, two steady vortices appear at its trailing edge. For $40<\text{Re}<10^3$ vortices are shed behind the cylinder and a von Karman street is formed. \cite{tansley2001flow,roshko1954development,bloor1964transition}

In order to compute a steady state solution that contains stable vortices, a Reynolds number of $40$ is chosen. The free-stream Mach number equals $M_\infty = 0.15$. The Prandtl number defined as Pr$=\dfrac{\mu c_p}{\kappa}$ where $\kappa$ is the Fourier heat conduction coefficient and $c_p$ the specific heat capacity. For this test case Pr is taken equal to $\text{Pr}=0.72$, the standard value for air.

The cylinder has a diameter of $D=1$ and is considered as an adiabatic no-slip wall. The computational domain is equal to $[-20,20]\times [-20,20]$. The outer circle that limits the computational domain is associated with a far field boundary condition. The free-stream parameters that are imposed at this boundary are given in table \ref{tab:freeStreamCylVisc}. The present test case is similar to the test case used to verify the SD solver in \cite{van2009development} and the FR solver in \cite{castonguay2011development}. The test case is run on grid 3 that was used for inviscid flow over a cylinder, shown in figure \ref{fig:CylinderMesh3}.

\begin{table}[H]
\caption{\label{tab:freeStreamCylVisc} Free-stream values}
\centering
\begin{tabular}{ccccc}
\hline
$M_\infty$ & $\rho_\infty\ [-]$ & $p_\infty\ [-]$ & $v_{x,\infty}\ [-]$ & $v_{y,\infty}\ [-]$\\\hline
0.15 & 1 & 1 & $0.15\sqrt{1.4}$ & 0\\
\hline
\end{tabular}
\end{table}

\begin{figure}[H]
    \centering
    \begin{subfigure}[b]{0.255\textwidth}
    \includegraphics[width=\textwidth]{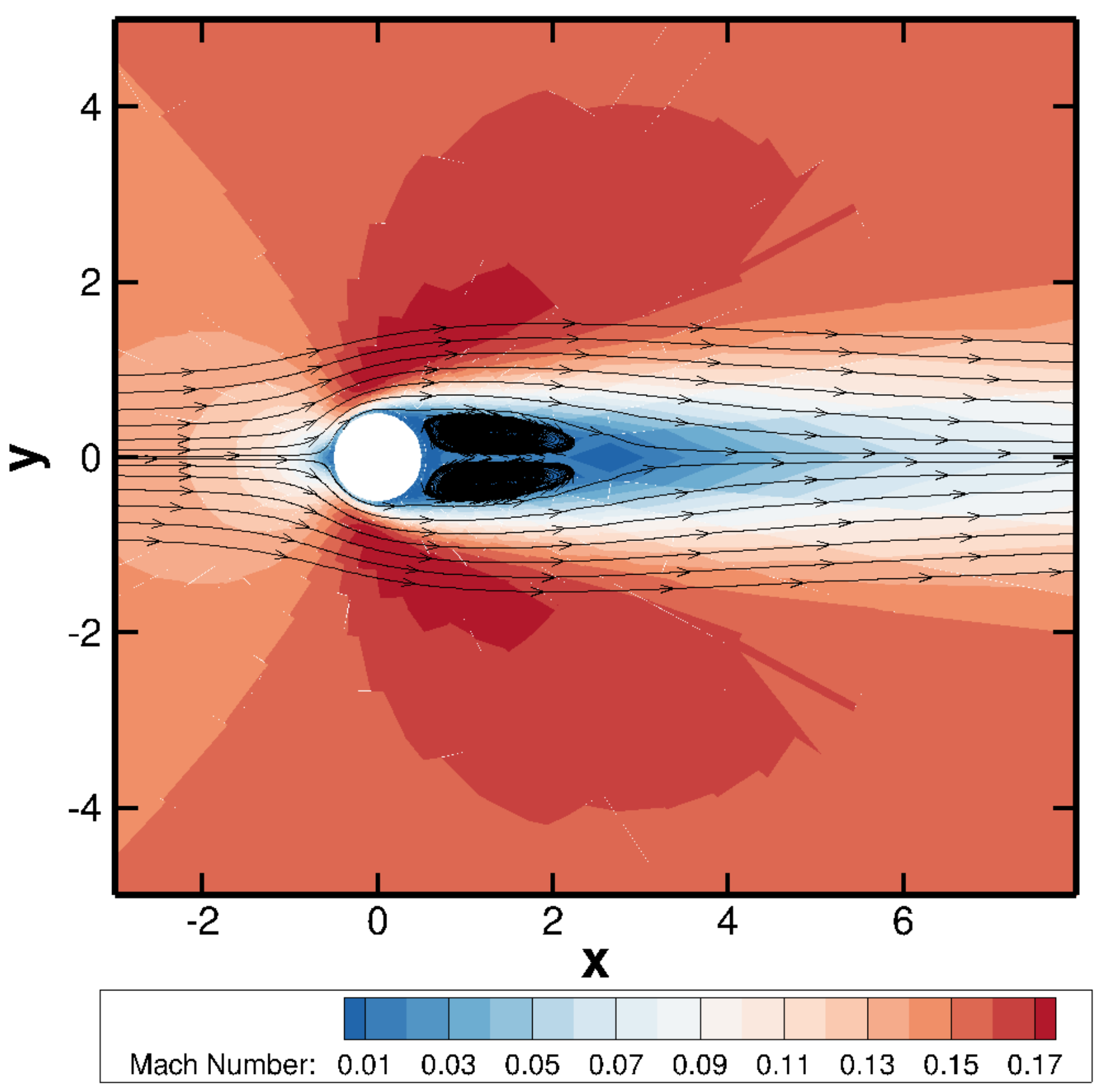}
        \caption{P1Q2, $\#$DOFs=4096}
    \end{subfigure}
    \qquad
    \begin{subfigure}[b]{0.25\textwidth}
    \includegraphics[width=\textwidth]{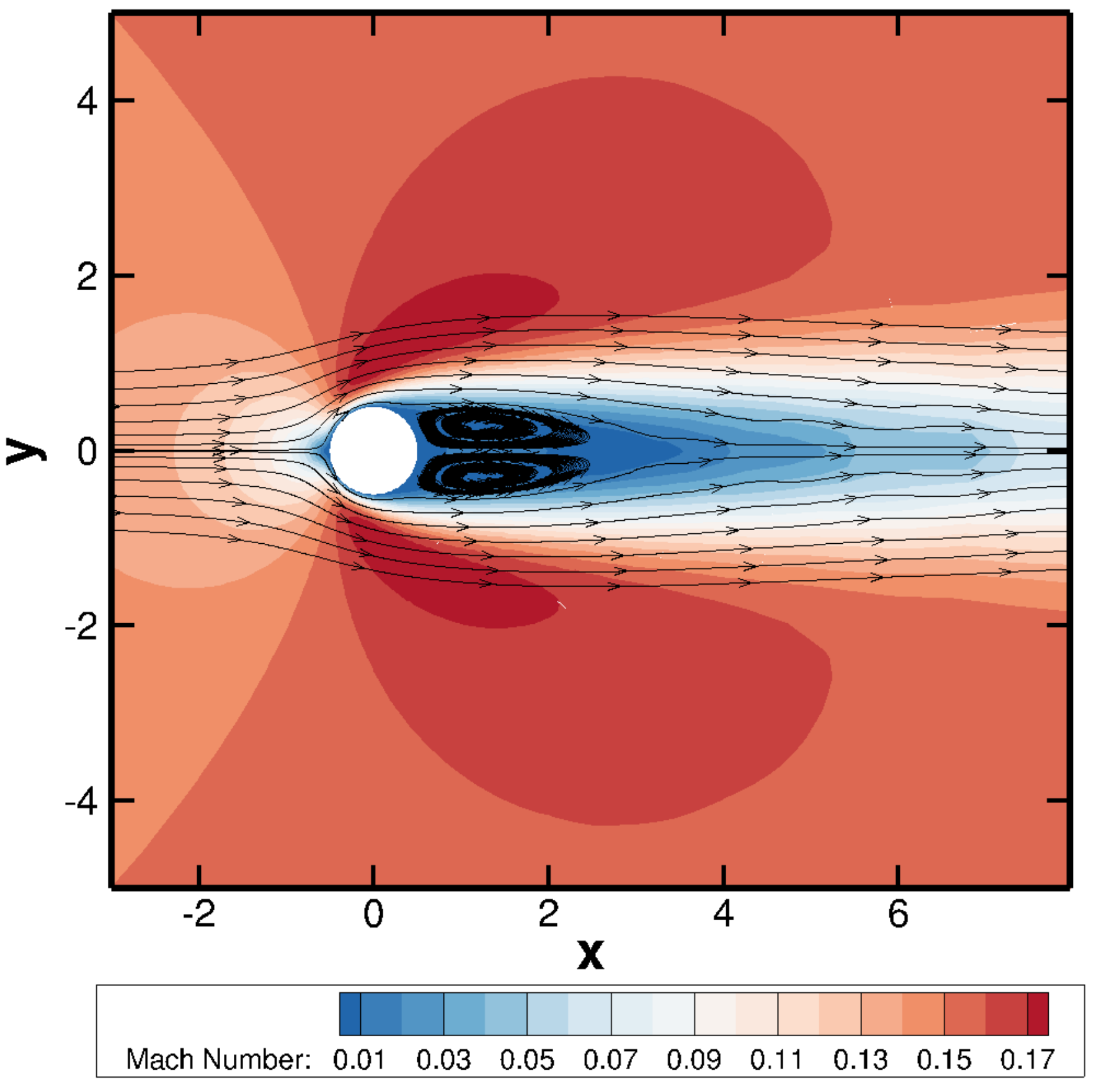}
        \caption{P3Q2, $\#$DOFs=16384}
    \end{subfigure}
    \qquad
    \begin{subfigure}[b]{0.262\textwidth}
    \includegraphics[width=\textwidth]{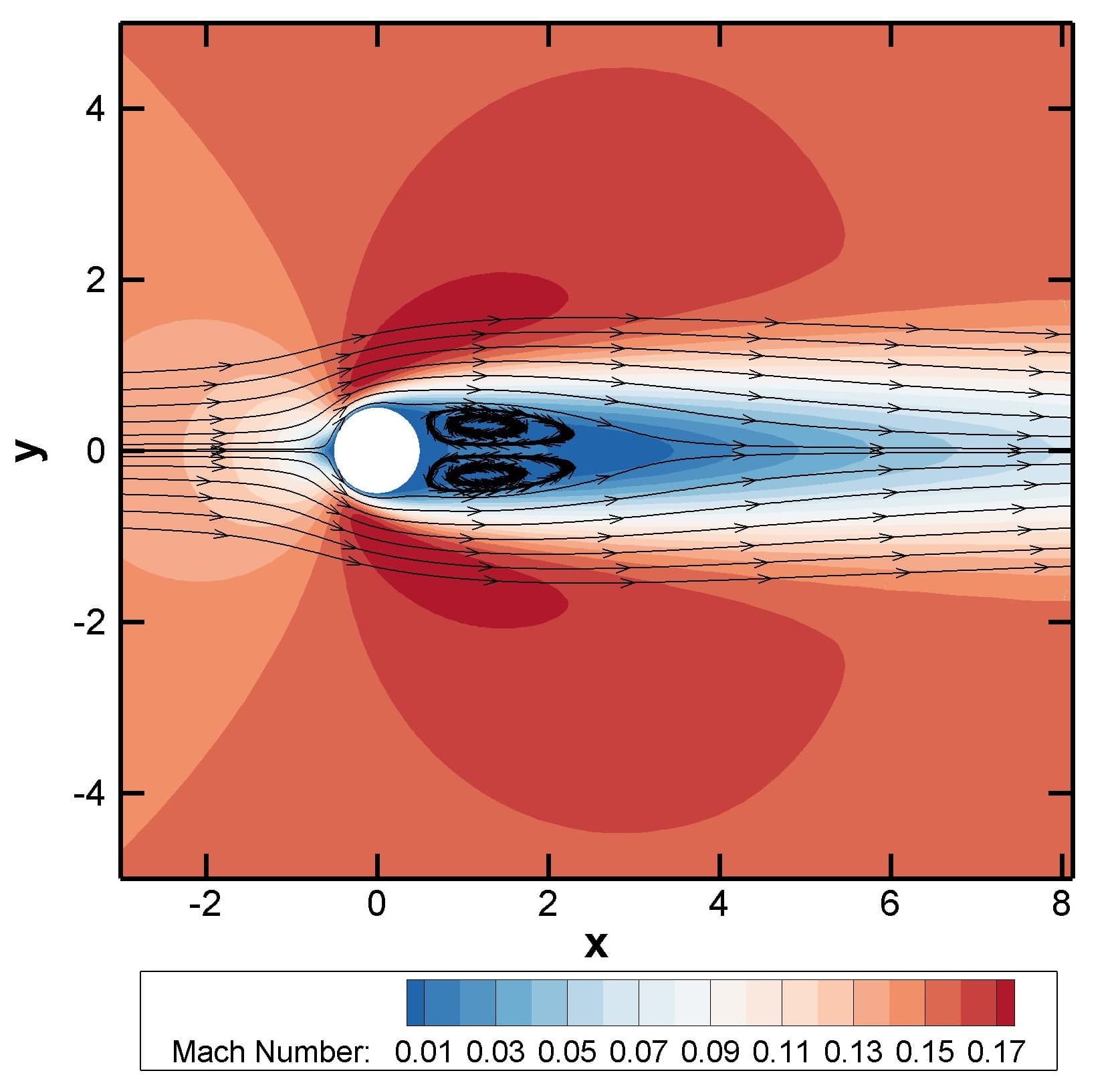}
        \caption{P5Q2, $\#$DOFs=36864}
    \end{subfigure}
    \caption{Mach contours for the viscous flow over a cylinder for the grid in figure \ref{fig:CylinderMesh3}}
    \label{fig:diffCylSol}
\end{figure}
\begin{figure}[H]
  \centering
  \includegraphics[width=0.35\textwidth]{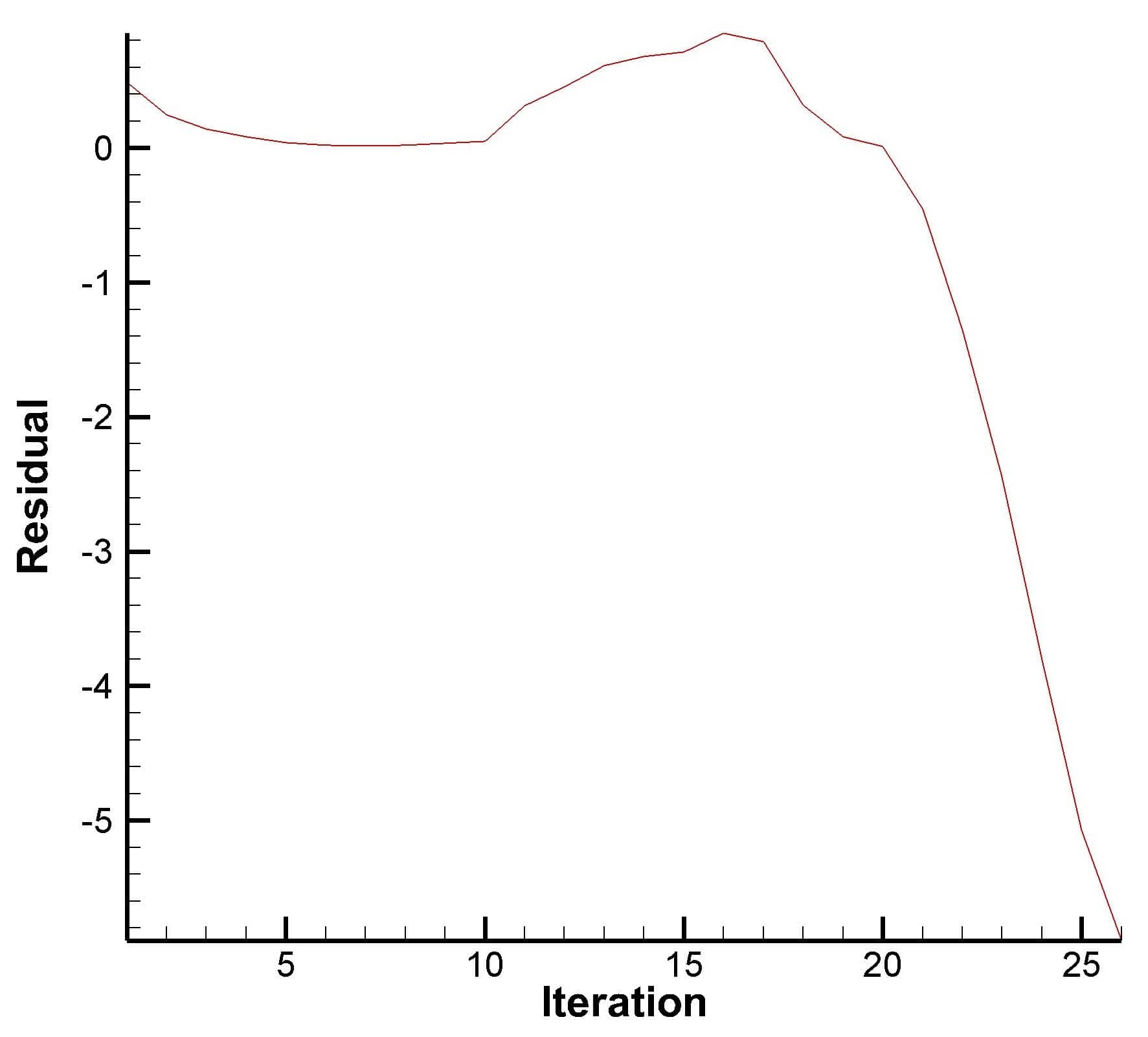}
  \caption{Logarithm of the residual of the density as a function of the iteration for the P3Q2 viscous 2D cylinder test case}
  \label{fig:cylRes}
\end{figure}

The Mach contours of the solution for order P1, P3 and P5 are shown in figure \ref{fig:diffCylSol}. Streamtraces are plotted as well to visualize the re-circulation zones. For each order, the same origin points of the streamtraces are used. It is clear that all orders predict the stable vortices behind the cylinder. It is also clear that towards higher orders, the quality of the solution increases: the resolution of the vortices is more precise. For order P1, a single streamtrace inside the vortex does not describe a clearly defined circular path. However, for increasing orders, the path becomes better defined. Note that these solutions are consistent to the ones obtained in \cite{castonguay2011development} and \cite{van2009development}. Figure \ref{fig:cylRes} shows the logarithm of the density residual as a function of the simulation iterations. The CFL law is the following:
\begin{equation}
    CFL(n) = \begin{cases}
    0.5, & \text{if $n\leq 10$,}\\
    \min(0.5\cdot 2^{n-10},10^4), & \text{otherwise}.
    \end{cases}
\end{equation}
where $n$ is the iteration. Using this CFL law, the present test case converges in 26 iterations. Once again if an explicit time stepping method is used, the CFL limit for FR would be $CFL_{max} = 0.25$ in this P3Q2 case with an even more severe limitation for higher orders. 

\subsubsection{Viscous Subsonic Flow Around a Cylinder in 3D}

Similarly to the approach used for the verification of inviscid 3D flow, an extrusion is made of the coarsest cylinder mesh shown in figure \ref{fig:CylinderMesh1}. In the newly formed $z$-direction, two elements are added of length one. The total dimension in the $z$-direction thus equals two. This results in a 3D mesh with $16 \times 5 \times 3$ nodes in the $x$-, $y$-, and $z$-direction respectively. The cylinder is still considered as an adiabatic no-slip wall. Similar to the 2D case, a far field boundary condition is imposed to the outer border of the computational domain. The newly created top and bottom faces that are parallel to the $xy$-plane are associated with a mirror boundary condition. This is because the flow should be symmetrical with respect to the $xy$-plane at these boundaries. The same CFL law as the 2D case is used.

\begin{figure}[ht!]
    \centering
    \begin{subfigure}[b]{0.37\textwidth}
    \includegraphics[width=\textwidth]{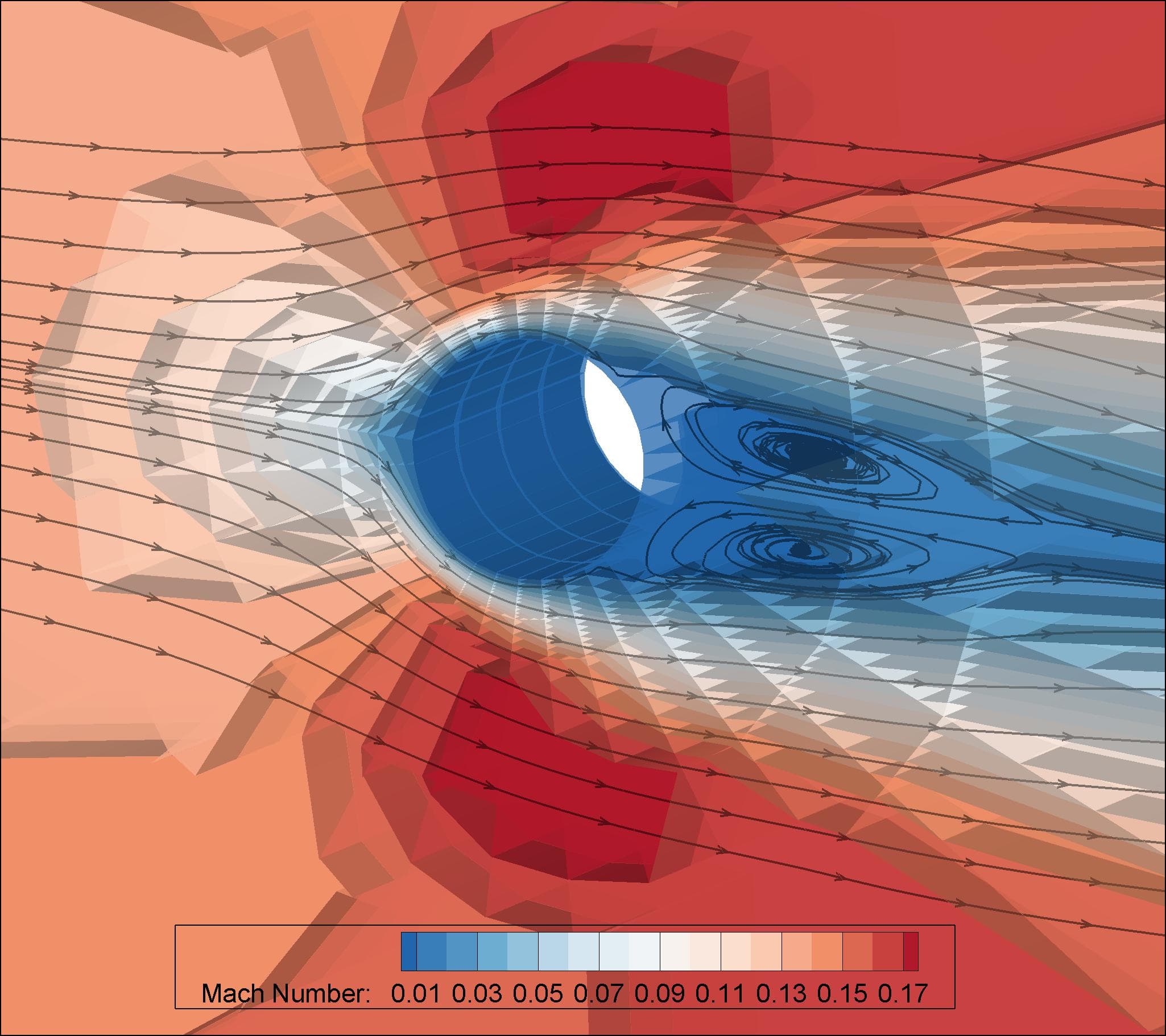}
        \caption{P2Q2, $\#$DOFs=3456}
    \end{subfigure}
    \qquad
    \begin{subfigure}[b]{0.37\textwidth}
    \includegraphics[width=\textwidth]{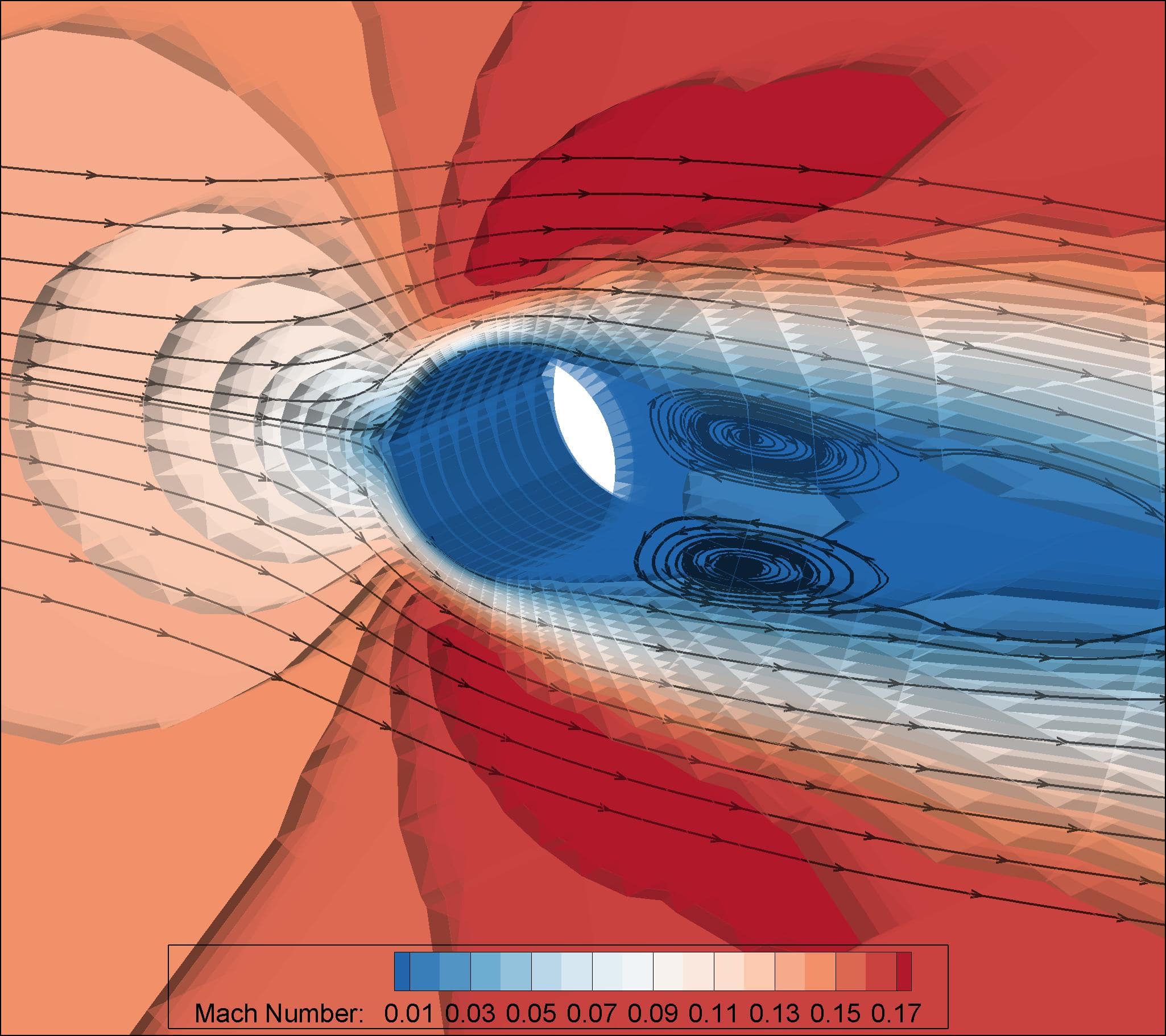}
        \caption{P4Q2, $\#$DOFs=54000}
    \end{subfigure}
    \caption{Mach contours for the viscous flow over a cylinder in 3D for orders P2Q2 and P4Q2 for the grid in figure \ref{fig:CylinderMesh1} extruded in the $z$-direction}
    \label{fig:diffCylSol3D}
\end{figure}

For 3D viscous flow around a cylinder, additional instabilities can occur that induce a transition from a 2D to a 3D flow. In \cite {williamson1988existence}, two modes of transition are proposed: one for $\text{Re}\approx 190$ and one for $\text{Re}\approx 260$ which are both higher than the considered Reynolds number. Figure \ref{fig:diffCylSol3D} shows the Mach contours and streamtraces for order P2 and P4. Indeed, it is clear that the flow is constant in the $z$-direction, as the streamtraces are confined in a vertical plane. In addition, the streamtraces are similar to the ones obtained in the 2D test case. The steady vortices behind the cylinder are thus accurately predicted in the 3D case as well. Although the same mesh is used for the P2 and P4 solution, it is clear that the latter results in a better resolution of the steady vortices, which is also consistent with the 2D case. 

\subsection{Viscous Hypersonic Flow over a Cylinder in 2D}\label{hypersonic}

The proposed shock capturing scheme was tested for a hypersonic cylinder test case, presented in \cite{gnoffo2004computational}. The considered geometry consists of a cylinder with radius 1m. The free-stream values imposed are presented in table \ref{tab:freeStreamHyperInvisc}.

\begin{table}[H]
\caption{\label{tab:freeStreamHyperInvisc} Free-stream values}
\centering
\begin{tabular}{ccccc}
\hline
$M_\infty$ & $\rho_\infty\ [kg/m^3]$ & $V_\infty\ [m/s]$ & $T_\infty\ [K]$ & $T_{wall}\ [K]$\\\hline
17.637 & 0.001 & 5000 & 200 & 500\\
\hline
\end{tabular}
\end{table}

\begin{figure}[H]
  \centering
  \includegraphics[width=0.7\textwidth]{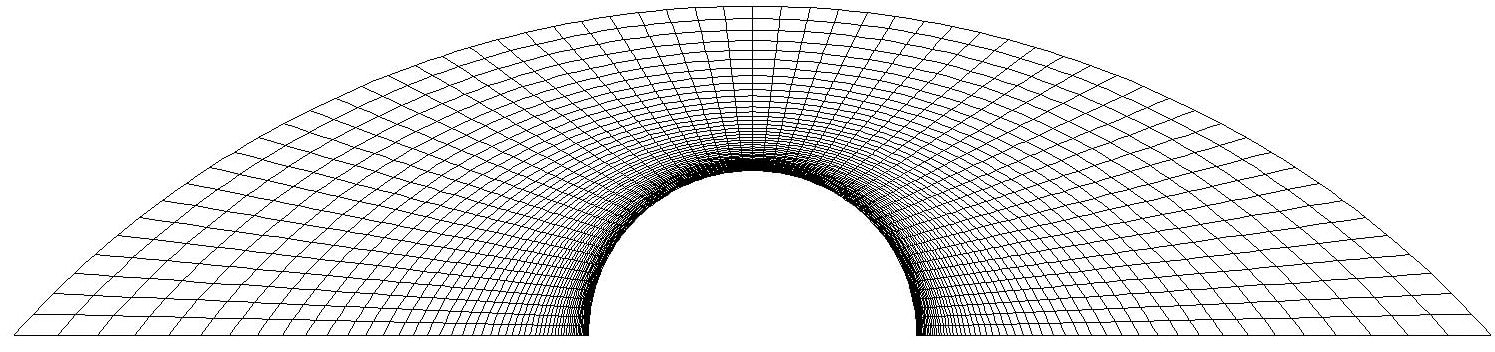}
  \caption{Mesh used for the hypersonic cylinder test case}
  \label{fig:mesh}
\end{figure}

The mesh used is presented in figure \ref{fig:mesh}. The mesh consists of curved quadrilateral elements of second-order and contains 4000 elements. The inlet boundary is imposed as a Dirichlet boundary condition that sets the states to the free-stream values. For the outlet, a supersonic outlet boundary condition is used. The cylinder is represented by an iso-thermal no-slip wall.

The FR method is used with a VCJH scheme for the correction functions. The distribution of both the solution and flux points is the Gauss-Legendre distribution. For the convective interface fluxes, the AUSM$^+$ scheme is used. For the LLAV scheme, $\kappa$ is set to 4.

\begin{figure}[ht!]
    \centering
    \begin{subfigure}[b]{0.45\textwidth}
        \includegraphics[width=\textwidth]{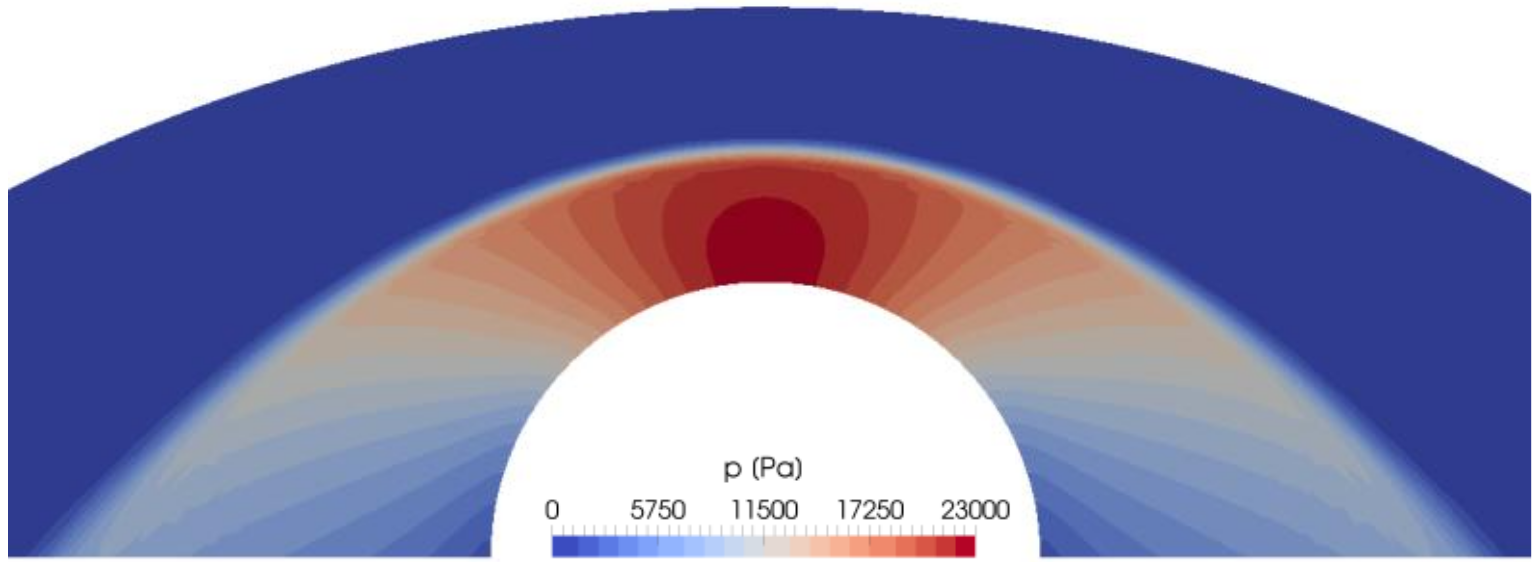}
        \caption{Pressure}
    \end{subfigure}
    \begin{subfigure}[b]{0.45\textwidth}
        \includegraphics[width=\textwidth]{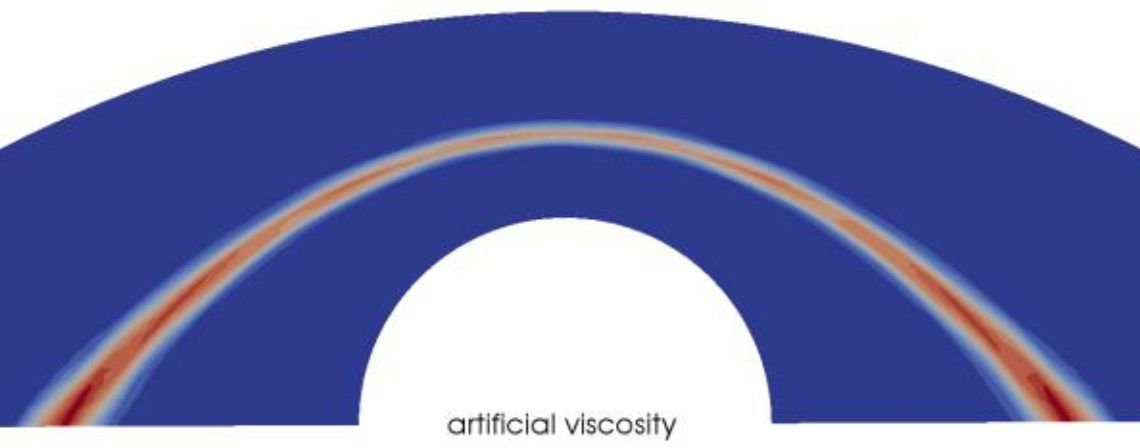}
        \caption{Artificial viscosity}
    \end{subfigure}
    \caption{Flow field of the hypersonic cylinder test case for third-order of accuracy}
    \label{fig:flowfield}
\end{figure}

\begin{figure}[ht!]
  \centering
  \includegraphics[width=0.4\textwidth]{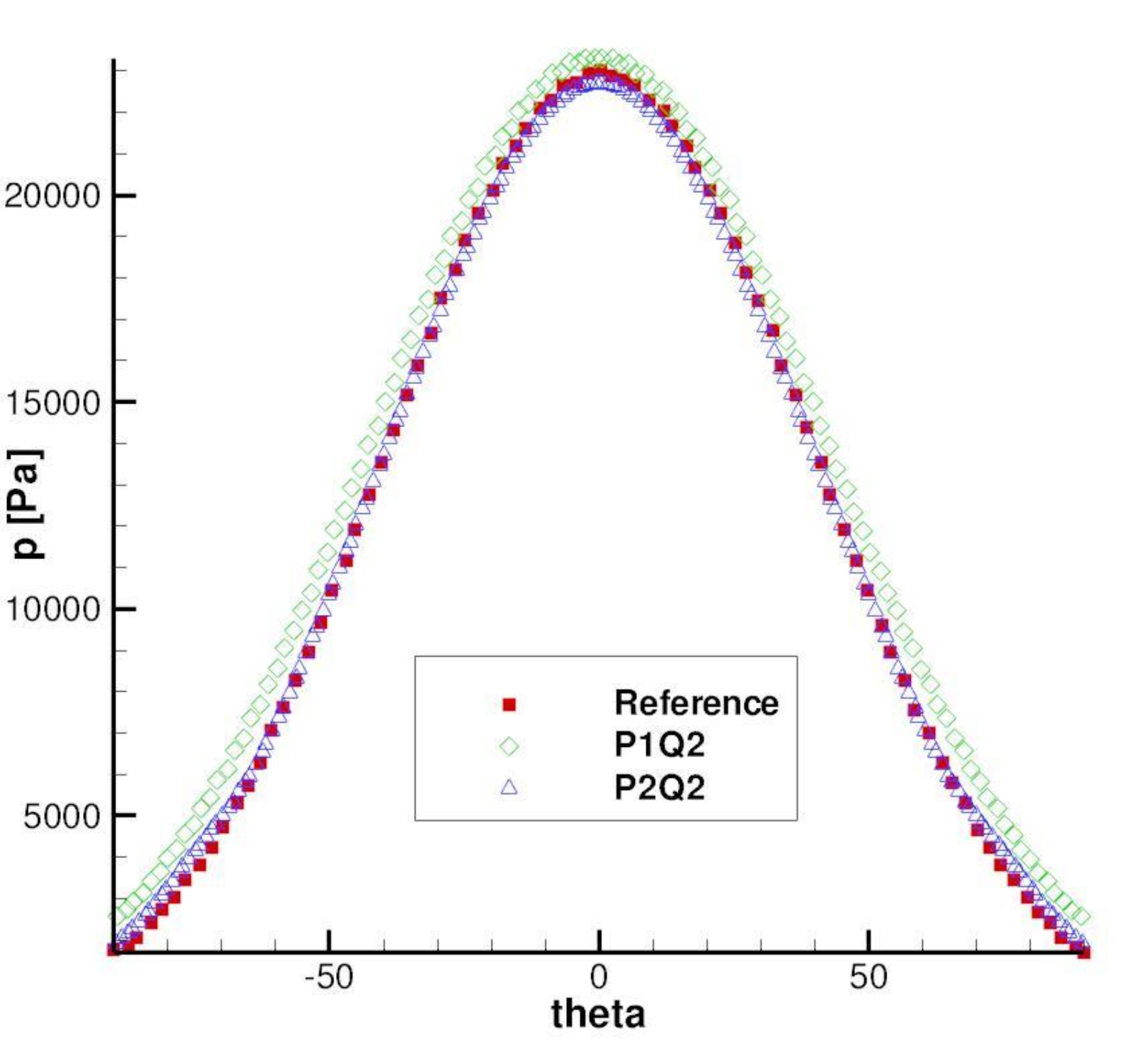}
  \caption{Wall pressure for the hypersonic cylinder test case}
  \label{fig:wallP}
\end{figure}

The pressure contours and artificial viscosity field for the P2Q2 solution are presented in figure \ref{fig:flowfield}. It is clear that the LLAV method is only active in the shock area. Figure \ref{fig:wallP} shows the wall pressure of the P1Q2 and P2Q2 solution compared to the pressure found in \cite{gnoffo2004computational}. For P1 the pressure is overestimated, while for P2 the pressure coincides with the reference.

\begin{figure}[H]
    \centering
    \begin{subfigure}[b]{0.163\textwidth}
        \includegraphics[width=\textwidth]{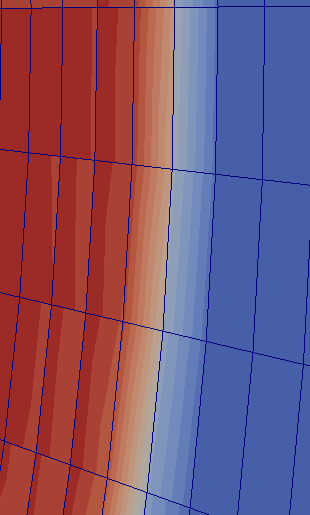}
        \caption{$c=0.05$}
        \label{fig:2}
    \end{subfigure}
    \quad
    \begin{subfigure}[b]{0.196\textwidth}
        \includegraphics[width=\textwidth]{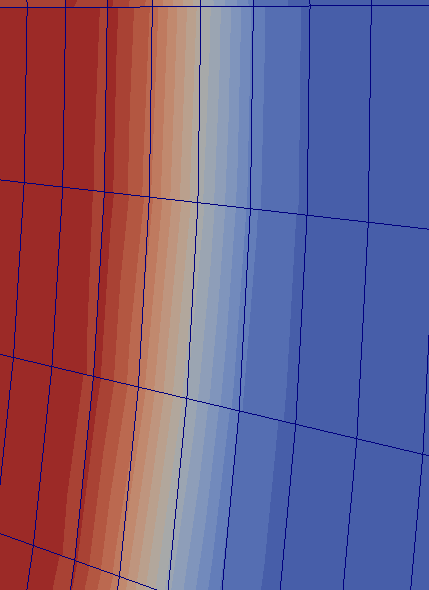}
        \caption{$c=0.1$}
        \label{fig:4}
    \end{subfigure}
    \quad
    \begin{subfigure}[b]{0.347\textwidth}
        \includegraphics[width=\textwidth]{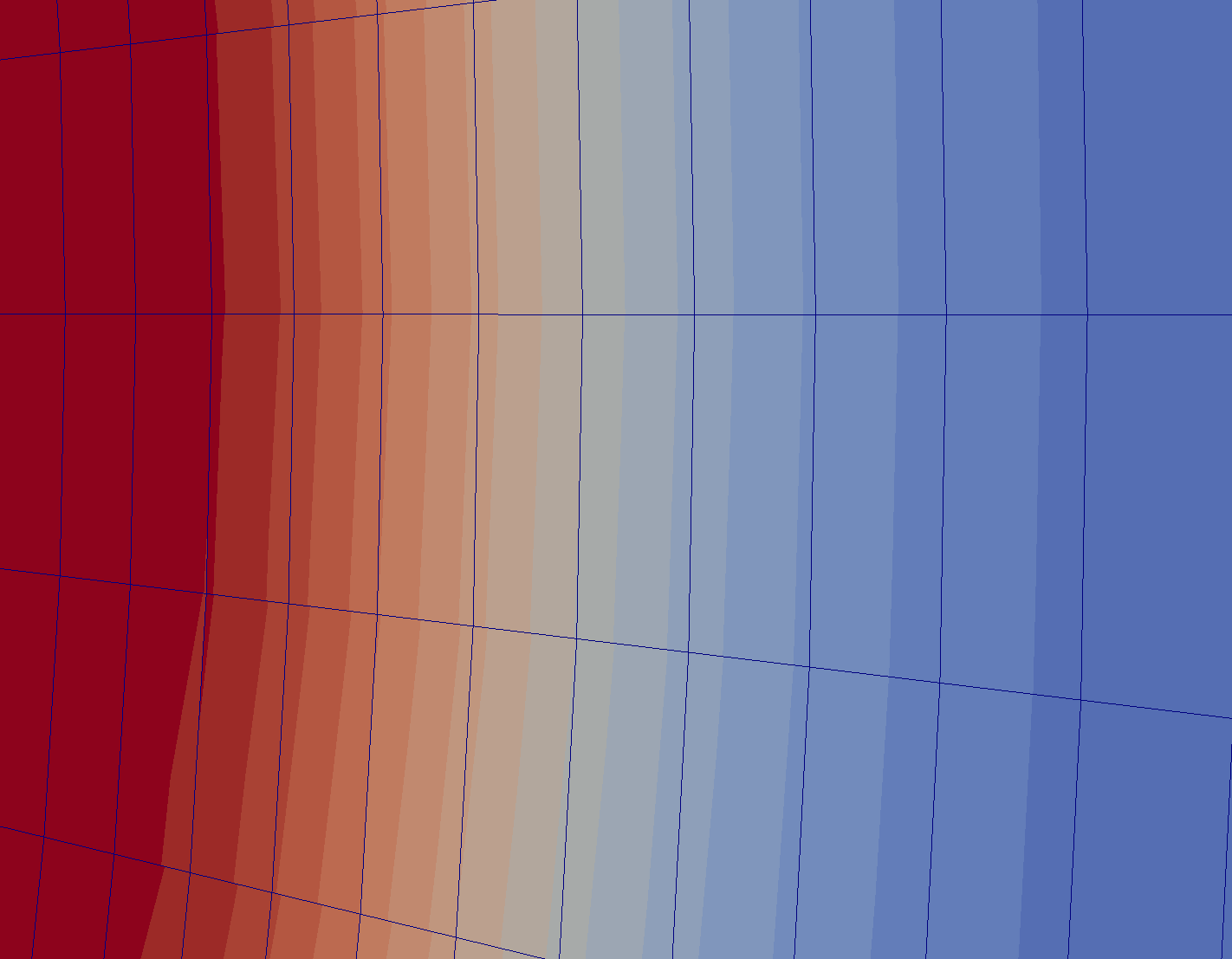}
        \caption{$c=0.2$}
        \label{fig:8}
    \end{subfigure}
    \caption{Shock thickness for different values of $c$, density $\rho$ is shown}
    \label{fig:thickness}
\end{figure}

Figure \ref{fig:thickness} shows the influence of the LLAV parameter $c$ on the computed shock thickness. This result is used to calibrate the relation between $c$ and $\delta_s^*$. From equation \ref{deltaCal}:
\begin{equation}\label{thicknessEq}
\alpha = \dfrac{c\ M_\infty}{\delta_s^* (M_\infty-1)}
\end{equation}
For $c=0.2$, the shock is spread over about 8 elements, as shown in figure \ref{fig:8}. Substituting $c=0.2$ and $\delta_s^* = 8$ in equation \ref{thicknessEq}, gives $\alpha = 0.027$. Figure \ref{fig:4} shows that for $c=0.1$, the shock is spread over about 4 elements, which gives the same result for $\alpha$. For $c=0.05$ the shock is spread over about two elements, as shown in figure \ref{fig:2}, which again confirms $\alpha = 0.027$. However, for this last value of $c$, the positivity preservation scheme remains active until convergence.

\subsection{Viscous Hypersonic Flow over a Cone in 2D}\label{hypersonicCone}

The second test case considered to test the LLAV shock capturing scheme is the flow of nitrogen gas over a rounded cone. The considered geometry consists of a $25\degree$ half-angle cone with a rounded nose of radius 1mm. The total length of the cone in the $x$-direction considered is 9cm. The free-stream values imposed are presented in table \ref{tab:freeStreamHyperCone}. This corresponds to Run 35 conditions as presented in \cite{candler2002cfd}.

\begin{table}[H]
\caption{\label{tab:freeStreamHyperCone} Free-stream values}
\centering
\begin{tabular}{ccccc}
\hline
$M_\infty$ & $\rho_\infty\ [kg/m^3]$ & $V_\infty\ [m/s]$ & $T_\infty\ [K]$ & $T_{wall}\ [K]$\\\hline
11.3 & 0.5515 & 2713 & 138.9 & 296\\
\hline
\end{tabular}
\end{table}

\begin{figure}[H]
  \centering
  \includegraphics[width=0.4\textwidth]{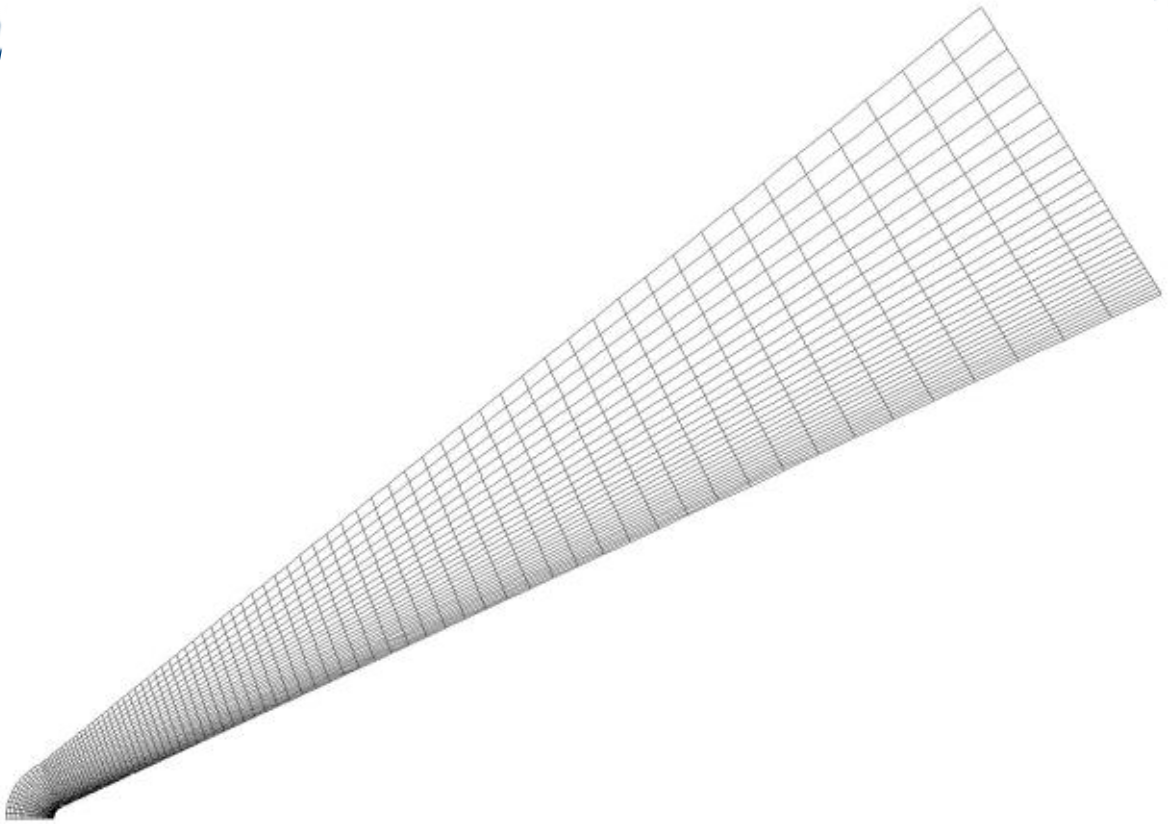}
  \caption{Mesh used for the hypersonic cone test case}
  \label{fig:meshCone}
\end{figure}

The mesh used is presented in figure \ref{fig:meshCone}. The mesh consists of curved quadrilateral elements of second-order and contains 2000 elements. The inlet boundary is imposed as a Dirichlet boundary condition that sets the states to the free-stream values. For the outlet, a supersonic outlet boundary condition is used. The cone is represented by an iso-thermal no-slip wall.

The FR method is used with a VCJH scheme for the correction functions. The distribution of both the solution and flux points is the Gauss-Legendre distribution. For the convective interface fluxes, the AUSM$^+$ scheme is used. For the LLAV scheme, $\kappa$ is set to 2. In order to simulate the flow over a cone on a 2D mesh, axisymmetric source terms are added to the Navier-Stokes equations. \ref{appendix} describes in detail how these source terms are discretized using the FR method and how they are evaluated for the axisymmetric case.

\begin{figure}[H]
    \centering
    \begin{subfigure}[b]{0.4\textwidth}
        \includegraphics[width=\textwidth]{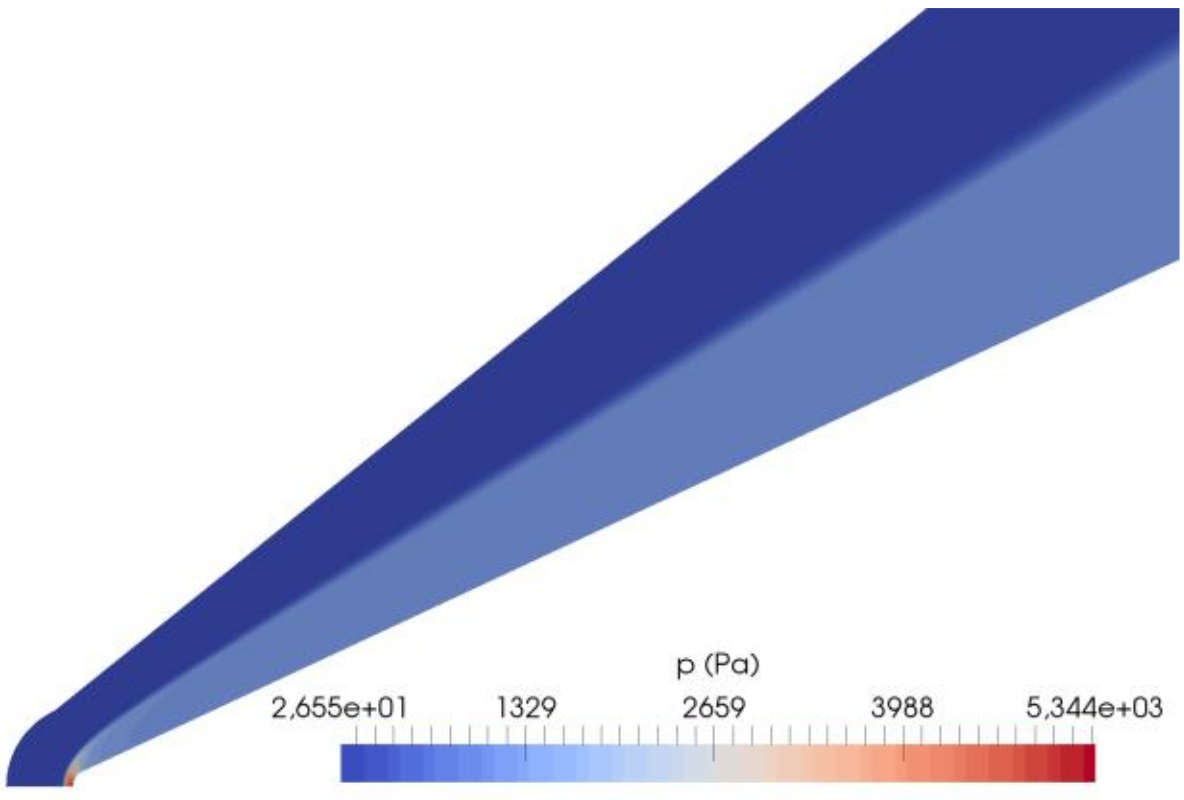}
        \caption{Pressure}
        \label{fig:flowFieldCone}
    \end{subfigure}
    \begin{subfigure}[b]{0.4\textwidth}
        \includegraphics[width=\textwidth]{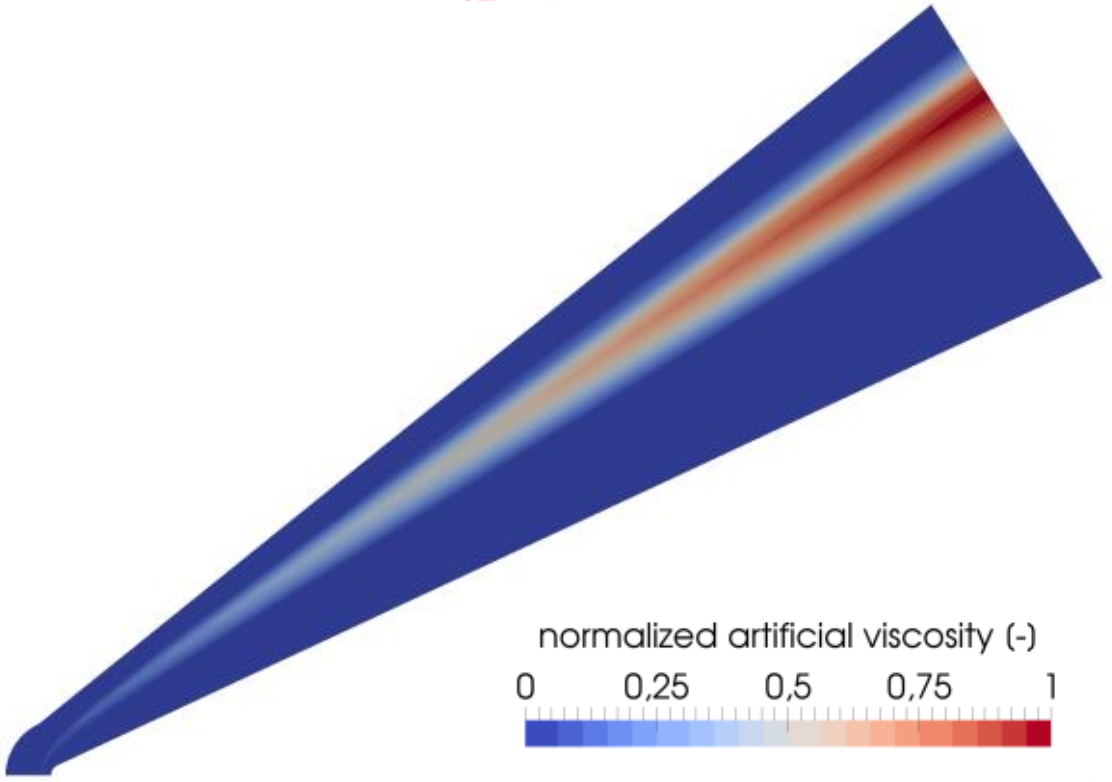}
        \caption{Artificial viscosity}
        \label{fig:artViscCone}
    \end{subfigure}
    \caption{Flow field of the hypersonic cone test case for fourth-order of accuracy}
\end{figure}

The pressure contours are presented in figure \ref{fig:flowFieldCone}. Figure \ref{fig:artViscCone} shows the artificial viscosity distribution. It is clear once again that the LLAV method is only active in the shock area.

\begin{figure}[H]
  \centering
  \includegraphics[width=0.35\textwidth]{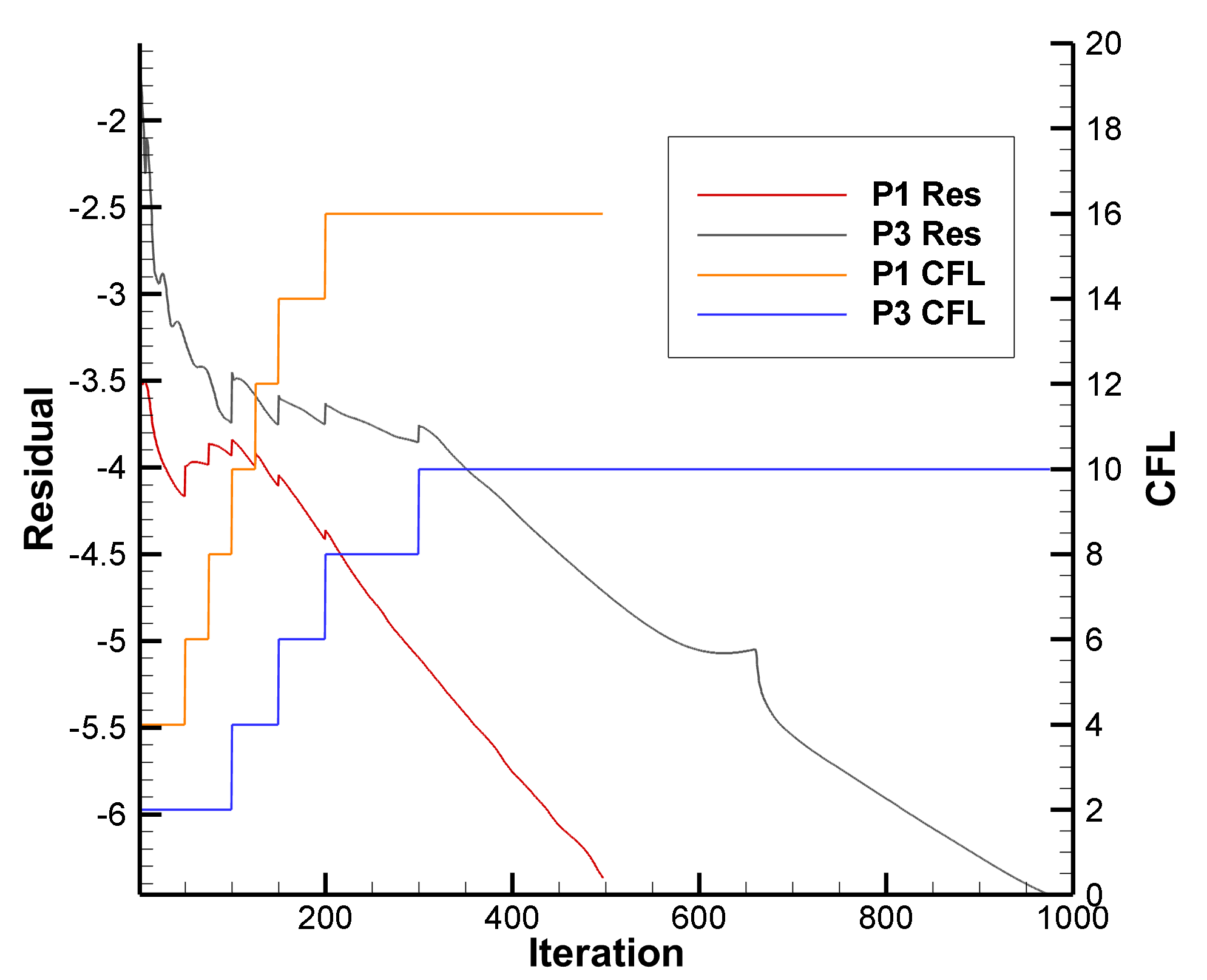}
  \caption{Logarithm of the density residual and CFL number as a function of the iteration for the P1 and P3 cone test case}
  \label{fig:CFLCone}
\end{figure}

Figure \ref{fig:CFLCone} presents the density residual history and the CFL number for both the P1 and P3 cases. For hypersonic test cases the CFL number is chosen smaller than for the previous subsonic cases in order to preserve stability. However, for both P1 and P3 the CFL number is still much larger than the CFL limit explicit time marching methods would impose: $CFL_{max} = 0.5$ and $CFL_{max}=0.25$ for respectively P1 and P3. For the P3 case it should be noted that at iteration 660 the AV field was frozen ($\varepsilon$ is no longer recomputed at each iteration) in order to speed up convergence.

\begin{figure}[H]
    \centering
    \begin{subfigure}[b]{0.32\textwidth}
        \includegraphics[width=\textwidth]{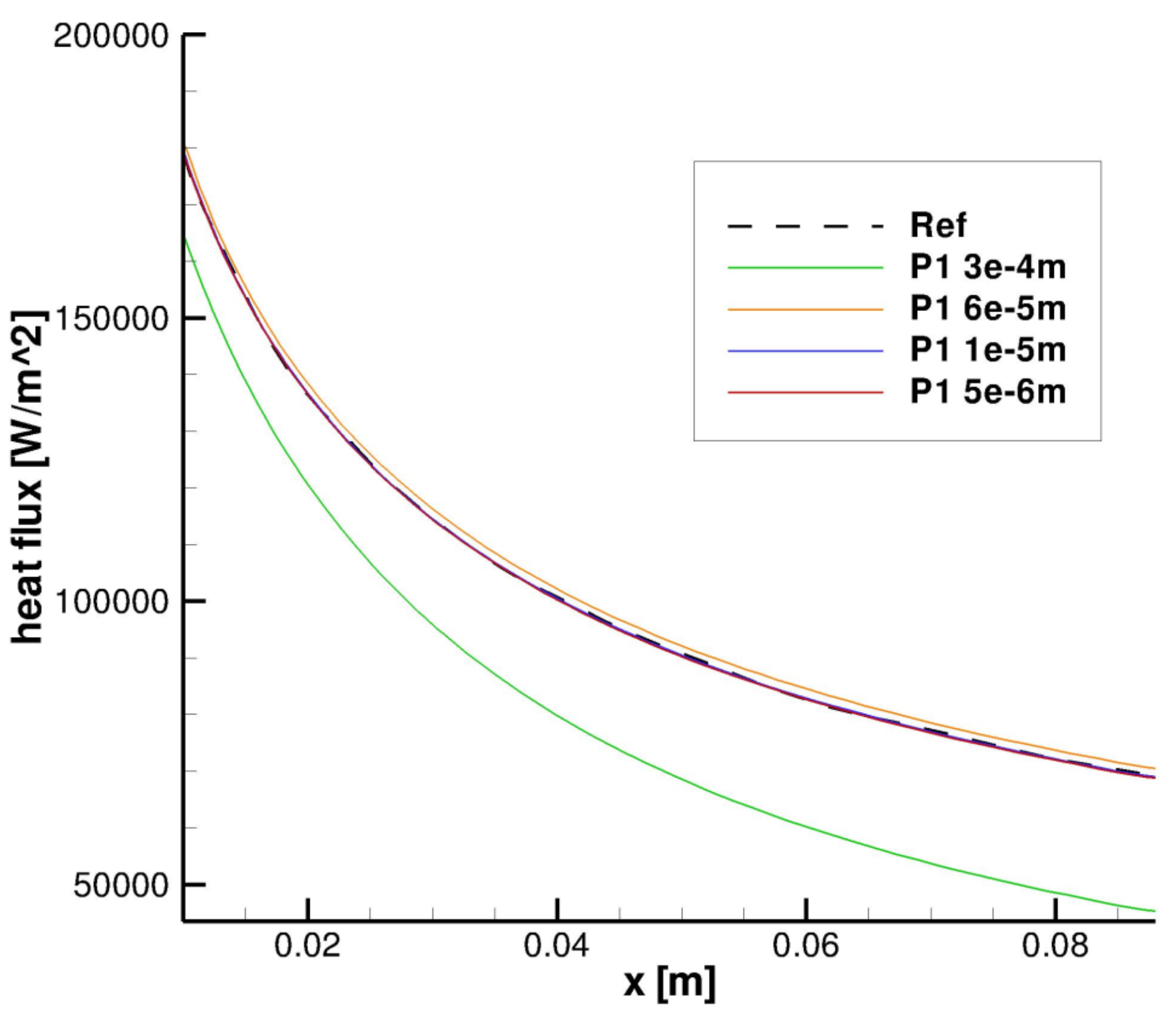}
        \caption{Different first cell height (P1)}
        \label{fig:HFConv}
    \end{subfigure}
    \begin{subfigure}[b]{0.35\textwidth}
        \includegraphics[width=\textwidth]{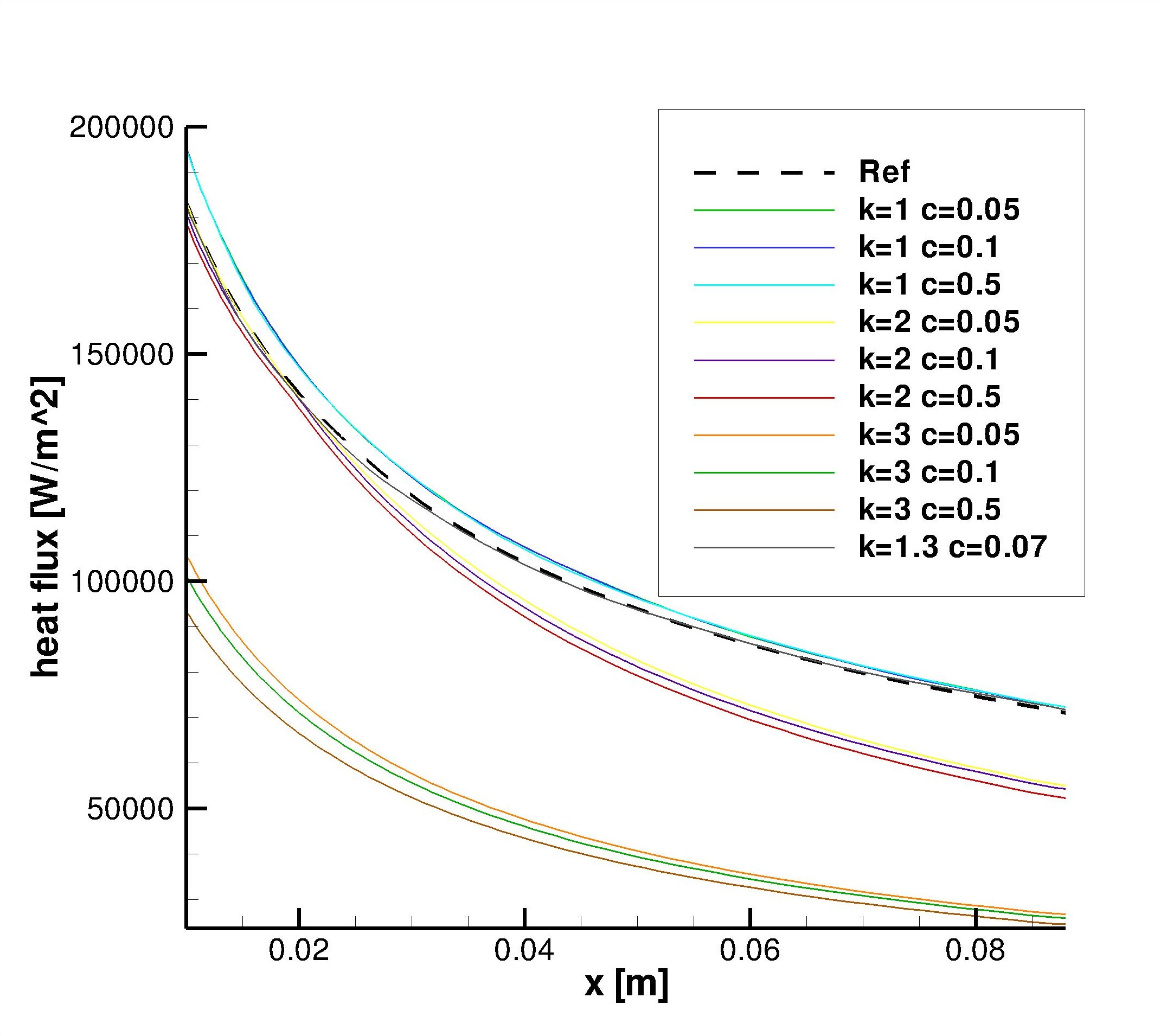}
        \caption{Different settings of LLAV (P1)}
        \label{fig:HFInfl}
    \end{subfigure}
    \begin{subfigure}[b]{0.315\textwidth}
    \includegraphics[width=\textwidth]{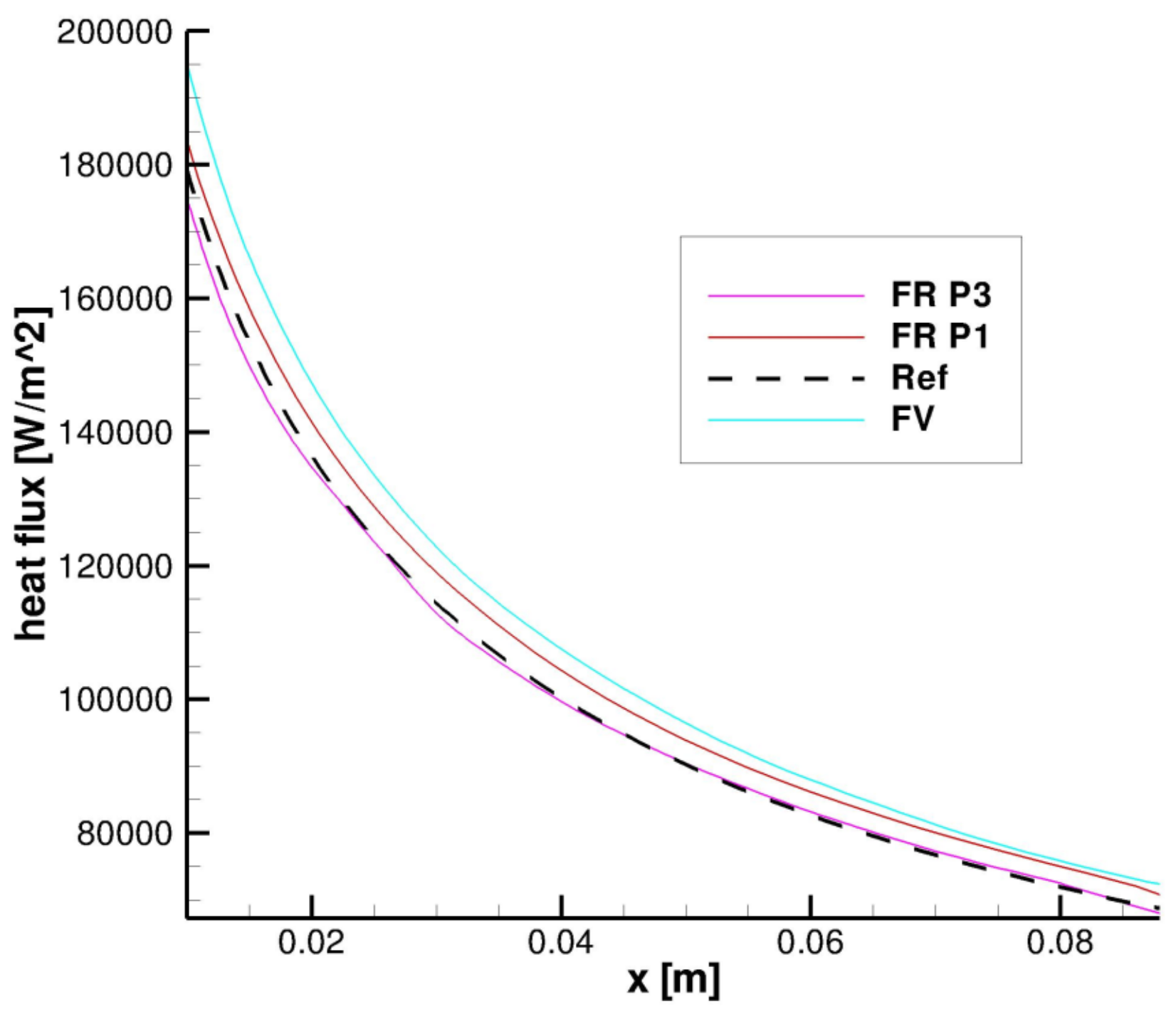}
    \caption{FR P1, P3 and FV (same DoFs)}
    \label{fig:HFIComp}
    \end{subfigure}
    \caption{Heat flux for different cases}
\end{figure}

Figure \ref{fig:HFConv} shows the heat flux to the surface of the cone for different heights of the first cell near the wall. This was done for P1 solutions. As the mesh gets finer near the wall, the heat flux coincides with the reference heat flux. The reference was obtained as the grid-converged heat flux solution for the second-order FV solver within COOLFluiD, which has been well validated for hypersonic test cases \cite{lani03}. As such it is clear that if the mesh is fine enough, the FR solver computes the correct heat flux. 

Figure \ref{fig:HFInfl} shows the influence of the LLAV $\kappa$- and $c$-parameters on the heat flux. It is clear that if $\kappa$ is increased, the heat flux that is found decreases dramatically. This is due to the fact that for large values of $\kappa$ AV is added outside of the shock area, and mainly in the boundary layer where strong gradients can be found. This causes a tendency to ``smear out'' the boundary layer, drastically decreasing the heat flux. The values for the $c$-parameter considered: (0.05;0.1;0.5), correspond to values of the characteristic shock thickness $\delta^*_s$ of respectively (2;4;20). For increasing the $c$-parameter, the same effect is found but much less dramatically. When $c$ is increased, more AV is added, as $\varepsilon_0$ scales with $c$, which causes the heat flux to drop slightly. Furthermore, it is possible to retrieve almost exactly the reference heat flux in this P1 test case with a specific set of ($\kappa$, $c$)-values: $\kappa=1.3$ and $c=0.07$. In that case the overestimation of the heat flux is countered by the spurious artificial viscosity. In this way these two errors cancel each other out. Finally, figure \ref{fig:HFIComp} shows three cases with the same amount of degrees of freedom, namely 8000 (for P1 the mesh in figure \ref{fig:meshCone} is used).  These cases are a P1 FR solution, a P3 FR solution and a second-order FV solution. Despite having the same amount of degrees of freedom, the P3 FR solution is the most accurate. This illustrates the higher computational efficiency per degree of freedom of high-order methods.

\section{Conclusion}
The implementation of a novel FR solver has been presented in detail in the present paper. The solver handles 2D and 3D flows governed by the Euler or Navier-Stokes equations on unstructured meshes of quadrilaterals or hexahedra. The solver is fully implicit and parallel. In particular, a novel positivity preserving shock capturing scheme based on the LLAV approach has been developed and implemented, allowing the method to tackle high-speed flow simulations up to Mach 17.6 for the first time to the authors' knowledge. An extensive verification of the developed code has been presented for 2D, 3D, inviscid, viscous, subsonic and hypersonic test cases. As the main achievement of this work the first open-source FR solver able to handle viscous hypersonic flows has been developed.

\section{Acknowledgments}

The research of Ray Vandenhoeck is supported by SB PhD fellowship 1S19918N of the FWO.

\appendix
\section{FR Method Including the Axisymmetric Source Term}\label{appendix}
This appendix describes the discretization of source terms within the FR solver as well as the implementation of the axisymmetric source term. This source term is added to the 2D Navier-Stokes equations for the hypersonic cone test case described in section \ref{hypersonicCone}.

\subsection{Discretization of Source Terms}
Consider the advection-diffusion equations governing the fluid motion with a source term $\textbf{s}(\textbf{u})$:
\begin{equation}
\label{eq:advection-diffusion}
    \frac{\partial \textbf{u}}{\partial t} = -\boldsymbol{\nabla} \cdot \textbf{f}(\textbf{u}) + \textbf{s}(\textbf{u}).
\end{equation}
Applying the space discretization as described in section \ref{sec:FR}, equation \ref{eq:advection-diffusion} is considered on the $i$-th solution point of an element of the computational domain:
\begin{equation}
\label{eq:decFR}
        \left(\frac{\partial \textbf{u}}{\partial t}\right)_i = - \left(\nabla \cdot \textbf{f}(\textbf{u})\right)_i + \left(\textbf{s}(\textbf{u})\right)_i.
\end{equation}
The approximate state vector, $\textbf{u}^{\delta}$ based on a Lagrange polynomial basis, is used to approximate $\textbf{u}$ in equation \ref{eq:decFR}:
\begin{equation}
\label{eq:decFR1}
        \left(\frac{\partial \textbf{u}^{\delta}}{\partial t}\right)_i = - \left(\nabla \cdot \textbf{f}(\textbf{u}^{\delta})\right)_i + \left(\textbf{s}(\textbf{u}^{\delta})\right)_i.
\end{equation}
Transforming the equation \ref{eq:decFR1} into the standard computational domain, using equation \ref{eq:ConservationLawTransform}, the following is found:
\begin{equation}
\label{eq:decHatFr}
        \left(\frac{\partial \hat{\textbf{u}}^{\delta}}{\partial t}\right)_i = - \left(\nabla \cdot \textbf{f}(\hat{\textbf{u}}^{\delta})\right)_i +\underline{\left(\textbf{s}(\hat{\textbf{u}}^{\delta})\right)_i}.
\end{equation}
The discretization of the source terms, i.e. the underlined term in equation \ref{eq:decHatFr}, is computed under the following assumption, which expresses that the source term can simply be computed based on the local value of the state:
\begin{equation}
\label{eq:decHatFrFinalVersion}
    \left(\textbf{s}(\hat{\textbf{u}}^{\delta})\right)_i = \textbf{s}(\hat{\textbf{u}}_i^{\delta}).
\end{equation}
Consequently, the FR solver computes the source terms using the approximate state vector $\hat{\textbf{u}}_i^{\delta}$ and adds them to the RHS of the equation \ref{eq:advection-diffusion}. In practise the source term is evaluated in each solution point using the approximate state polynomial in that solution point and added to the RHS in that solution point. For the jacobian, the same numerical flux approach is used as for the convective and diffusive flux jacobian. 

\subsection{Axisymmetric Source Term}

The first formulation presented in \cite{lani2008object} is used to compute axisymmetric test cases. In this formulation, the convective and diffusive fluxes remain the unchanged. However, a source term is added to the conservative form of the equations (\ref{cons law}) that has the following form:
\begin{equation}
    \mathbf{s} = (0,0,p-\tau_{\theta\theta},0)^T,
\end{equation}
where $\tau_{\theta\theta}$ is the viscous stress component in the circumferential direction $\theta$. The pressure is evaluated using expression \ref{pressure}. $\tau_{\theta\theta}$ is computed using the following relation:
\begin{equation}\label{axist}
    \tau_{\theta\theta} = -\frac{2}{3}\mu\left(\dfrac{\partial v_x}{\partial x} + \dfrac{\partial v_r}{\partial r} - 2\frac{v_r}{r}\right),
\end{equation}
where $r$ is the radius, which is chosen equal to the $y$-coordinate. The partial derivatives in equation \ref{axist} can simply be computed using the corrected gradient $\mathbf{q}^{\delta}$. The corrected gradient is calculated for the diffusive flux evaluation and as such is already known when evaluating the axisymmetric source term. 

\bibliography{CPC_FR}

\begin{thebibliography}{10}
\expandafter\ifx\csname url\endcsname\relax
  \def\url#1{\texttt{#1}}\fi
\expandafter\ifx\csname urlprefix\endcsname\relax\def\urlprefix{URL }\fi
\expandafter\ifx\csname href\endcsname\relax
  \def\href#1#2{#2} \def\path#1{#1}\fi

\bibitem{sheshadri2016analysis}
A.~Sheshadri, An analysis of stability of the flux reconstruction formulation
  with applications to shock capturing, Ph.D. thesis, Stanford University
  (2016).

\bibitem{vincent2011facilitating}
P.~E. Vincent, A.~Jameson, Facilitating the adoption of unstructured high-order
  methods amongst a wider community of fluid dynamicists, Mathematical
  Modelling of Natural Phenomena 6~(3) (2011) 97--140.

\bibitem{huynh2014high}
H.~T. Huynh, Z.~J. Wang, P.~E. Vincent, High-order methods for computational
  fluid dynamics: a brief review of compact differential formulations on
  unstructured grids, Computers \& Fluids 98 (2014) 209--220.

\bibitem{geuzaine2009gmsh}
C.~Geuzaine, J.-F. Remacle, Gmsh: A {3-D} finite element mesh generator with
  built-in pre-and post-processing facilities, International journal for
  numerical methods in engineering 79~(11) (2009) 1309--1331.

\bibitem{lele1992compact}
S.~K. Lele, Compact finite difference schemes with spectral-like resolution,
  Journal of computational physics 103~(1) (1992) 16--42.

\bibitem{pulliam2011high}
T.~Pulliam, High order accurate finite-difference methods: as seen in
  {OVERFLOW}, in: 20th AIAA Computational Fluid Dynamics Conference, 2011, p.
  3851.

\bibitem{henshaw2006moving}
W.~D. Henshaw, D.~W. Schwendeman, Moving overlapping grids with adaptive mesh
  refinement for high-speed reactive and non-reactive flow, Journal of
  Computational Physics 216~(2) (2006) 744--779.

\bibitem{lani2013variable}
A.~Lani, B.~Sj{\"o}green, H.~C. Yee, W.~D. Henshaw, Variable high-order
  multiblock overlapping grid methods for mixed steady and unsteady multiscale
  viscous flows, part {II}: hypersonic nonequilibrium flows, Communications in
  Computational Physics 13~(2) (2013) 583--602.

\bibitem{barth1990higher}
T.~Barth, P.~Frederickson, Higher order solution of the {Euler} equations on
  unstructured grids using quadratic reconstruction, in: 28th aerospace
  sciences meeting, 1990, p.~13.

\bibitem{barth2013high}
T.~J. Barth, H.~Deconinck, High-order methods for computational physics,
  Vol.~9, Springer Science \& Business Media, 2013.

\bibitem{delanaye1999quadratic}
M.~Delanaye, Y.~Liu, Quadratic reconstruction finite volume schemes on {3D}
  arbitrary unstructured polyhedral grids, in: 14th Computational Fluid
  Dynamics Conference, 1999, p. 3259.

\bibitem{harten1987uniformly}
A.~Harten, B.~Engquist, S.~Osher, S.~R. Chakravarthy, Uniformly high order
  accurate essentially non-oscillatory schemes, {III}, Journal of computational
  physics 71~(2) (1987) 231--303.

\bibitem{abgrall1994essentially}
R.~Abgrall, On essentially non-oscillatory schemes on unstructured meshes:
  analysis and implementation, Journal of Computational Physics 114~(1) (1994)
  45--58.

\bibitem{ollivier1997quasi}
C.~F. Ollivier-Gooch, Quasi-{ENO} schemes for unstructured meshes based on
  unlimited data-dependent least-squares reconstruction, Journal of
  Computational Physics 133~(1) (1997) 6--17.

\bibitem{liu1994weighted}
X.-D. Liu, S.~Osher, T.~Chan, Weighted essentially non-oscillatory schemes,
  Journal of computational physics 115~(1) (1994) 200--212.

\bibitem{hu1999weighted}
C.~Hu, C.-W. Shu, Weighted essentially non-oscillatory schemes on triangular
  meshes, Journal of Computational Physics 150~(1) (1999) 97--127.

\bibitem{friedrich1998weighted}
O.~Friedrich, Weighted essentially non-oscillatory schemes for the
  interpolation of mean values on unstructured grids, Journal of computational
  physics 144~(1) (1998) 194--212.

\bibitem{reed1973triangularmesh}
W.~H. Reed, T.~Hill, Triangular mesh methods for the neutron transport
  equation, Los Alamos Report LA-UR-73-479.

\bibitem{cockburn1998local}
B.~Cockburn, C.-W. Shu, The local discontinuous {Galerkin} method for
  time-dependent convection-diffusion systems, SIAM Journal on Numerical
  Analysis 35~(6) (1998) 2440--2463.

\bibitem{peraire2008compact}
J.~Peraire, P.-O. Persson, The compact discontinuous {Galerkin} ({CDG}) method
  for elliptic problems, SIAM Journal on Scientific Computing 30~(4) (2008)
  1806--1824.

\bibitem{cockburn2009unified}
B.~Cockburn, J.~Gopalakrishnan, R.~Lazarov, Unified hybridization of
  discontinuous {Galerkin}, mixed, and continuous {Galerkin} methods for second
  order elliptic problems, SIAM Journal on Numerical Analysis 47~(2) (2009)
  1319--1365.

\bibitem{kopriva1996conservative}
D.~A. Kopriva, J.~H. Kolias, A conservative staggered-grid {Chebyshev}
  multidomain method for compressible flows, Journal of computational physics
  125~(1) (1996) 244--261.

\bibitem{liu2006spectral}
Y.~Liu, M.~Vinokur, Z.~Wang, Spectral difference method for unstructured grids
  {I}: basic formulation, Journal of Computational Physics 216~(2) (2006)
  780--801.

\bibitem{liang2009spectral}
C.~Liang, A.~Jameson, Z.~J. Wang, Spectral difference method for compressible
  flow on unstructured grids with mixed elements, Journal of Computational
  Physics 228~(8) (2009) 2847--2858.

\bibitem{liang2009large}
C.~Liang, S.~Premasuthan, A.~Jameson, Z.~Wang, Large eddy simulation of
  compressible turbulent channel flow with spectral difference method, in: 47th
  AIAA Aerospace Sciences Meeting Including the New Horizons Forum and
  Aerospace Exposition, 2009, p. 402.

\bibitem{huynh2007flux}
H.~T. Huynh, A flux reconstruction approach to high-order schemes including
  discontinuous {Galerkin} methods, AIAA paper 4079 (2007) 2007.

\bibitem{huynh2009reconstruction}
H.~T. Huynh, A reconstruction approach to high-order schemes including
  discontinuous {Galerkin} for diffusion, AIAA paper 403 (2009) 2009.

\bibitem{huynh2013flux}
H.~T. Huynh, Flux reconstruction / correction procedure via reconstruction, a
  unified approach to high-order accuracy, 37th Advanced VKI CFD Lecture
  Series: Recent Developments in Higher Order Methods and Industrial
  Application in Aeronautics.

\bibitem{allaneau2011connections}
Y.~Allaneau, A.~Jameson, Connections between the filtered discontinuous
  {Galerkin} method and the flux reconstruction approach to high order
  discretizations, Computer Methods in Applied Mechanics and Engineering
  200~(49) (2011) 3628--3636.

\bibitem{de2014connections}
D.~De~Grazia, G.~Mengaldo, D.~Moxey, P.~Vincent, S.~Sherwin, Connections
  between the discontinuous {Galerkin} method and high-order flux
  reconstruction schemes, International journal for numerical methods in fluids
  75~(12) (2014) 860--877.

\bibitem{zwanenburg2016equivalence}
P.~Zwanenburg, S.~Nadarajah, Equivalence between the energy stable flux
  reconstruction and filtered discontinuous {Galerkin} schemes, Journal of
  Computational Physics 306 (2016) 343--369.

\bibitem{lopez2014verification}
M.~R. L{\'o}pez, A.~Sheshadri, J.~R. Bull, T.~D. Economon, J.~Romero, J.~E.
  Watkins, D.~M. Williams, F.~Palacios, A.~Jameson, D.~E. Manosalvas,
  Verification and validation of {HiFiLES}: a high-order {LES} unstructured
  solver on multi-{GPU} platforms, in: 32nd AIAA Applied Aerodynamics
  Conference, 2014, p. 3168.

\bibitem{vincent2015pyfr}
P.~Vincent, F.~D. Witherden, A.~M. Farrington, G.~Ntemos, B.~C. Vermeire, J.~S.
  Park, A.~S. Iyer, {PyFR}: Next-generation high-order computational fluid
  dynamics on many-core hardware, in: 22nd AIAA Computational Fluid Dynamics
  Conference, 2015, p. 3050.

\bibitem{witherden2015heterogeneous}
F.~D. Witherden, B.~C. Vermeire, P.~E. Vincent, Heterogeneous computing on
  mixed unstructured grids with {PyFR}, Computers \& Fluids 120 (2015)
  173--186.

\bibitem{huynh2011high}
H.~T. Huynh, High-order methods including discontinuous {Galerkin} by
  reconstructions on triangular meshes, AIAA Paper 44 (2011) 2011.

\bibitem{castonguay2012new}
P.~Castonguay, P.~E. Vincent, A.~Jameson, A new class of high-order energy
  stable flux reconstruction schemes for triangular elements, Journal of
  Scientific Computing 51~(1) (2011) 224--256.

\bibitem{gao2009high}
H.~Gao, Z.~Wang, A high-order lifting collocation penalty formulation for the
  {Navier-Stokes} equations on {2D} mixed grids, ratio 1 (2009) 2.

\bibitem{wang2009unifying}
Z.~Wang, H.~Gao, A unifying lifting collocation penalty formulation including
  the discontinuous {Galerkin}, spectral volume/difference methods for
  conservation laws on mixed grids, Journal of Computational Physics 228~(21)
  (2009) 8161--8186.

\bibitem{yu2013connection}
M.~Yu, Z.~Wang, On the connection between the correction and weighting
  functions in the correction procedure via reconstruction method, Journal of
  Scientific Computing 54~(1) (2013) 227--244.

\bibitem{haga2011high}
T.~Haga, H.~Gao, Z.~J. Wang, A high-order unifying discontinuous formulation
  for the {Navier-Stokes} equations on {3D} mixed grids, Mathematical Modelling
  of Natural Phenomena 6~(3) (2011) 28--56.

\bibitem{haga2010high}
T.~Haga, H.~Gao, Z.~Wang, A high-order unifying discontinuous formulation for
  {3D} mixed grids, AIAA paper 540 (2010) 2010.

\bibitem{vincent2011new}
P.~E. Vincent, P.~Castonguay, A.~Jameson, A new class of high-order energy
  stable flux reconstruction schemes, Journal of Scientific Computing 47~(1)
  (2011) 50--72.

\bibitem{jameson2010proof}
A.~Jameson, A proof of the stability of the spectral difference method for all
  orders of accuracy, Journal of Scientific Computing 45~(1-3) (2010) 348--358.

\bibitem{castonguay2011development}
P.~Castonguay, D.~M. Williams, P.~E. Vincent, M.~Lopez, A.~Jameson, On the
  development of a high-order, multi-{GPU} enabled, compressible viscous flow
  solver for mixed unstructured grids, AIAA paper 3229 (2011) 2011.

\bibitem{castonguay2011application}
P.~Castonguay, P.~Vincent, A.~Jameson, Application of high-order energy stable
  flux reconstruction schemes to the {Euler} equations, in: 49th AIAA Aerospace
  Sciences Meeting including the New Horizons Forum and Aerospace Exposition,
  2011, p. 686.

\bibitem{ou2011high}
K.~Ou, P.~Vincent, A.~Jameson, High-order methods for diffusion equation with
  energy stable flux reconstruction scheme, in: 49th AIAA Aerospace Sciences
  Meeting Including the New Horizons Forum and Aerospace Exposition, 2011, pp.
  4--7.

\bibitem{williams2011extension}
D.~M. Williams, P.~Castonguay, P.~E. Vincent, A.~Jameson, An extension of
  energy stable flux reconstruction to unsteady, non-linear, viscous problems
  on mixed grids, in: 20th AIAA Computational Fluid Dynamics Conference, 2011,
  pp. 27--30.

\bibitem{lani13}
A.~Lani, N.~Villedie, K.~Bensassi, L.~Koloszar, M.~Vymazal, S.~M. Yalim,
  M.~Panesi, {COOLFluiD}: an open computational platform for multi-physics
  simulation and research, in: 21st AIAA Computational Fluid Dynamics
  Conference, 2013, p. 2589.

\bibitem{lani2006reusable}
A.~Lani, T.~Quintino, D.~Kimpe, H.~Deconinck, S.~Vandewalle, S.~Poedts,
  Reusable object-oriented solutions for numerical simulation of {PDE}s in a
  high performance environment, Scientific Programming 14~(2) (2006) 111--139.

\bibitem{lani1}
A.~Lani, T.~Quintino, D.~Kimpe, H.~Deconinck, S.~Vandewalle, S.~Poedts, The
  {COOLFluiD} framework: Design solutions for high-performance object oriented
  scientific computing software, in: V.~S. Sunderan, G.~D. van Albada, P.~M.~A.
  Sloot, J.~J. Dongarra (Eds.), Computational Science ICCS 2005, Vol.~1 of LNCS
  3514, Emory University, Springer, Atlanta, GA, USA, 2005, pp. 281--286.

\bibitem{bassi1997high}
F.~Bassi, S.~Rebay, A high-order accurate discontinuous finite element method
  for the numerical solution of the compressible {Navier-Stokes} equations,
  Journal of computational physics 131~(2) (1997) 267--279.

\bibitem{roe1981approximate}
P.~L. Roe, Approximate riemann solvers, parameter vectors, and difference
  schemes, Journal of computational physics 43~(2) (1981) 357--372.

\bibitem{sheshadri2014shock}
A.~Sheshadri, A.~Jameson, Shock detection and capturing methods for high order
  discontinuous-{Galerkin} finite element methods, in: 32nd AIAA Applied
  Aerodynamics Conference, 2014, p. 2688.

\bibitem{park2014higher}
J.~S. Park, T.~K. Chang, C.~Kim, Higher-order multi-dimensional limiting
  strategy for correction procedure via reconstruction, in: 52nd Aerospace
  Sciences Meeting, 2014, p. 0772.

\bibitem{du2015simple}
J.~Du, C.-W. Shu, M.~Zhang, A simple weighted essentially non-oscillatory
  limiter for the correction procedure via reconstruction ({CPR}) framework,
  Applied Numerical Mathematics 95 (2015) 173--198.

\bibitem{yu2015localized}
M.~Yu, F.~X. Giraldo, M.~Peng, Z.~J. Wang, Localized artificial viscosity
  stabilization of discontinuous {Galerkin} methods for nonhydrostatic
  mesoscale atmospheric modeling, Monthly Weather Review 143~(12) (2015)
  4823--4845.

\bibitem{li2017convergent}
Y.~Li, Z.~J. Wang, A convergent and accuracy preserving limiter for the
  {FR/CPR} method, in: 55th AIAA Aerospace Sciences Meeting, 2017, p. 0756.

\bibitem{persson2006sub}
P.-O. Persson, J.~Peraire, Sub-cell shock capturing for discontinuous
  {Galerkin} methods, in: 44th AIAA Aerospace Sciences Meeting and Exhibit,
  2006, p. 112.

\bibitem{park2014comparative}
J.~S. Park, M.~Yu, C.~Kim, Z.~Wang, Comparative study of shock-capturing
  methods for high-order {CPR}: {MLP} and artificial viscosity, Proceedings of
  the 8th International Conference on CFD, ICCFD8-2014-0067.

\bibitem{zhang2017positivity}
X.~Zhang, On positivity-preserving high order discontinuous {Galerkin} schemes
  for compressible {Navier-Stokes} equations, Journal of Computational Physics
  328 (2017) 301--343.

\bibitem{knight2017assessment}
D.~Knight, O.~Chazot, J.~Austin, M.~A. Badr, G.~Candler, B.~Celik, D.~de~Rosa,
  R.~Donelli, J.~Komives, A.~Lani, et~al., Assessment of predictive
  capabilities for aerodynamic heating in hypersonic flow, Progress in
  Aerospace Sciences 90 (2017) 39--53.

\bibitem{panesi07}
M.~Panesi, A.~Lani, T.~Magin, J.~Molnar, O.~Chazot, H.~Deconinck, Numerical
  investigation of the non equilibrium shock-layer around the {EXPERT} vehicle,
  in: 18th AIAA Computational Fluid Dynamics Conference, 2007, p. 4317.

\bibitem{degrez09}
G.~Degrez, A.~Lani, M.~Panesi, O.~Chazot, H.~Deconinck, Modelling of
  high-enthalpy, high-mach number flows, J. Phys. D: App. Phys 41.

\bibitem{knight12}
D.~Knight, J.~Longo, D.~Drikakis, D.~Gaitonde, A.~L. et~al., Assessment of
  {CFD} capability for prediction of hypersonic shock interactions, Prog.
  Aerosp. Sci. 48-49 (2012) 8--26.

\bibitem{munafo13}
A.~Munafo, A.~Lani, A.~Bultel, M.~Panesi, Modeling of non-equilibrium phenomena
  in expanding flows by means of a collisional-radiative model, Physics of
  Plasma 20~(7).

\bibitem{panesi13}
M.~Panesi, A.~Lani, Collisional radiative coarse-grain model for ionization in
  air, Physics of Fluids 25 (2013) 057101.

\bibitem{lani13b}
A.~Lani, P.~Duarte~Santos, A.~Sanna, An efficient {Monte Carlo} method for
  radiation transport in aerothermodynamic simulations, in: 44th AIAA
  Thermophysics Conference, 2013, p. 2893.

\bibitem{lani2008object}
A.~Lani, An object oriented and high performance platform for
  aerothermodynamics simulation, Ph.D. thesis, Universit{\'e} Libre de
  Bruxelles (2008).

\bibitem{quintino2008component}
T.~Quintino, A component environment for high-performance scientific computing:
  design and implementation, Ph.D. thesis, Ph. D. Thesis, Katholike
  Universiteit Leuven, von Karman Institute for Fluid Dynamics, Leuven (2008).

\bibitem{wang2013high}
Z.~J. Wang, K.~Fidkowski, R.~Abgrall, F.~Bassi, D.~Caraeni, A.~Cary,
  H.~Deconinck, R.~Hartmann, K.~Hillewaert, H.~T. Huynh, et~al., High-order
  {CFD} methods: current status and perspective, International Journal for
  Numerical Methods in Fluids 72~(8) (2013) 811--845.

\bibitem{van1987comparison}
B.~Van~Leer, J.~L. Thomas, P.~L. Roe, R.~W. Newsome, A comparison of numerical
  flux formulas for the {Euler} and {Navier-Stokes} equations, AIAA Paper.

\bibitem{rusanov1961calculation}
V.~V. Rusanov, Calculation of interaction of non-steady shock waves with
  obstacles, J. Comput. Math. Phys. USSR 1 (1961) 267--279.

\bibitem{liou1996sequel}
M.-S. Liou, A sequel to {AUSM}: {AUSM}+, Journal of computational Physics
  129~(2) (1996) 364--382.

\bibitem{liou2006sequel}
M.-S. Liou, A sequel to {AUSM}, part {II}: {AUSM+-up} for all speeds, Journal
  of Computational Physics 214~(1) (2006) 137--170.

\bibitem{laniPart}
A.~Lani, T.~Quintino, Design techniques for high performance multi-physics
  simulations, VKI Lecture Series on High Performance Computing of Industrial
  Flows.

\bibitem{saad1986gmres}
Y.~Saad, M.~H. Schultz, Gmres: A generalized minimal residual algorithm for
  solving nonsymmetric linear systems, SIAM Journal on scientific and
  statistical computing 7~(3) (1986) 856--869.

\bibitem{krivodonova2006high}
L.~Krivodonova, M.~Berger, High-order accurate implementation of solid wall
  boundary conditions in curved geometries, Journal of computational physics
  211~(2) (2006) 492--512.

\bibitem{bassi1997euler}
F.~Bassi, S.~Rebay, High-order accurate discontinuous finite element solution
  of the {2D} {Euler} equations, Journal of computational physics 138~(2)
  (1997) 251--285.

\bibitem{golub1969calculation}
G.~H. Golub, J.~H. Welsch, Calculation of {Gauss} quadrature rules, Mathematics
  of computation 23~(106) (1969) 221--230.

\bibitem{luo2001influence}
X.~Luo, M.~S. Shephard, J.-F. Remacle, The influence of geometric approximation
  on the accuracy of high order methods, Rensselaer SCOREC report 1.

\bibitem{van2009development}
K.~Van~den Abeele, Development of high-order accurate schemes for unstructured
  grids, Ph.D. thesis, Vrije Universiteit Brussel (2009).

\bibitem{jameson2012non}
A.~Jameson, P.~E. Vincent, P.~Castonguay, On the non-linear stability of flux
  reconstruction schemes, Journal of Scientific Computing 50~(2) (2012)
  434--445.

\bibitem{tansley2001flow}
C.~E. Tansley, D.~P. Marshall, Flow past a cylinder on a $\beta$ plane, with
  application to {Gulf Stream} separation and the {Antarctic Circumpolar
  Current}, Journal of Physical Oceanography 31~(11) (2001) 3274--3283.

\bibitem{roshko1954development}
A.~Roshko, On the development of turbulent wakes from vortex streets, National
  Advisory Committee for Aeronautics.

\bibitem{bloor1964transition}
M.~S. Bloor, The transition to turbulence in the wake of a circular cylinder,
  Journal of Fluid Mechanics 19~(02) (1964) 290--304.

\bibitem{williamson1988existence}
C.~Williamson, The existence of two stages in the transition to
  three-dimensionality of a cylinder wake, The Physics of fluids 31~(11) (1988)
  3165--3168.

\bibitem{gnoffo2004computational}
P.~Gnoffo, J.~White, Computational aerothermodynamic simulation issues on
  unstructured grids, in: 37th AIAA Thermophysics Conference, 2004, p. 2371.

\bibitem{candler2002cfd}
G.~Candler, I.~Nompelis, M.-C. Druguet, M.~Holden, T.~Wadhams, I.~Boyd, W.-L.
  Wang, {CFD} validation for hypersonic flight-hypersonic double-cone flow
  simulations, in: 40th AIAA Aerospace Sciences Meeting \& Exhibit, 2002, p.
  581.

\bibitem{lani03}
A.~Lani, Development of an object oriented framework for {PDE} solvers on
  unstructured grids, Travail r\'ealis\'e pour obtenir le {DEA} en sciences
  appliqu\'ees, Universit\'e Libre de Bruxelles, Facult\'e des Sciences
  Appliqu\'ees (Sep 2003).

\end{thebibliography}

\end{document}